\documentclass[aps,prx,twocolumn,floatfix,superscriptaddress]{revtex4-1}

\usepackage{mathrsfs}
\usepackage[export]{adjustbox} 	
\usepackage{soul}
\usepackage[colorlinks=true,citecolor=blue,breaklinks=true,linkcolor=red,linktocpage=true,urlcolor=blue,pagebackref=false]{hyperref} 

\usepackage{graphicx}	
\usepackage{dcolumn}	
\usepackage{bm}	
\usepackage{epsfig}
\usepackage{amsmath}
\usepackage{amssymb}
\usepackage{xcolor}
\usepackage{bbm}
\usepackage{braket}

\usepackage{float}
\usepackage[normalem]{ulem}
\usepackage{lineno}

\usepackage[utf8]{inputenc}
\usepackage[T1]{fontenc}

\newcommand{\addGR}[1]{\textcolor{black}{#1}}

\usepackage{pifont}

\begin{document}
\title{Entangled photon-pair emission in waveguide circuit QED from a Cooper pair splitter}
\author{Michele  Governale}
\affiliation{School of Chemical and Physical Sciences and MacDiarmid Institute for Advanced Materials and Nanotechnology, Victoria University of Wellington, PO Box 600, Wellington 6140, New Zealand}
\author{Christian Sch{\"o}nenberger}
\affiliation{Department of Physics, University of Basel, Klingelbergstrasse 82, CH-4056, Switzerland}
\author{Pasquale Scarlino}
\affiliation{Institute of Physics and Center for Quantum Science and Engineering,Ecole Polytechnique F{\'e}d{\'e}rale de Lausanne, CH-1015 Lausanne, Switzerland}
\author{Gianluca Rastelli}
\affiliation{Pitaevskii Center on Bose-Einstein Condensation, CNR-INO and Dipartimento di Fisica dell'Universit{\'a} di Trento, Via Sommarive 14, 38123 Trento, Italy}
\affiliation{INFN-TIFPA, Trento Institute for Fundamental Physics and Applications, Via Sommarive 14, I-38123 Trento, Italy}
\email{Corresponding author: gianluca.rastelli@ino.cnr.it}
\date{\today}
%
%
\begin{abstract}
As a waveguide circuit QED architecture,  we investigate \addGR{theoretically} the single-photon pair emission of a Cooper pair splitter composed of two
double quantum dots ,  each coupled to a microwave transmission line. 
We 
\addGR{find that this system can} generate frequency-entangled photon pairs in the left and right transmission lines, 
specifically a superposition of two photon wavepackets at different frequencies.
The frequency entanglement of the two photons arises from the particle-hole coherent superposition (i.e. Andreev bound states) 
involving the delocalized entangled spin singlet.
We also estimate a lower bound for the efficiency of entangled photon-pair generation, accounting for the presence of non-radiative processes such as phonon emissions.
Our proposal is realistic and achievable with state-of-the-art techniques in quantum 
microwave engineering with electostatically defined semiconducting quantum dots.

\end{abstract}
%
%
%
\maketitle 
%
%
%
%

\section{Introduction}

Research on quantum mesoscopic systems tests quantum theory on scales larger than atomic dimensions. 
This has led to the rise of a host of quantum technologies that are based on our ability to create, detect and manipulate individual
quantum states in different setups, such as electronic, photonic and superconducting nanostructures. 
In this perspective, hybrid quantum devices combining  components with different physical properties and 
with complementary functionalities \cite{Wallquist:2009,Xiang:2013,Clerk:2020} are at the forefront of 
future advancements in quantum nanotechnologies.

Entanglement, the non classical correlations that are possible between multipartite quantum systems, is  an important  physical resource for quantum technologies \cite{Nielsen:2000}.
For example, photon  entanglement  is at the heart of quantum information   
and further advancements rely on the ability to generate entangled-photon pairs with high efficiency \cite{Yuana:2010}.
Optical techniques such as spontaneous parametric down-conversion in nonlinear media \cite{Hong:1985,Rubin:1994} have been the go-to methods for entangled-photon generation. However, in the last decade or so, alternative approaches based on  solid-state devices have also been investigated.

One of the most promising systems is made of an optically active (self-assembled) semiconductor quantum dot embedded in  an optical microcavity 
(i.e., within the visible frequency range). 
We will henceforth refer to these as {\sl optical quantum dots}.
For such systems, the basic scheme for generating polarization-entangled photon pairs relies on a single quantum dot \cite{Benson:2000}, and is based on the biexciton-exciton cascade process \cite{Orieux:2017}. 
The experimental demonstration of this scheme was later achieved \cite{Akopian:2006} in a quantum dot coupled to a planar microcavity, where photon emission is enhanced due to the Purcell effect, with the quantum dot continuously excited by a laser.
Polarization-entangled photon pairs at optical frequencies have even been generated using electrically driven quantum dots in semiconductor light-emitting diodes \cite{Salter:2010}.
Theoretical models have also investigated the dephasing effects and the influence of the environment on the transfer protocol from entangled exciton states (spin-entangled electron-hole pairs) to polarized photons \cite{Budich:2010}, as well as the possibility of generating polarization-entangled microwaves through intraband transitions \cite{Emary:2005}.
Overall, in contrast to the probabilistic nature of parametric down-conversion sources, a key advantage of optical quantum dots is the ability to achieve a brightness (the probability of emitting a photon pair per excitation pulse) close to unity, enabling the generation of entangled photon pairs {\sl on-demand} \cite{Muller:2014}.

Remarkably, superconductors can serve as natural sources of entanglement in nanoscale solid-state systems 
\cite{Torres:1999,Choi:2000,Recher:2001,Lesovik:2001,Chtchelkatchev:2002,Bouchiat:2003}.
The ground state of a conventional BCS superconductor is composed of spin-singlet Cooper pairs and these spin-entangled pairs can potentially be manipulated. 
This requires ideally a device that is able to extract a pair and split the singlet state into two different spatially-separated electronic
orbitals so that they can eventually be manipulated locally and interfaced with other degrees of freedom.
In nanostructures, the pair splitting is
often synonymous of  non-local or crossed Andreev reflection in the sub-gap regime \cite{Beckmann:2004,Russo:2005,Cadden-Zimansky:2009,Kleine:2009,Wei:2010}.
If we consider a superconductor contacted with two normal leads, 
 electrons  
with energies within the superconducting gap can only enter (exit) the superconductor as Cooper pairs. If the two electrons forming a Cooper pair come from (end up into) the same lead, we have standard Andreev reflection. In contrast, if the two electrons come from (end up into) different normal leads, the process is known as non-local Andreev reflection.

Electronically tunable devices operating as Cooper pair splitters (CPS) have been implemented using semiconducting hybrid nanostructures, where a superconducting nanocontact is interposed between two electrostatically defined quantum dots (hereafter referred to simply as QDs) \cite{Hofstetter:2009,Herrmann:2010}.
Several experiments reported 
Cooper pair splitting based on QDs realised in low-dimensional nanostructures, such
as carbon nanotubes \cite{Herrmann:2010,Schindele:2012,Schindele:2014},  
nanowires \cite{Hofstetter:2009,Hofstetter:2011,Das:2012,Fulop:2014,Fulop:2015}  
graphene \cite{Tan:2015,Borzenets:2016,Tan:2016,Pandey:2021} 
and parallel QDs with interdot interaction \cite{Kurtossy:2022}.
Although these nanodevices are made of different materials all of them share some common features: 
(i) each QD behaves as an artificial atom with tunable energy levels; 
(ii) the electrons forming a Cooper pair tunnel out from the superconductor into localized orbitals  on the two spatially separated QDs.
Many theoretical works have investigated the electrical transport properties and coherence effects 
 of CPSs \cite{Sauret:2004,Morten:2006,Futterer:2009,Eldridge:2010,Chevallier:2011,Rech:2012,Trocha:2015,Dominguez:2016,Amitai:2016,Hussein:2017,Walldorf:2020,Buka:2022}, 
exploring the possibility of 
time-dependent optimal control \cite{Brange:2021,Tam:2021}.

Past experiments have probed the charge correlations of the two outputs of a CPS. 
These experiments have been based, for example, 
on nonlocal measurements of the conductance, current correlations 
\cite{Hofstetter:2009,Herrmann:2010,Schindele:2012}, and noise correlations and cross-correlations  
\cite{Wei:2010,Das:2012} which allow to infer the presence of the pair splitting.
A recent experiment has reported the real-time detection with high fidelity of the concomitant tunneling 
of paired electrons 
using a single-charge detector  \cite{Ranni:2021}.

To detect spin correlations in a CPS device, one has to go beyond charge sensing adding sensitivity to the spin degree of freedom. With this in mind spin-spectroscopy in curved carbon nanotubes in an in-plane magnetic field has been developed~\cite{Hels:2016}.
In another recent experiment spin sensitivity in a CPS device was achieved with the aid of ferromagnetic side gates ~\cite{Bordoloi:2022}. Here, the stray field of the gates results in a polarization dependent tunneling 
through the QDs. 
Experiment of Ref.~\onlinecite{Bordoloi:2022}
demonstrated, for the first time, a clear spin-spin anti-correlation in a CPS device as expected from the spin-singlet nature of Cooper pairs. 
Interestingly, a closely related experiment demonstrated 
spin-triplet correlations in a CPS device where the triplet character was caused by strong spin-orbit interaction in the semiconducting 
\addGR{nanowires (NWs)} used as material platform~\cite{Wang:2022}.

The direct observation and measurement of the spin-entanglement in a CPS has not been reported so far and still remains an open challenge. 
In the literature, several theoretical proposals have investigated ways to detect the spin entanglement of a single Cooper pair.
A consistent group of these focused on testing Bell-type inequalities for current-current cross-correlations in transport experiments on a CPS, where the paired electrons escape into drain reservoirs 
\cite{Chtchelkatchev:2002,Kawabata:2001,Samuelsson:2003,Braunecker:2013,Soller:2013,Klobus:2014,Busz:2017}. 
These theoretical proposals are hard to realize emperimentally. 
For continuous variables, such as the electrical current in mesoscopic transport, 
entanglement can be tested as the violation of a classically derived inequality using higher-order cumulants \cite{Bednorz:2011} 
which are experimentally challenging to measure. 
An alternative approach is to bypass electron transport measurements and instead, for example, employing specific techniques to detect electrons - and their spins - locally on the QDs \cite{Scherubl:2014}.

Taking inspiration from the biexciton-exciton cascade scheme \addGR{in  optical quantum dots}, theoretical proposals have suggested to convert electron spin entanglement, originating from a superconductor, into photon polarization entanglement  using 
{\it optical quantum dots}  \cite{Cerletti:2005,Nigg:2015,Schroer:2015}.
These proposals examine a system of optical quantum dots, either directly coupled to a superconducting nanocontact \cite{Cerletti:2005,Nigg:2015} or embedded in a superconducting p-n junction \cite{Schroer:2015}, and integrated into optical microcavities to emit polarization-entangled photon pairs via cavity leakage.
Specifically, Ref.~\onlinecite{Cerletti:2005} investigated spin light-emitting diodes that emit photon pairs into free space, with the degree of entanglement depending on the emission angle.
The theoretical scheme in Ref.~\onlinecite{Nigg:2015}, which involves photonic crystal cavities coupled to optical quantum dots, requires a drive laser pulse to illuminate the two quantum dots immediately after Cooper pair injection. 
This necessitates precise timing and synchronization of optical and electrical signals.
However, 
the most significant challenge for these proposals is that, to date, CPSs with optical quantum dots — specifically, superconducting nanocontacts interfaced with optical quantum dots — have not yet been experimentally realized.

{\it  
In contrast to previous studies, we present a theoretical proposal with a detailed experimental analysis to 
\addGR{generate  photon frequency entanglement in the microwave regime from a single Cooper pair.} 
Our approach relies entirely on waveguide circuit QED (Quantum Electrodynamics) architecture, using electrostatically defined QDs (encoding semiconductor qubits \cite{Chatterjee:2021,Zhang:2019,Burkard:2023}) as implemented in current experimental CPS devices \cite{Hofstetter:2009,Herrmann:2010,Schindele:2012,Schindele:2014,Hofstetter:2011,Das:2012,Fulop:2014,Fulop:2015,Tan:2015,Borzenets:2016,Tan:2016,Pandey:2021,Ranni:2021,Bordoloi:2022}.
}
\addGR{Before describing our approach, we briefly summarize the state of the art of these systems.}

Circuit QED with semiconductor QDs \cite{Cottet:2017review,Burkard:2020} 
can be considered as 
{\it photonics on a chip}, where the QDs  are coupled to microwave resonators (such as coplanar waveguides or microstrip lines), which serve as microwave photon cavities, in cryogenic systems  \cite{Frey:2011,Delbecq:2011,Petersson:2012,Basset:2013,Braakman:2014,Rossler:2015,Viennot:2015,Ranjan:2015,Viennot:2016,Beaudoin:2016,Bruhat:2016}.
They provide a platform to study quantum correlations between charge transport and emitted radiation \cite{Brandes:2003gu,Childress:2004kt,Trif:2008,Cottet_Kontos:2010,Jin:2012,Hu:2012,Bergenfeldt:2013,Contreras-Pulido:2013,Kloeffel:2013,Bergenfeldt:2014,Cottet:2015a,Dmytruk:2016,Cottet:2020}.
When single-electron transport is switched off, these systems act as charge or spin qubits, as in double quantum dot (DQD) systems in which 
the strong coupling regime with the microwave photon cavity has been achieved for both charge qubits \cite{Mi:2016,Stockklauser:2017,Mi:2017,Scarlino2019,Li_Li_Gao:2018} and spin qubits \cite{Samkharadze:2018,Mi:2018ip,Landig:2018,Zihlmann:2023}.
Such strong coupling regime for the coherent interaction is crucial for quantum state transfer.
Supporting our proposal, experiments have shown that it is possible to couple CPS with electrostatically defined QDs to  microwave resonators, used primarily as detectors for charge spectroscopy  \cite{Bruhat:2018,deJong:2023}. In particular, the recent work of Ref.~\onlinecite{deJong:2023} demonstrated the capability to detect and control Cooper-pair splitting using microwave resonators for dispersive gate charge measurements, though spin and entanglement detection were not addressed. 

\mbox{}\\
\addGR{In this work,} 
we go beyond previous theoretical proposals and existing experiments by exploring the possibility of generating frequency-entangled photon pairs in a waveguide circuit QED setup. These photons propagate through separate transmission lines ({flying qubits}), without involving photon cavities and their generation does not require charge currents.

\begin{figure*}[t!]
\begin{center}
\includegraphics[scale=0.27]{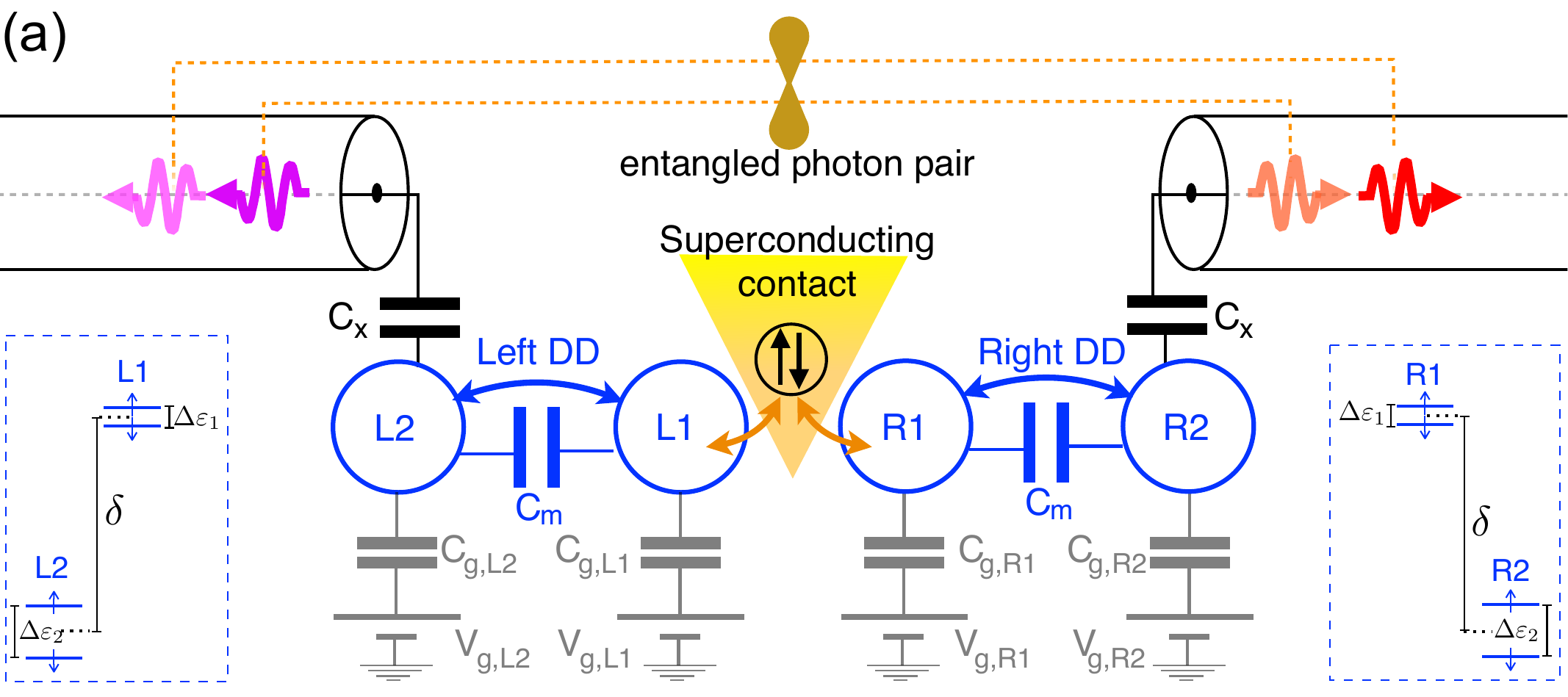} \hspace{1cm} \includegraphics[scale=0.175]{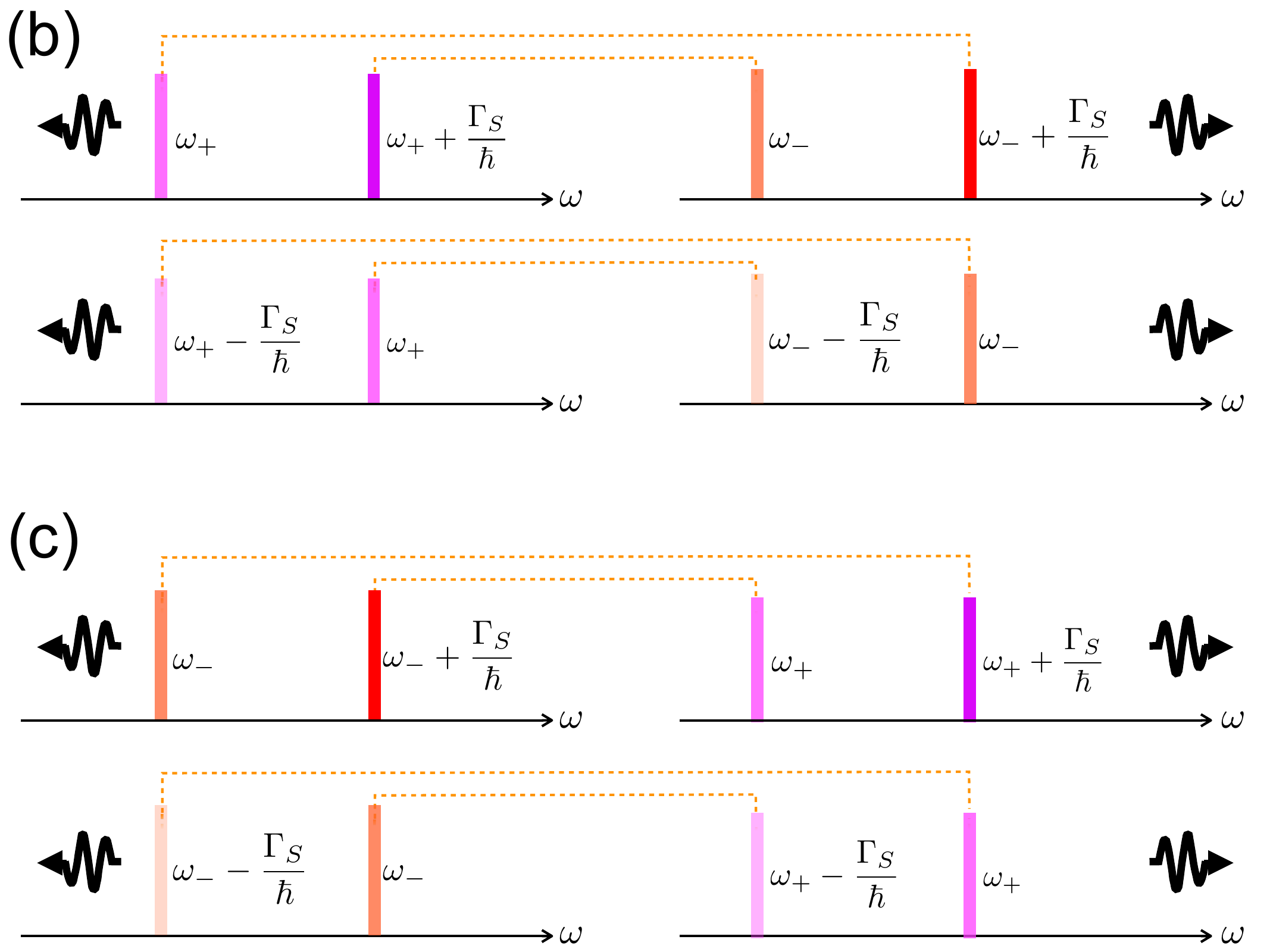}
\end{center}
\vspace{-2.5mm}
\caption{
{\bf (a)}
Schematic picture of the CPS based on two DQDs and example of the entangled photon pair. 
The superconducting nanocontact is tunnel-coupled to the two QDs $L1$ and $R1$.
The QDs $L2$ and $R2$ are capacitively coupled to microwave transmission lines in which photons are emitted.
In the transmission lines we give a pictorial sketch of the emitted two-photon entangled state.
The two lateral insets show the spin-resolved energy levels in each DQD for large detuning $\delta>0$ and in absence of tunnel coupling with the superconductor, where 
$\Delta\varepsilon_1$ and $\Delta\varepsilon_2$ are the local Zeeman splittings.
{\bf (b)} and  {\bf (c)} 
illustrate the frequency entanglement of the two photons propagating along the left and right transmission lines.
The frequencies of the photons can take the values $\omega_{\pm}$  and 
$\omega_{\pm}\pm \Gamma_S/\hbar$ with 
$\hbar\omega_{\pm} = \delta \pm \Delta\varepsilon_m$,  
$\Delta\varepsilon_m = \left( \Delta\varepsilon_2-  \Delta\varepsilon_1 \right)/2$ and 
$\Gamma_S$ the nonlocal pair amplitude of the CPS (see text).
The high-frequency photons are $\omega_{+} + q_s \Gamma_S/\hbar$ 
with $q_s=1,0,-1$ (dark magenta, magenta and light magenta) 
whereas 
the low-frequency photons are $\omega_{-} + q_s \Gamma_S/\hbar$ with $q_s=1,0,-1$ (dark red, red and light red).
The system can emit four possible frequency-entangled photon pairs.
Depending on the initial nonlocal Andreev bound states, 
the two photon state is characterized by  a frequency shift of $+\Gamma_S/\hbar$ or 
$-\Gamma_S/\hbar$ for one photon. 
Additionally, the final electronic states  in the two DQDs after the photons emission determine the directions of the high-frequency $\omega_{+}$ and 
low-frequency $\omega_{-}$ photons.
{\bf (b)} 
The two frequency-entangled photon pairs 
corresponding 
to the final electronic ground state 
with 
the spin down in $L2$  of the left DQD 
and 
the spin up in $R2$ of the right DQD.  
{\bf (c)} 
The two frequency-entangled photon pairs 
corresponding 
to the final electronic ground state 
with 
the spin up in $L2$ 
and 
the spin down in $R2$.
}
\label{fig:1}
\end{figure*}

The scheme is shown in Fig.~\ref{fig:1}(a), where the CPS is realized with two separate DQDs, one on the left and the other on the right 
of a superconducting nanocontact.
Each of the DQDs is capacitively coupled to a transmission line.
When the electronic system is initialized in an excited state - as described in detail in the next sections -   
photon emission occurs. 
Since the QDs are not connected to normal electrodes,  
the two electrons, 
delocalised over the two DQDs, remains in the system even after the system relaxes by emitting entangled photons.

Our proposal
\addGR{ 
uniquely combines cutting-edge components of nanoscale quantum hardware by requiring the integration of superconducting nanocontacts, semiconducting (electrostatically defined) QDs, and microwave quantum devices. 
It offers} several advantages:
(i) the entangled photons frequencies are largely tunable as they are determined by the electronic levels of the two DQDs controlled by the electrical gates;
(ii) \addGR{no measurement of current or noise correlations is required to reveal entanglement};   
(iii) no optical drives and signals are needed, neither time-dependent control of optical, electrical or magnetic pulses, 
or synchronized switching.
Furthermore, 
(iv) \addGR{no resonators (cavities) are required as the photons are emitted directly and irreversibly into the continuum spectrum of the two separate transmission lines}, thus avoiding the experimental problem of 
enhanced cross-talking between two \addGR{cavities}; 
(v) it has optimized detection, as emitted photons are confined in one-dimensional lines.
In summary, our scheme leverages the unique properties of hybrid quantum superconductor-semiconductor devices. As discussed in the following sections, its implementation is feasible with current state-of-the-art techniques.

\addGR{In our scheme, 
frequency-entangled photons arise from the particle-hole coherent superposition (i.e. Andreev bound states) involving the delocalized entangled spin singlet. 
Spin entanglement is a prerequisite for photon entanglement, as we explain further below. 
Therefore, photon entanglement acts as a witness for the pair entanglement of a $s$-wave BCS superconductor.
Equally important, }
our scheme has the potential to become a powerful resource for quantum technologies since 
it enables the generation of entangled propagating photons 
in the microwave domain using electronically driven devices integrated on a chip.
Indeed, manipulating individual entangled photons will elevate quantum microwave nanophotonics, still in its early stages, to the level of optical quantum nanophotonics. 
This could enable the testing of Bell inequalities in the frequency domain as in optics \cite{Ramelow:2009,Guo:17,Lingaraju:2022} or utilizing frequency bins for quantum information processing in nanophotonics \cite{Lu:2023apt}.

%
%
%
%
%
%
%
%
%

\mbox{}\\ 
 The paper is structured as follows. In Sec. \ref{sec:overview}, we summarise the main results of the paper and describe the working principle of the device. The operational regime, that is the range of tunable parameters used to operate the device, is discussed in \ref{sec:operational-regime}. 
In Sec. \ref{sec:theoretical-model}, we introduce the theoretical model of the system and the main results in the absence of non-radiative decay processes, charge and spin dephasing, and spin relaxation. The effect of non-radiative decay processes (phonon emission) is taken into account in Sec. \ref{sec:non-radiative}. While the detrimental effect of charge relaxation and spin relaxation and dephasing are discussed, respectively, in  Sec.~\ref{sec:charge-dephasing} and Sec.~\ref{sec:spin-relaxation}.
In Sec.~\ref{sec:feasibiltiy} we discuss the experimental feasibility and implementation.
Finally, we draw some concluding remarks and speculate on possible future applications in Sec.~\ref{sec:conclusions}.

%
%
%
%
%
%
%
%
%
\section{Overview of the main results}
\label{sec:overview}

Here we give a qualitative summary of the working principle of the device. A more detailed analysis is provided in the rest of the paper. 
We choose a symmetric setup to simplify the discussion and notation. 
For the analysis presented in this overview section we focus on the ideal case of a CPS coupled to two transmission lines.
We neglect the effect of energy relaxation and charge or spin decoherence in the dots, which  will be discussed in Sections 
\ref{sec:non-radiative},
\ref{sec:charge-dephasing} and  \ref{sec:spin-relaxation}.

Below, we introduce the characteristic energies of the system [see also inset of Fig.~\ref{fig:1}(a)] .
\begin{description}
    \item[$\delta$] %
    the  electrically-defined 
     energy detuning between the orbital level of QD $1$ and QD $2$ in a DQD ($\delta>0$ for the initial electronic state); 
    \item[$\Delta\varepsilon_{1(2)}$] the Zeeman splitting on the dots $L1, R1$ $(L2, R2)$;
    \item[$ \Delta\varepsilon_m$] the difference between the Zeeman splittings in the two DQDs, that is   $ \Delta\varepsilon_m = \left( \Delta\varepsilon_2-  \Delta\varepsilon_1 \right)/2$ 
    (we assume  $\Delta\varepsilon_m>0$); 
    \item[$t_{12}$] the tunneling matrix element between the two dots in a DQD (we assume $t_{12}\in \mathbb{Re}$);
    \item[$\Gamma_S $] the tunneling rate with the superconducting lead, which acts as a nonlocal pairing amplitude in the CPS.
\end{description}
For large DQD detuning, $|\delta| \gg |t_{12}|$,
the states of each DQD are well approximated by the localised states in the dots $L1$ and $R1$ 
and in the dots $L2$ and $R2$, see inset of Fig.~\ref{fig:1}(a). 
We define the operator $\hat{d}^{\dagger}_{\nu,n_D,\sigma}$ that creates an electron with spin $\sigma \in\{\uparrow, \downarrow\}$ in the QD $n_D \in\{1,2 \}$ of the DQD $\nu\in\{L,R\}$.

In the CPS, two nonlocal Andreev bound states $\ket{A_{\pm}}$ are formed, corresponding to quantum-coherent superpositions 
of the non-local singlet, 
$\ket{S_{11}}
=
\frac{1}{\sqrt{2}} 
\Big( \hat{d}^\dagger_{L,1,\uparrow} \hat{d}^\dagger_{R,1,\downarrow}  - \hat{d}^\dagger_{L,1,\downarrow}  \hat{d}^{\dagger}_{R,1,\uparrow}  \Big)\ket{0}$ and the vacuum state, $\ket{0}$.
In particular, when the energy of the non-local singlet is in resonance with the empty state, the two Andreev bound states have an energy splitting $2 \Gamma_S$.

For the initial electronic state, we assume the following hierarchy of energy scales $|\delta| > \Delta\varepsilon_m$ and $|\delta| \gg  \Gamma_S$.  
As it will be shown later, 
this hierarchy results in an emission of correlated or entangled 
photons, with an average frequency determined by $\delta/\hbar$.
The four characteristic photon frequencies are 
$\hbar\omega_{\pm} = \delta \pm \Delta\varepsilon_m$ 
and
$\hbar\omega_{\pm} \pm \Gamma_S$, see 
Fig.~\ref{fig:1}(b).
These frequencies are in the microwave domain, roughly between $f\sim 2-8$ GHz.

Due to coupling with the transmission lines, energy relaxation processes in the electronic system lead to two-photon emission, which is characterized by specific emission rates.
We denote 
$\gamma_{\pm}$ as the photon emission rates when the electronic system decays from one of  the two nonlocal Andreev bound states $\ket{A_{\pm}}$.
On the other hand, 
$\Gamma_{\pm}$  
correspond to the photon emission rates of the single-electron  relaxation processes within each DQD
(see Appendix \ref{app:F} for the theoretical derivation).
These rates also determine the linewidths of the emitted photon wavepackets and, as it will be discussed later, they are much smaller than the frequencies of the emitted photons.

In the next subsections, we discuss the decay processes for the photon-pair emission and we distinguish  different regimes, see Figs.~\ref{fig:2},\ref{fig:3}, and \ref{fig:4}.

%
%
%
%
%
%
\subsection{Large tunnel coupling with the superconductor and zero Zeeman splitting}
\begin{figure}[btp]
\begin{flushleft}
\includegraphics[scale=0.225]{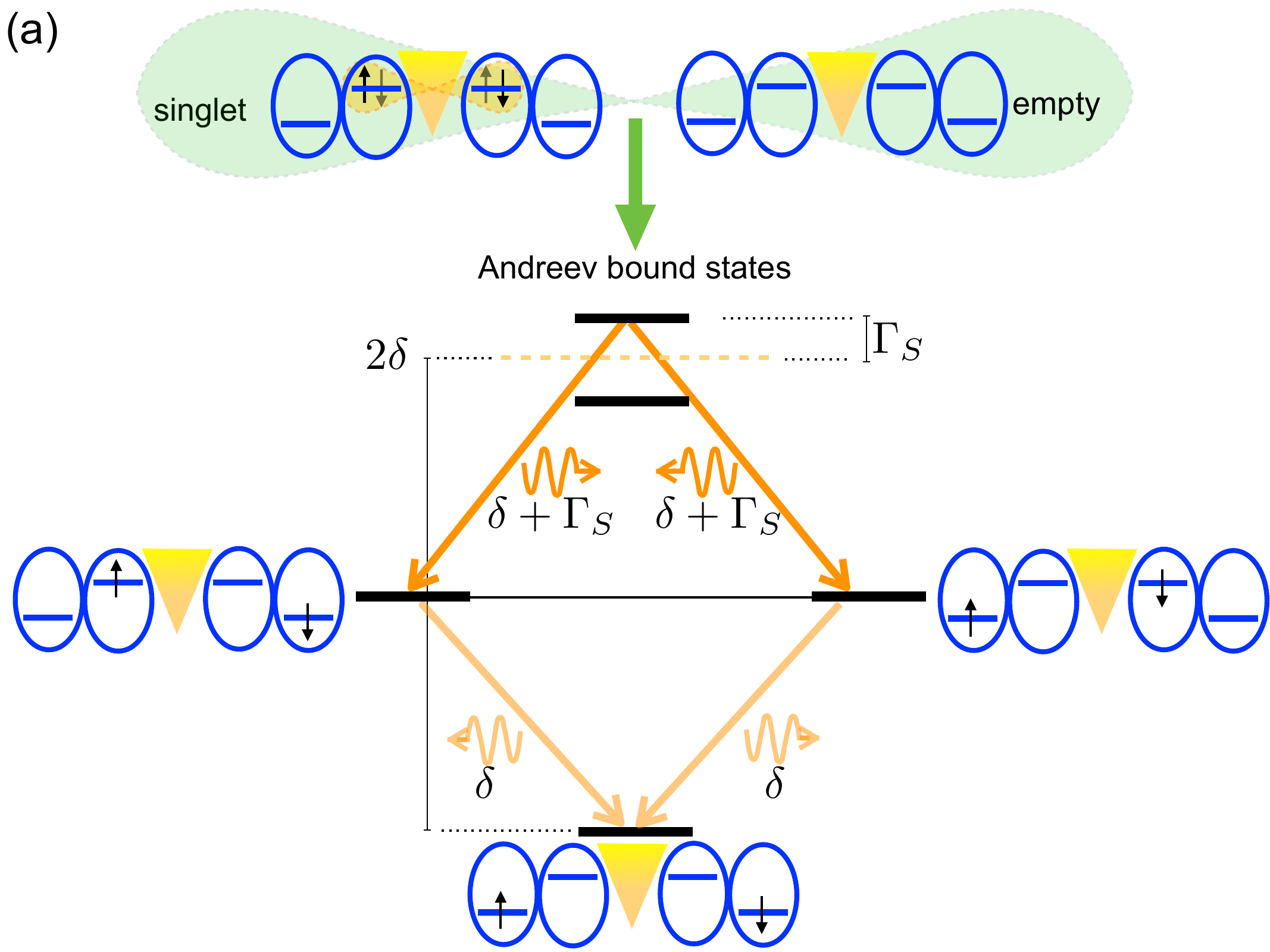}\\[5mm]
\includegraphics[scale=0.225]{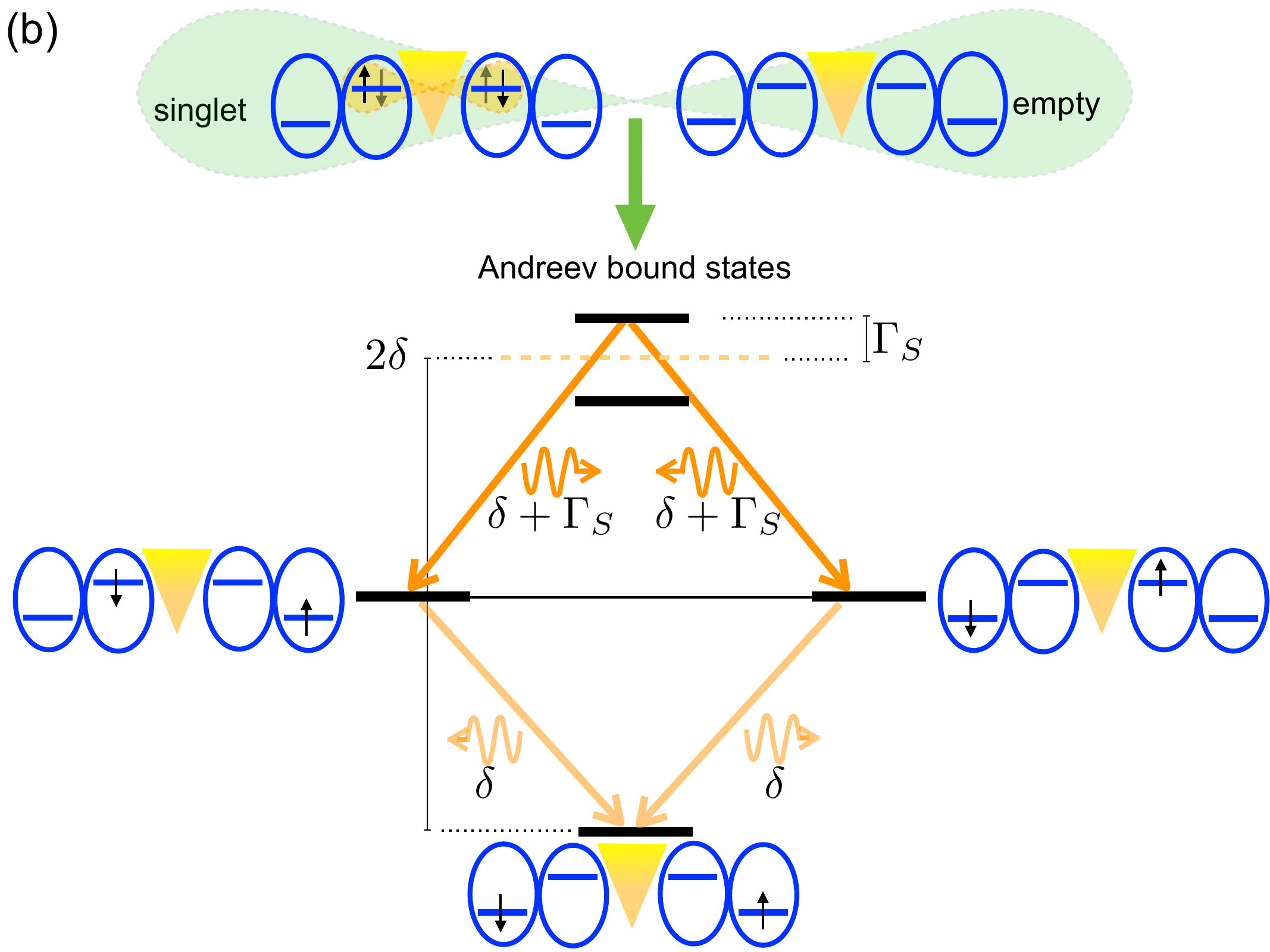}\\[5mm]
\includegraphics[scale=0.225]{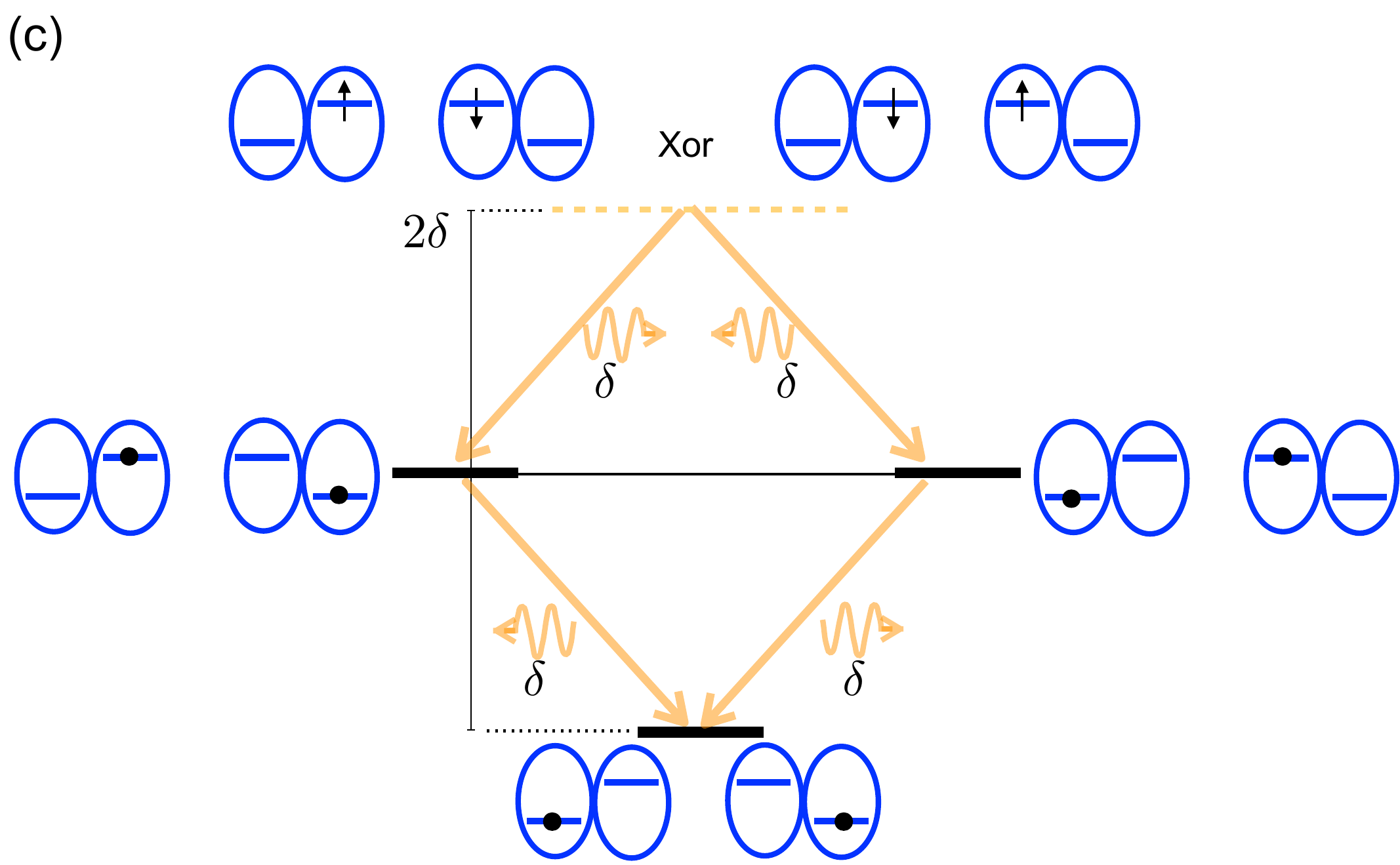}
\end{flushleft}
\vspace{-2.5mm}
\caption{
\addGR{
    In panels {\bf (a)} and {\bf (b)}, we show a scheme to explain the entangled photon pair emission  in the absence of Zeeman splitting. As an example, we consider $\ket{A_{+}}$, 
one of the two nonlocal Andreev bound states, as the initial state.
The two decay paths are symmetric and they do not depend on the final  spin configuration 
but the final photon states are different, hence the emitted photons are entangled.
In {\bf (c)}, we show, for comparison, the case of two doubly excited (uncorrelated) DQDs without the superconducting contact, 
where the initial electronic state can be an arbitrary spin  statistical mixture. Here, the filled circle on a level indicates either a spin-up or a spin-down electron. 
In this case, the two emitted photons are clearly not entangled.
}
}
\label{fig:2}
\end{figure}

Here we consider vanishing Zeeman splitting $\Delta\varepsilon_2= \Delta\varepsilon_1=0$.
In this limit the frequencies of the two emitted photons are $\left( \delta/\hbar, (\delta\pm\Gamma_S\right)/\hbar)$, Fig.~\ref{fig:2}.
We assume that the tunnel coupling to the superconductor is sufficiently large, such that $\Gamma_S \gg \gamma_{\pm}$, allowing to resolve the frequency difference between the two emitted photons.

%
%
%
%
%
%
%
%
%

In the limit of an isolated system and in the absence of interaction with the photon transmission lines, the system should display coherent oscillations.  
Including the interaction with the transmission lines, the electronic  system can decay from the subspace spanned by $\{\ket{A_{+}},\ket{A_{-}} \}$. 
Depending on which initial state ($\ket{A_{+}}$ or $\ket{A_{-}}$) the system decays from, the frequency-entangled photon pair has a 
frequency splitting $+\Gamma_S/\hbar$ or $-\Gamma_S/\hbar$, see  Fig.~\ref{fig:2}.

As elucidated in Fig.~\ref{fig:2}(a) and  Fig.~\ref{fig:2}(b), 
\addGR{
starting from an initial nonlocal Andreev bound state, two possible decay processes arise, 
each with a different final spin configuration. 
For each process, we have two symmetric decay paths that lead to the same final electronic orbital state but to different final photon states. 
}
This corresponds to a frequency-entangled photon pair 
with $\left( \hbar \omega_{L} = \delta, \hbar \omega_{R} = \delta \pm \Gamma_S \right)$ and $\left( \hbar \omega_{L} = \delta  \pm \Gamma_S , \hbar \omega_{R} = \delta \right) $, 
with $\omega_{L}$ and $\omega_{R}$ denoting the frequencies of the left 
$ (L) $ or right $(R)$ photon.
 
\addGR{
We remark that, beside destroying spin entanglement, spin decoherence also disrupts the Andreev bound states, 
which rely on the coherent tunneling of the electron pair between the QDs and the BCS ($s$-wave) condensate.
Indeed, as depicted in Fig.~\ref{fig:2}(c),  two simply excited (uncorrelated) DQDs (not in a CPS), starting from an arbitrary spin statistical mixture, 
simply emit two un-entangled photons with the same frequency.
}

%
%
%
%
%
%
\subsection{Small tunnel coupling with the superconductor and finite inhomogeneous Zeeman splitting}
%
%
%

%
%
%
%
\begin{figure}[btp]
\begin{flushleft}
\includegraphics[scale=0.225]{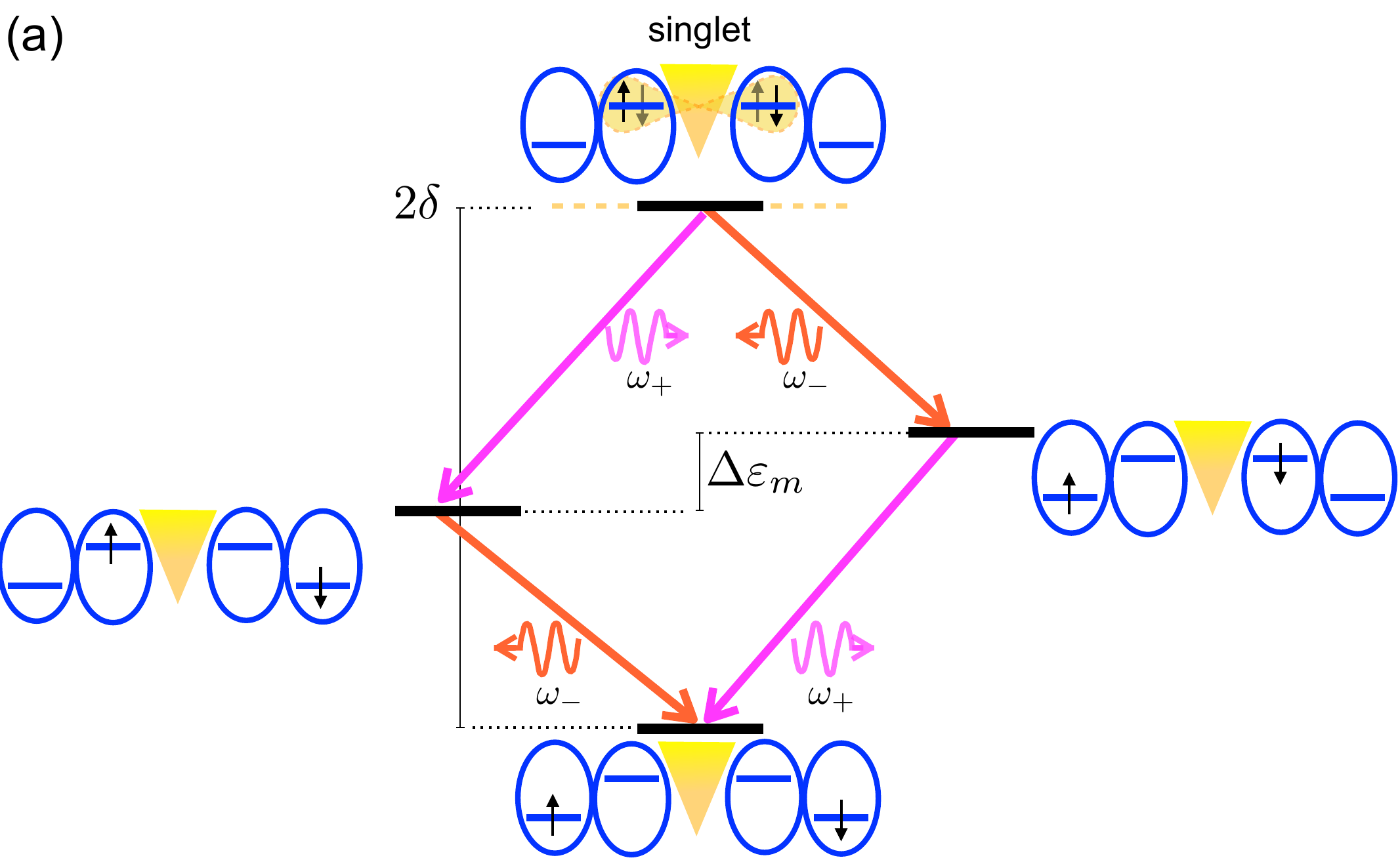}\\[5mm]
\includegraphics[scale=0.225]{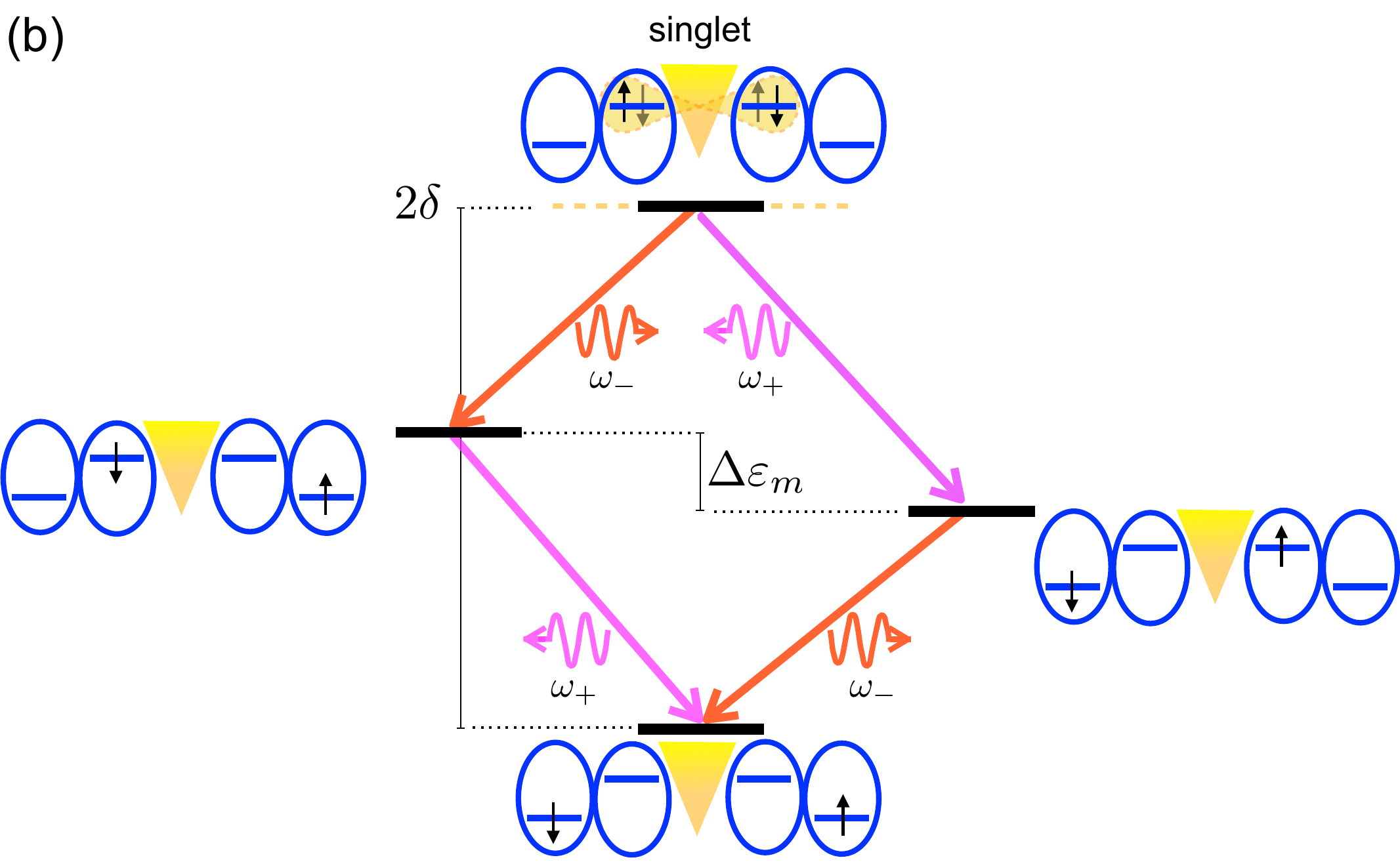}
\end{flushleft}
\vspace{-2.5mm}
\caption{
Schemes to explain 
the correlated photon pair emission 
in the limit of small tunnel coupling with the superconductor 
and in presence of inhomogeneous Zeeman splitting. 
For each final electronic state, the two paths lead to the same  photonic state: 
The photons are not entangled but  one can measure the spin-correlation of the injected Cooper pair. 
{\bf (a)} Decay paths with the final electronic state corresponding to the spin  up in the left DQD  and the spin  down in the  right DQD.
{\bf (b)} Decay paths with the final electronic state 
having inverted spins in the left and right DQD.
}
\label{fig:3}
\end{figure}
%
%
%
%
%

Here we consider the case of small $\Gamma_S$, which means   $\Gamma_S\ll \Delta \varepsilon_m$ 
and 
$\Gamma_S \ll \gamma_{\pm}$.
In this limit we can  approximate  the frequencies of the two emitted photons as $\hbar \omega_{\pm} = \delta\pm\Delta\varepsilon_m$, see Fig.~\ref{fig:3}, 
and the rates $\gamma_{+}\simeq\gamma_{-}$. 
At the same time, we consider a sizeable
$\Delta \varepsilon_m$ 
such that it is larger than all the photon emission rates, 
$\Delta \varepsilon_m   \gg \gamma_{\pm}$ 
and 
$\Delta \varepsilon_m \gg \Gamma_{\pm}$.
This ensures that the frequency difference of the two emitted photons can be resolved.

When the nonlocal Andreev bound states are not resolved, 
the superconducting contact simply acts as a source of single Cooper pairs that are split between the two spacially separated DQDs.
In this regime,  as shown in Fig.~\ref{fig:3}, 
for each final electronic state, the two paths lead to the same  photonic state: 
The emitted photons are not entangled but  one can measure the spin-correlation of the injected Cooper pair. 
Fig.~\ref{fig:3}(a) and Fig.~\ref{fig:3}(b) describe the two-photon emission for the two possible final electronic states - with inverted spins - in the DQDs:  the frequencies of the left and right propagating correlated photons are simply exchanged.

For example, in Fig.~\ref{fig:3}(a), 
the initial state associated to the singlet can decay by emitting photons to the two lines via two possible paths.
The final photonic 
state associated to these two decay paths is the same, e.g.
$\omega_R=\omega_{+}$ (magenta arrow) and $\omega_L=\omega_{-}$ (red arrow) 
despite the fact that 
the intermediate DQDs states have different energy  
$\Delta\varepsilon_m \neq 0$.
Thus the emitted photons are not entangled.
Moreover, the electronic system can still decay towards two distinct final states owing to the coherent superposition of the spin-singlet:
(i) in Fig.~\ref{fig:3}(a)  the final electronic state has spin up in the left DQD and spin down in the right DQD
or 
(ii) in Fig.~\ref{fig:3}(b), the final spin configuration is inverted. 
Therefore the frequencies of the two emitted photons for the two cases 
are inverted.

In this case, 
by measuring 
the frequency correlation 
of 
the single photons propagating along the left and the right lines,
one can detect the spin correlation of the singlet Cooper pair.

%
%
%
%
%
%
\subsection{Large tunnel coupling with the superconductor and finite inhomogeneous Zeeman splitting}
\begin{figure}[t]
\begin{flushleft}
\includegraphics[scale=0.225]{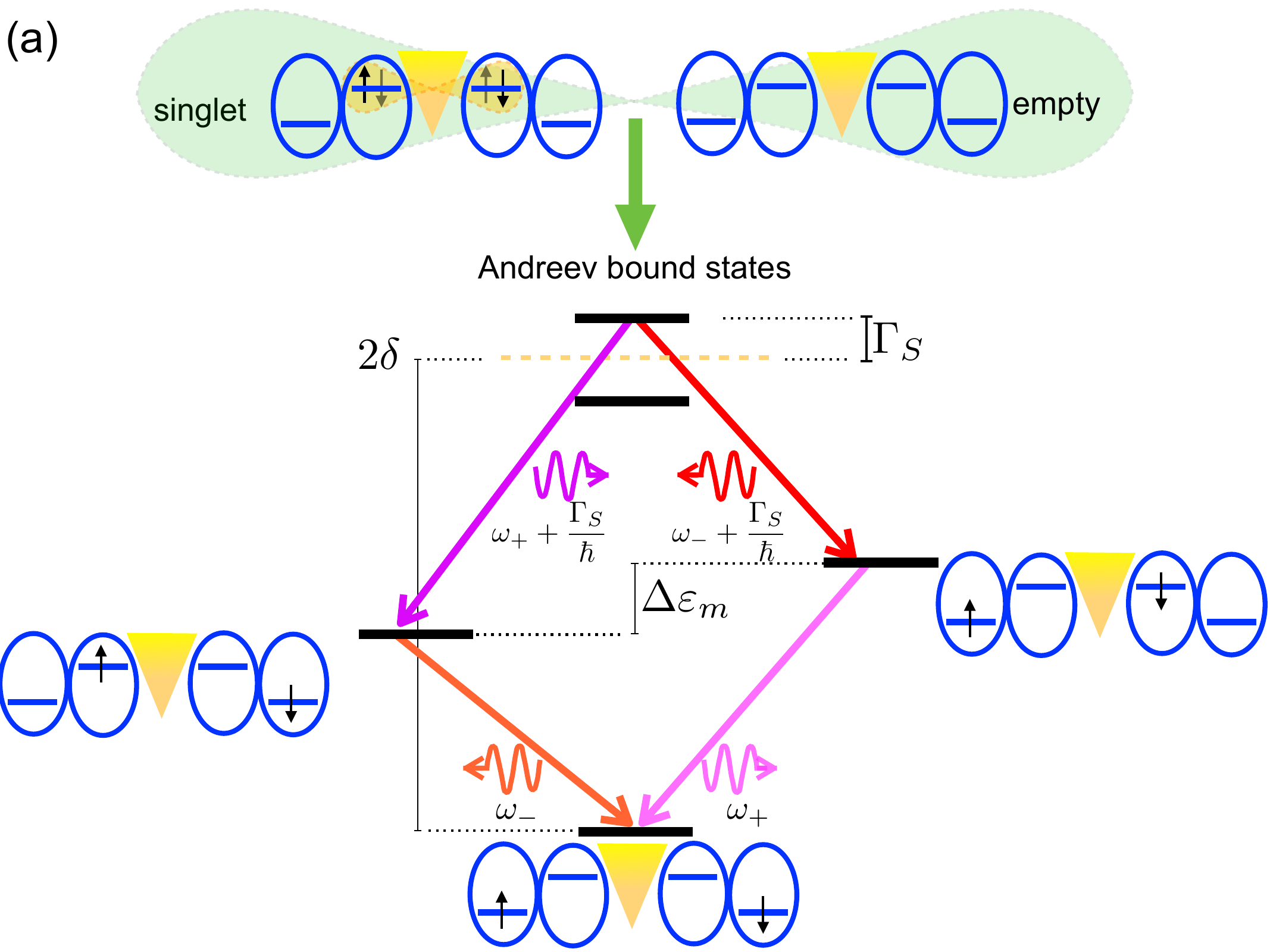}\\[5mm]
\includegraphics[scale=0.225]{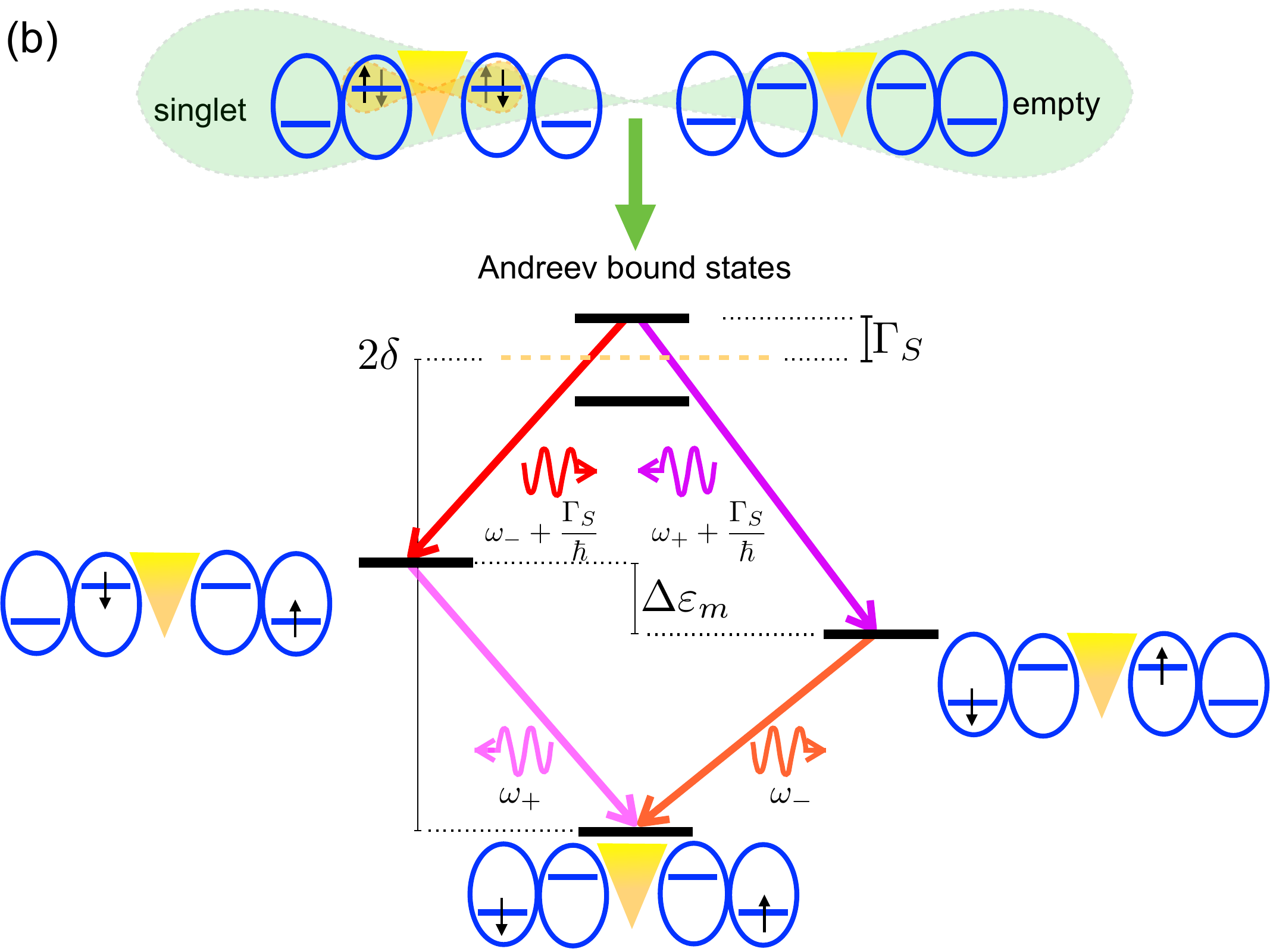}
\end{flushleft}
\vspace{-2.5mm}
\caption{
Scheme to explain the entangled photon pair emission for large tunnel  coupling with the superconductor and in presence of inhomogeneous Zeeman splitting.
Here it is shown an example with the initial state corresponding to $\ket{A_{+}}$, one of the two nonlocal Andreev bound states.
Two possible paths lead to the same final electronic state but to a different final photonic state: this corresponds to a frequency entangled photon pair.
{\bf (a)} Decay paths with the final electronic state corresponding to the spin  up in the left DQD and  spin  down in the right DQD.
{\bf (b)} Decay paths with the final electronic state 
having inverted spin configuration in the two DQDs.
}
\label{fig:4}
\end{figure}

Finally we discuss the general case $\Gamma_S \gg \gamma_{\pm}$, $\Delta\varepsilon_m \gg \gamma_{\pm}$ and 
$\Delta\varepsilon_m \gg  \Gamma_{\pm}$.
The photon emission processes for this case are shown in Fig.~\ref{fig:4}. 
A schematic description of the paired emitted-photon frequencies is shown in  Fig.~\ref{fig:1}(b).

%
%
%
%
%
%
%
%
%

In the photon emission processes shown in Fig.~\ref{fig:4}, 
the electronic system is initially in the higher-energy Andreev bound state $\ket{A_{+}}$.
Fig.~\ref{fig:4}(a) and Fig.~\ref{fig:4}(b) describe the two-photon emission for the two possible final electronic states - with
inverted spins - in the DQDs: the frequencies of the left
and right propagating correlated photons are simply exchanged.

For example, in Fig.~\ref{fig:4}(a), we consider one of the possible final electronic states.
The decay with photon emission has two possible paths corresponding 
to two different final photonic states: 
the 
photons emitted in the left path have frequencies 
$ \omega_{R}= \omega_{+}  + \Gamma_S/\hbar$ 
and
$\omega_{L}= \omega_{-}$ 
(dark magenta arrow and   red arrow); 
the
photons emitted in the right path have frequencies 
$ \omega_{L}=\omega_{-}  +  \Gamma_S/\hbar $ 
and 
$ \omega_{R}=  \omega_{+} $ 
(dark red arrow and magenta arrow).
As the final electronic state is still the same for the two paths, 
the final photonic state \addGR{results in} a coherent superposition of the final states for the two possible paths, namely
a superposition of states at different frequencies or frequency entangled pair. 

Similarly, Fig.~\ref{fig:4}(b) shows the case with a final state of the DQDs with an opposite spin configuration with respect to  Fig.~\ref{fig:4}(a).
For this case the  entangled-pair photon state has interchanged frequencies between the left and right propagating photons. 
In  Fig.~\ref{fig:4}(b)  
the 
photons emitted in the left path have frequencies 
$ \omega_{R}= \omega_{-}  + \Gamma_S/\hbar$
and 
$\omega_{L}= \omega_{+}$ 
(dark red arrow and magenta arrow)
and  
the
photons emitted in the right path have frequencies 
$ \omega_{L}=\omega_{+}  +  \Gamma_S/\hbar $ 
and 
$ \omega_{R}=  \omega_{-} $ 
(dark magenta arrow and  red arrow).
Similar processes \addGR{occur} when the system starts from the 
low-energy Andreev bound state $\ket{A_{-}}$ (not shown in Fig.~\ref{fig:4}).

\mbox{}\\
In summary, the system can emit four possible photonic states of frequency-entangled pairs, due to two possible initial states and two final states.
These are summarized in 
Fig.~\ref{fig:1}(b) and Fig.~\ref{fig:1}(c).
\addGR{In Sec.~\ref{subsec:distillation} we explain how to detect and separate different entangled states. 
In the general case, frequency-entangled states involve four distinct frequencies. The method in Ref.~\cite{Imany:2018} can be used to perform a Bell test in the frequency domain using a microwave-phase modulator to generate frequency sidebands.}

%
%
%
%
%
%
\section{Operational regime}
\label{sec:operational-regime}
\subsection{Cyclical preparation of the excited state in the CPS}
%
%
%
%
%
%
%
\begin{figure}[tbph!]
\begin{flushleft}
\includegraphics[scale=0.22]{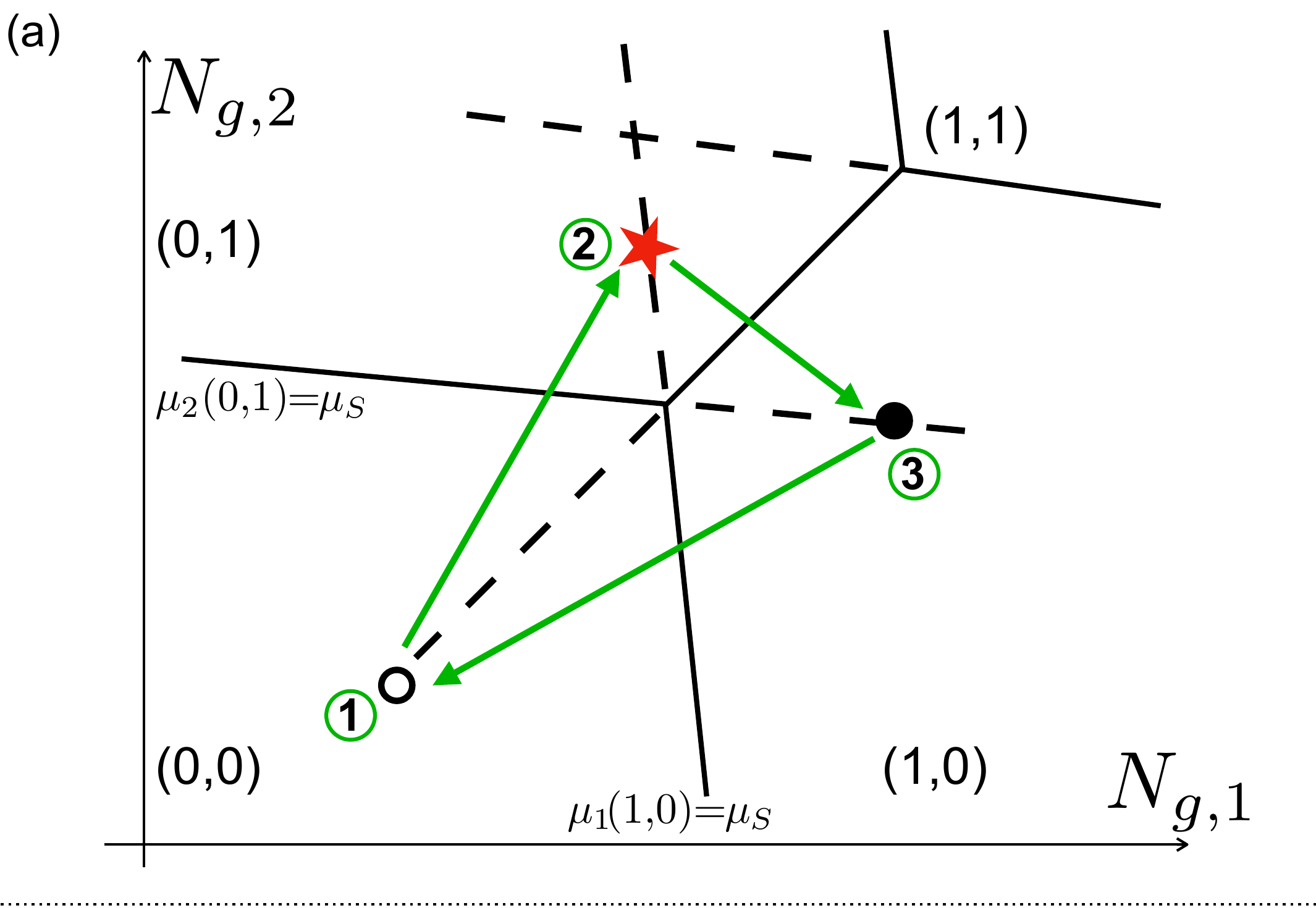} \\[2mm]
\includegraphics[scale=0.22]{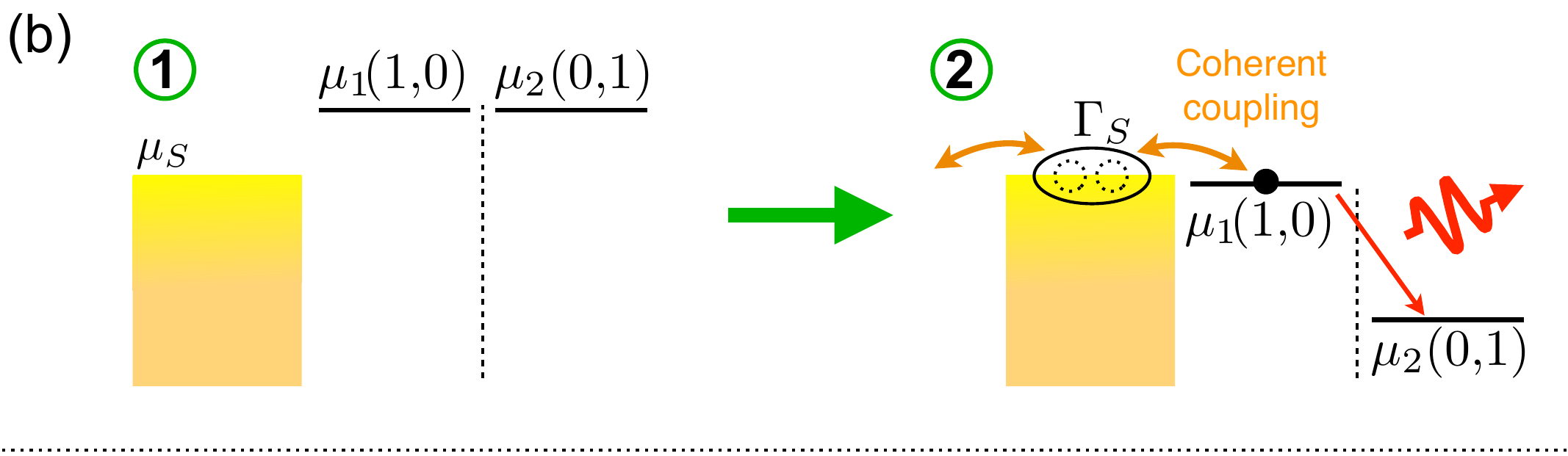} \\
\includegraphics[scale=0.22]{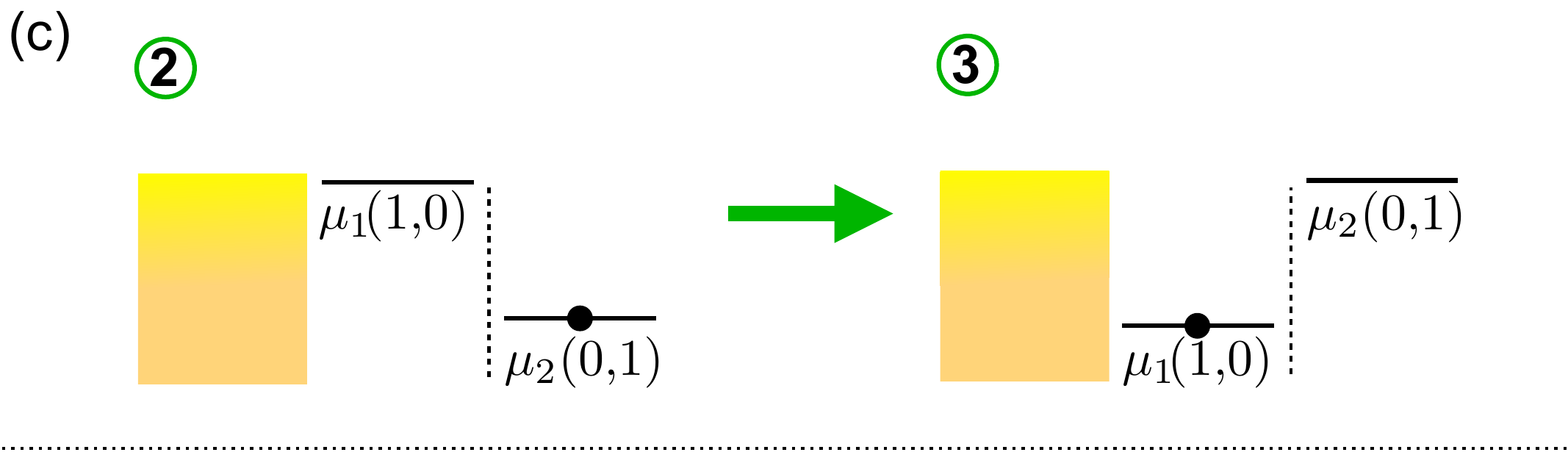} \\
\includegraphics[scale=0.22]{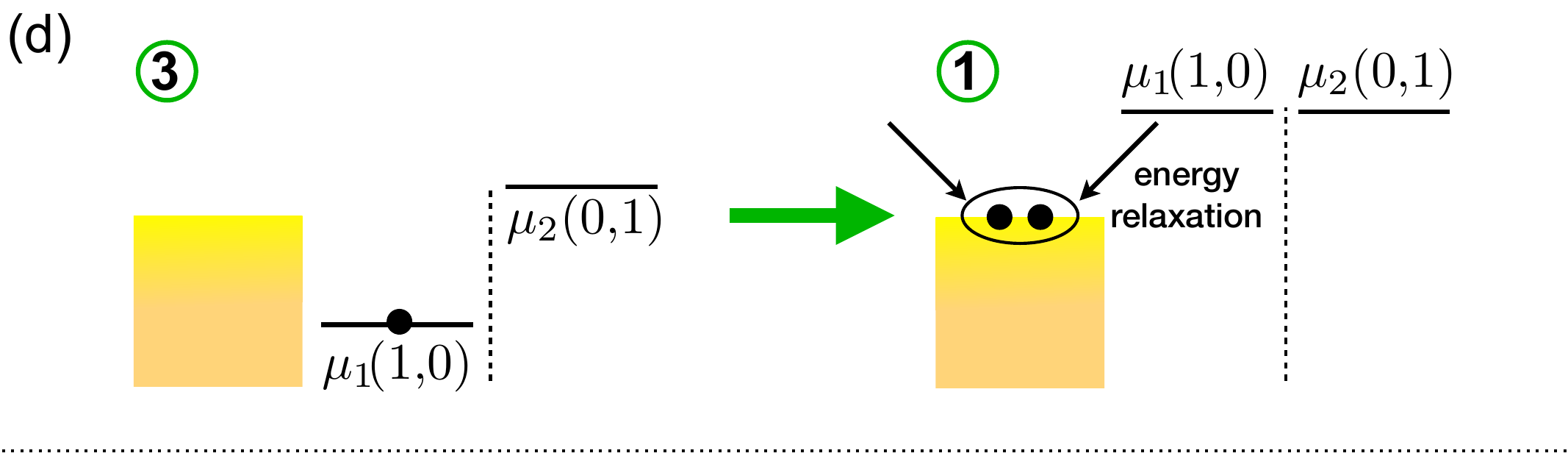}
\end{flushleft}
\vspace{-5mm}
\caption{ 
We sketch only the processes occurring in a single DQD.
$\mu_1\left(1,0\right)$ and $\mu_2\left(0,1\right)$ 
are the electrochemical potentials of the QD1 and of the QD2. 
The chemical potential of the superconductor is $\mu_S$.
{\bf(a)}
Sketch of stability phase diagram of a single DQD 
as function of the scaled gate voltages 
$N_{g,1} = C_g V_{g,1}/|e| $ and  $N_{g,2} = C_g V_{g,2}/|e| $.
%
%
%
%
The occupation numbers $(N_1, N_2)$ for
$(0, 0)$, $(1, 0)$, $(0, 1)$ and $(1, 1)$ denote the ground states.
{\bf(b)} 
One moves quickly (see text) the system 
from the initial point \ding{172}
with $\mu_1\left(1,0\right) > \mu_s$ and no electron 
to the point \ding{173} with $\mu_1\left(1,0\right)=\mu_s$ such that 
Cooper pair splitting can occur with one electron entering  the DQD.
At this point \ding{173}, we have also a finite detuning 
$\mu_1\left(1,0\right) > \mu_2\left(0,1\right)$: the nonlocal spin singlet 
can relax by emitting photons as discussed in the text.
{\bf(c)} 
After the photons have been emitted, each DQD is in its 
ground state with one electron in the QD2.
Then 
one moves the system 
from this point \ding{173}  to the point \ding{174}
with 
$\mu_1\left(1,0\right) < \mu_2\left(0,1\right)$ and the electron moves back from the  QD2 to the QD1.
{\bf(d)} At last step, moving along the line 
from the point \ding{174} to the initial point \ding{172}, 
the two electrons on the QD1s of the two DQDs can tunnel back into the superconductor (crossed Andreev reflection) as 
long as $\mu_1\left( 1,0 \right) \geq \mu_S$.
In this way, each DQD reaches again the initial  state 
corresponding to empty state.
}
\label{fig:5}
\end{figure}
%
%
%
%
%

Here we propose a possible scheme to prepare the electronic system in the excited state that can decay by emitting two entangled photons.
Our scheme is based on controlled extraction and injection of Cooper pairs in the splitter.
We again emphasize that, in our proposal, the electrons are prevented from leaving the  QDs towards  normal leads and hence no charge transport occurs.
A similar idea of trapping the entangled singlet with no charge transport
has been recently demonstrated in the experiment of  Ref.~\onlinecite{deJong:2023}.
Other initialization schemes  can be conceived. For example, it has been recently reported that reliable Cooper-pair injection can be obtained by employing an adiabatic cycle which is resilient to noise \cite{Brange:2024}.

In  Fig.~\ref{fig:5} we sketch only the processes occurring in one of the two DQDs forming the CPS.
The stability phase diagram of a single DQD  is shown in Fig.~\ref{fig:5}(a). 
We set $\mu_S$ as the chemical potential of the superconductor  whereas 
$\mu_1\left(1,0\right)$ and $\mu_2\left(0,1\right)$ 
are the electrochemical potentials of the QD1 and of the QD2. 
The electrochemical potential $\mu_{1} (N_1 ,N_2 )$   of QD1 is defined as the energy needed to add the $N_1$th electron to dot 1, while having $N_2$ electrons on QD2 \cite{vanderWiel:2003}.
An analogous definition applies for $\mu_2(N1 ,N2 )$. 
In this section we consider $\Delta \varepsilon_{1(2)}=0$ to simplify the notation.
Similar analysis holds for finite Zeeman splittings.

In Fig.~\ref{fig:5}(b), starting from the empty state $(0,0)$ in each DQD, 
corresponding to the point \ding{172},  
we tune quickly the plunger-gate voltages to move the system to the 
operational point \ding{173} (indicated by a star). 
This point is defined by the following conditions:
(i) 
the Cooper pair can split from the superconductor to the QD1s,  $\mu_S=\mu_1\left(1,0\right)$; 
(ii) we have a positive and fixed detuning between the QD1 and QD2, 
$\mu_1\left(1,0\right) - \mu_2\left(0,1\right) = \delta$, that sets the 
typical frequency of the emitted photons.

\addGR{ 
At the point \ding{173} the energy of the non-local singlet $\ket{S_{11}}$ becomes degenerate with the empty state. 
As a result of the coherent coupling with the superconductor, two nonlocal Andreev bound states, $\ket{A_+}$ and $\ket{A_-}$, with an energy splitting of $2\,\Gamma_S$ are formed.
Starting from the empty state, tunneling from the superconducting contact injects a Cooper pair and prepares the system in a coherent superposition of the two nonlocal Andreev bound states.  
The preparation of one of the two Andreev bound states could be achieved with a more fine-tuned control of the gate voltages. 
However, in Sec.~\ref{subsec:distillation}, we demonstrate that such fine-tuned control of the CPS is not strictly necessary. 
The entanglement can still be extracted (by distillation) even without precise knowledge of the initial electronic state, as long as it belongs to the subspace spanned by
$\ket{A_+}$ and $\ket{A_-}$.
}

In this way we have prepared the system in an initial \addGR{excited} state \addGR{from which} 
subsequently, the relaxation via two-photon emission can occur as discussed in the previous section.

We require that the time of switching $\Delta \tau_s$ between the empty state 
and the operational point must be fast enough, explicitly 
$1/\Delta \tau_s$ must be much larger than the rates of 
photon emission into the lines and larger than the rates of other (radiationless) energy relaxation  processes in the detuned DQD.
This prevents photons of arbitrary frequency from being emitted in the transmission lines during the adjustment of the voltages, ensuring that the electrons are prepared in the QD1s of the two DQDs as desired.

After the photon emission,  
the two DQDs have one electron in the QD2s, as shown in Fig.~\ref{fig:5}(c). 
We remark that, at the operational point \ding{173} (indicated by a star),  
a second electron from the superconductor cannot enter the QD1 
as $\mu_1\left(1,1\right) > \mu_S$ with $\mu_1\left(1,1\right) $
the electrochemical potential from the ground state $(0,1)$ to the excited state $(1,1)$ (not shown in the figure). 
Then we change   
the  plunger gate voltages to make the electron 
hop from the QD2 to the QD1, that is from point \ding{173} to \ding{174}. 
We achieve this by moving perpendicular to the diagonal line and approaching the point  $\mu_1\left(1,0 \right)<\mu_S$ and $\mu_2\left(0,1 \right)=\mu_S$.

As last step, Fig.~\ref{fig:5}(d), we need to remove the two electrons residing  in the QD1s of the two DQDs in order to reset the system to the initial empty state.
Hence, 
as shown in  Fig.~\ref{fig:5}(d), we move   the system from the point 
\ding{174}   to point \ding{172}.

\addGR{
Along this path, the electrochemical potential of QD1 increases and reaches $\mu_1\left(1,0\right) > \mu_S$ at  \ding{172}. 
At this point, a nonlocal Andreev reflection event can occur, during which electrons are injected into the superconductor.
Unlike the previous step, this process is irreversible and incoherent, as the excited state (with one electron on L1  and another on R1) 
is not degenerate with the empty state (corresponding now tothe ground state) but is instead significantly detuned. 
This is analogous to the DC transport regime, where two electrons with opposite spins can irreversibly enter the superconductor, forming a Cooper pair within the condensate.   
Indeed, such energy relaxation processes can occur via energy release into the environment (for example generating phonons in the substrate or photons in the transmission lines) 
which absorbs the energy difference between the initial and final states.
}

\addGR{In this way}, at  \ding{172},  the DQD system has relaxed
in its empty ground state.

In our analysis, we have assumed that the temperature is sufficiently low
so that thermal excitation of the excited levels of the DQD charge states can be neglected.
We provide an estimate of the energies involved in the next section.

\subsection{Charge states levels of a single DQD}
Using the electrostatic model for capacitively coupled QDs \cite{vanderWiel:2003} (see also Appendix \ref{app:A}),
we can estimate the electrostatic charging energies of the different charge occupations and the electrochemical potentials of a single DQD.

The dots are realized by small semiconducting islands that behave as artificial atoms with discrete levels in which 
a few electrons can be confined. 
In order to provide an estimate, we neglect the effect of orbital confinement, namely the discrete quantum levels of the QDs. 
We also assume that the total capacitance of the isolated QDs is $C_g$ and that the two QDs are equal (symmetric DQD).
To fix the notation, we set $(N_1,N_2)$ as the effective charge occupancy of QD$1$ and QD$2$.

By properly tuning the gate voltages, we choose the operational point \ding{173}, denoted by a start in Fig.~\ref{fig:5}, in 
which the ground state is $(0,1)$ with energy $E_{(0,1)}$ 
and the two excited states $(0,0)$ and $(1,0)$ are degenerate
$E_{(1,0)}=E_{(0,0)}$, see dashed line in Fig.~\ref{fig:5}(a). 
The DQD states with two electrons in a single QD, $(2,0)$ and $(0,2)$, or two electrons in a single DQD, $(1,1)$, 
have much higher energies, and therefore do not participate in the system dynamics.

For example,by setting $C_m=0.2 C_g$, with $C_m$ the interdot capacitance, we fix  the scaled gate voltages to $N_{g,1}=C_g V_{g,1}/|e| = 0.45$ and $N_{g,2}=C_g V_{g,2}/|e| = 0.5$ 
to achieve a degeneracy point with $E_{(1,0)}= E_{(0,0)}$
(see star in Fig.~\ref{fig:5}a). 
Then, setting $C_g \approx 1$ fF, namely a charging 
energy scale of $e^2/(2C_g)  \approx 80 \, \mu$eV 
($\approx 20$ GHz),  
the energy difference between the two degenerate excited states 
$(1,0)$ and $(0,0)$ and the ground state $(0,1)$,
that corresponds to the DQD detuning at this working point,
is 
$\delta^{\star} = E_{(1,0)} - E_{(0,1)} =E_{(0,0)} - E_{(0,1)}  \approx 7.5 \, \mu$eV  ($\approx 2$ GHz).
This sets the frequency scale 
of the emitted photons. 
The state with double occupation on two different dots has an energy difference with respect to the ground states 
 $E_{(1,1)} - E_{(0,1)} \approx 16 \, \mu$eV. 
Finally,  the states with two electrons on the same dot  $(2,0)$  and $(0,2)$
are even higher energy states respect to the ground state   
$E_{(2,0)} - E_{(0,1)} \simeq 167 \mu$eV 
and 
$E_{(0,2)}-E_{(0,1)} \simeq 157 \mu$eV. 

In summary, this analysis supports the the feasibility of the  initialization scheme presented in the previous section.
Furthermore, it 
provides a solid justification for considering  the CPS as an ideal source of delocalized, spin-entangled electrons, 
while allowing for the exclusion of other background processes such as cotunneling and local pair injection.

\subsection{Range for Zeeman splitting}
The Zeeman splitting is given by $g \mu_B B_{n_D}$  ($n_D=1,2$) where $g$ is the Land{\'e} g-factor, $\mu_B$ the Bohr magneton and $B_{n_D}$ the magnetic field 
on the QD, $n_D=1,2$. 

To create the local magnetic field and inhomogeneous Zeeman splitting, one can make use of micromagnets with optimzed shapes built nearby the DQD or of ferromagnetic split gates. A strong magnetic stray field can develop 
in the narrow gap etched in a long strip of a ferromagnetic metal (Permalloy)  \cite{Bordoloi:2022}.
These stray fields (order of tens of mT) do not suppress the superconductivity in the nearby superconducting contact.
A magnetic field gradient of the order of $1-10 $ mT/nm can be achieved with an optimized 
 micromagnet geometry \cite{Legrand2023,Philips2022,Chang2024,Bersano2023}.
Looking at the different semiconducting materials that can be used in our proposal (silicon, germanium, InGaAs, etc.)
we estimate the typical Zeeman splitting
$\Delta\varepsilon_1, \Delta\varepsilon_2$ in 
a range between zero and a
maximum value of $500$ MHz ($\sim 2 \mu$eV), given a typical DQD dimension.

Alternatively, a different Zeeman splitting between the two QDs of a DQD can be engineered by making use of the g-factor anysotropy present in semiconducting materials with strong spin-orbit coupling, such as in planar Ge or InAs \cite{Hendrickx2023}.

\subsection{Tunnel coupling between the superconductor and the 
two DQDs}

Real nanodevices implementing the CPS host 
competing processes, e.g.  the local tunneling of a Cooper pair into a single orbital (i.e. normal Andreev reflection), 
the cotunneling between the two sides mediated by the superconductor 
and 
the sequential tunneling 
of the two electron from a broken Cooper pair into the same dot.
These  processes reduce the efficiency of the splitting and hinder the observation of the correlations of the split pair.
In the previous section, we have discussed how the Coulomb interaction can efficiently suppress local Andreev reflection and elastic cotunneling. 
Here we discuss the range of values for $\Gamma_S$ to suppress sequential pair tunneling.

So far, in CPS devices where charge is transported from a central superconductor through two laterally connected QDs the splitting efficiency was limited in most experiments by $\Gamma_S$ being too large \cite{Hofstetter:2009,Herrmann:2010,Schindele:2012,Schindele:2014,Hofstetter:2009,Hofstetter:2011,Das:2012,Fulop:2014,Fulop:2015}. In the initial step, the splitting of Cooper pairs is controlled by the single-electron charging energy which forces the two electrons to tunnel into different QDs. This requirement has always been met. However, if $\Gamma_S\sim \Delta$, the splitting can be circumvented by local, sequential pair tunneling. This can be seen as follows: one electron of the Cooper pair tunnels, e.g over to the left QD and then leaves from there to a normal lead. The remaining electron is now able to co-tunnel through the same QD. This is allowed if the tunnel rates are fast, both the rate from the QD to the normal drain contact and the rate from the superconductor to the QD.

In the current design, there are no drain contacts so that Coulomb blockade effectively suppresses the tunneling of the second electron to the same DQD as discussed before. However, we still have to consider the magnitude of $\Gamma_S$. As said, in early experiments where the superconductor was evaporated onto typically a semiconducting nanowire, one arrived at $\Gamma_S\sim \Delta$ with $\Delta$ the induced superconducting gap. Since the superconductor was made from Al, $\Delta \approx 200$\,$\mu$eV, corresponding to $50$\,GHz. For the current proposal, 
this is way too large, since $\Gamma_S$ determines the shift in frequency of the emitted photons.

In a recent study, semiconducting nanowires with multiple gates were demonstrated to function as a CPS featuring three islands~\cite{deJong:2023}. 
This setup includes three plunger gates that regulate the charge on two QDs and on a larger central island, which is proximitized by a thin layer of superconducting aluminum.
Additionally, extra gates enable the individual tuning of the tunnel couplings. 
The authors reported a typical value for the tunnel coupling to the superconductor $\Gamma_S\approx 20$\,GHz. 
Although this value is noticeably smaller than the superconducting gap $\Delta$ it remains significant for the current proposal.
For our proposal, we target a much weaker tunnel coupling 
$\Gamma_S \leq 1$\,GHz, namely $\Gamma_S<\delta$ (see next section). 
Considering the typical operational bandwidth of the superconducting circuit QED technology, $\delta $ should be of the order of a few GHz.

However, achieving a weak enough coupling to the superconductor is challenging. On the positive side, gate-tunable tunnel barriers have been demonstrated that can be fully pinched off, offering a potential solution.
On the other hand, precise control is essential when operating in the pinch-off region, as the low-transparency potential barrier makes this section of the wire susceptible to the formation of uncontrolled localized states, which can behave like spurious additional QDs.

%
%
%
%
%
%
\section{Theoretical model}
\label{sec:theoretical-model}

%
%
%
%
%
%
\subsection{CPS formed by two DQDs}
As explained before, we assume a working point, with large DQD detuning $\delta$, 
in which, for each DQD, we have only three relevant energy states:
the ground state $(0,1)$ with occupied state on QDs 2,
the two degenerate excited state $(0,0)$ and $(1,0)$ 
corresponding respectively to 
the empty state and the occupied state on QDs 1.

In the presence of an externally applied magnetic field, 
the spin degeneracy is removed.
We assume an inhomogeneous Zeeman splitting with $\Delta\varepsilon_{1,2} < \delta$.
The effective Hamiltonian of a single DQD is formed by four spin-dependent nondegenerate levels, see inset of Fig.\ref{fig:1}(a).
In presence of tunneling between the QDs  $1$ and $2$, with amplitude $t_{12}$, 
the DQD eigenstates are hybridized ($\nu=L,R$ and $\sigma=\uparrow,\downarrow$) 
%
%
%
%
%
%
%
%
%
%
\begin{equation}
\ket{\nu, \pm,\sigma}
= 
\left(
\begin{array}{c}
\cos\left(\frac{\theta_{\sigma}}{2}\right) \\
\sin\left(\frac{\theta_{\sigma}}{2}\right) 
\end{array}
\right)
\ket{\nu,1,\sigma} 
\mp
\left(
\begin{array}{c}
\sin\left(\frac{\theta_{\sigma}}{2}\right)  \\
 \cos \left(\frac{\theta_{\sigma}}{2}\right) 
\end{array}
\right)
\ket{\nu,2,\sigma} \, .
\label{eq:1nu}
\end{equation}
We define the  annihilation operators corresponding to these single-particle states as 
\begin{equation}
\hat{d}_{\nu, \pm,\sigma}
= 
\left(
\begin{array}{c}
\cos\left(\frac{\theta_{\sigma}}{2}\right) \\
\sin\left(\frac{\theta_{\sigma}}{2}\right) 
\end{array}
\right)
\hat{d}_{\nu,1,\sigma} 
\mp
\left(
\begin{array}{c}
\sin\left(\frac{\theta_{\sigma}}{2}\right)  \\
 \cos \left(\frac{\theta_{\sigma}}{2}\right) 
\end{array}
\right)
\hat{d}_{\nu,2,\sigma} \, .
\end{equation}

%
%
%
%
%
%
%
%
%
%
For large DQD detuning $\delta \gg t_{12}$, 
the mixing angles are  
$\sin\theta_{\uparrow} 
\approx 
\theta_{\uparrow}
\approx t_{12}/(\delta - \Delta\varepsilon_m)
$ 
and
$\sin\theta_{\downarrow} 
\approx 
\theta_{\downarrow}
\approx t_{12}/(\delta + \Delta\varepsilon_m)
$.
For $\theta_{\sigma}\ll 1$,  
the excited state can be approximated as the level of QD$1$, $\ket{\nu, +,\sigma} \simeq \ket{\nu, 1,\sigma}$ 
whereas  
the ground state can be approximated as the level of QD$2$, $\ket{\nu, -,\sigma} \simeq \ket{\nu, 2,\sigma}$.

The superconducting contact is a $s$-wave BCS superconductor  tunnel-coupled to the QDs $L1$ and $R1$.
As discussed in the previous section, in the appropriate parameter regime of the system,  
the double electron occupation of a single QD, $(2,0)$ and $(0,2)$, 
as well as  
the simultaneous electron occupation on different QDs, $(1,1)$, 
are  
highly improbable as energetically forbidden.
In this way,  
the local Andreev reflections and 
the cotunneling between the two DQDs mediated by the superconductor 
are suppressed.

Energy relaxation processes are related to charge interaction and therefore they conserve the total spin.
As consequence, the states of the two DQDs can be expressed in terms of the triplet states with total spin $S=1$ and singlet states with  total spin $S=0$.
As the superconductor is a s-wave BCS, the triplet subspace  $S=1$ is decoupled.
In a simple picture the superconducting contact behaves as an infinite reservoir:  a single Cooper pair can split between the two QDs (crossed Andreev reflection).
In this regime the effective low-energy Hamiltonian 
(i.e. large superconducting gap limit)
%
%
%
%
for the CPS is
$ \hat{H}_{CPS} $ which connects only the empty state with $\ket{S_{11}}
=
\frac{1}{\sqrt{2}} 
\Big( \hat{d}^\dagger_{L,1,\uparrow} \hat{d}^\dagger_{R,1,\downarrow}  - \hat{d}^\dagger_{L,1,\downarrow}  \hat{d}^{\dagger}_{R,1,\uparrow}  \Big)\ket{0}$, the delocalized singlet in the QDs $1$ \cite{Rozhkov2000,Recher:2001,Meng2009,Eldridge:2010,Trocha:2015}
%
%
%
%
%
%
%
%
%
%
\begin{equation}
\label{eq:2_H_CPS}
\hat{H}_{CPS} 
=
-  \Gamma_S 
\ket{0}\bra{S_{11}} + \,\, \mbox{H.c.}
\, .
\end{equation}
%
%
%
%
%
%
%

The coherent coupling  of Eq.~(\ref{eq:2_H_CPS}) connects the empty state with all possible spin singlets
$\{\ket{S_{++} },\ket{S_{+-}},\ket{S_{-+}},\ket{S_{-- }} \}$ with coupling amplitudes determined by the mixing angles in Eq.~(\ref{eq:1nu}), 
where  $(s,s'=\pm)$
\begin{align}
\label{eq:singles_ss}
\ket{S_{ss'}}
=
\frac{1}{\sqrt{2}} 
\Big( \hat{d}^\dagger_{L,s,\uparrow} \hat{d}^\dagger_{R,s',\downarrow}  - \hat{d}^\dagger_{L,s',\downarrow}  \hat{d}^{\dagger}_{R,s,\uparrow}  \Big)\ket{0}.
\end{align}
However, for $|\delta| \gg |t_{12}|$, 
we can approximate $\ket{S_{11}} \simeq \ket{S_{++}}$ and the 
Hamiltonian reduces to 
%
%
%
%
%
%
%
%
%
\begin{equation}
\hat{H}_{CPS} 
\simeq
-
\Gamma_S 
\ket{0}\bra{S_{++}} + \,\, \mbox{H.c.}
\end{equation}
%
%
%
%
%
%
%
%
%
At resonance, when the two states $\ket{0}$ and $\ket{S_{++}}$ are degenerate, this yields two non-local Andreev bound state  
$\ket{A_{\pm}}=\frac{1}{\sqrt{2}}\left(\ket{0}\mp\ket{S_{++}} \right) $ 
with  energies  $\Delta E_{A_{\pm}} = 2 \delta  \pm \Gamma_S$.
%

%
%
%
%
\begin{figure}[tbp!]
\begin{flushleft}
\includegraphics[scale=0.25]{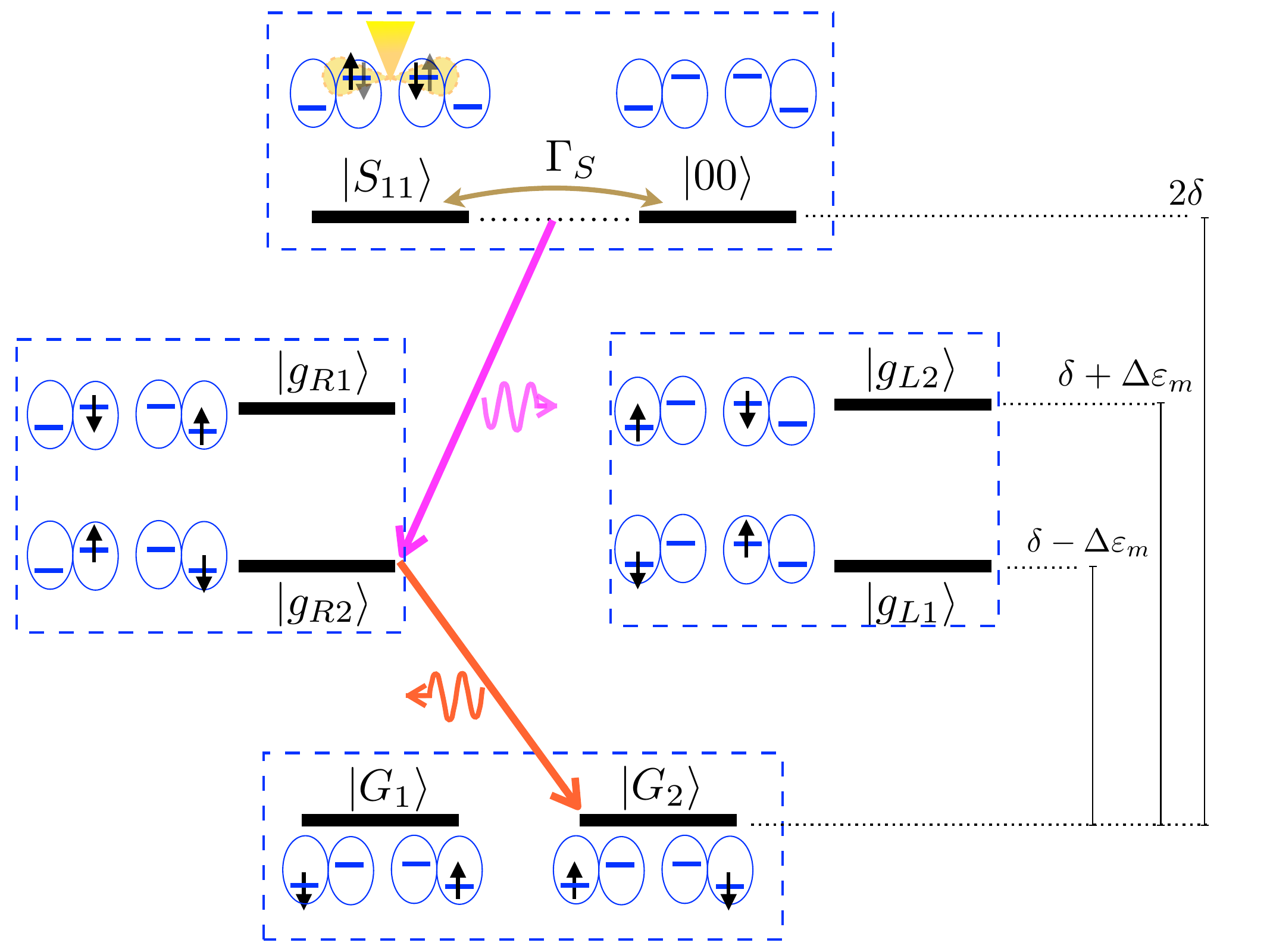} 
\end{flushleft}
\vspace{-2.5mm}
\caption{
Diagram of the relevant energy levels for two-photon emission in the CPS formed by two DQDs tunnel-coupled to a superconductor via QDs $1$ (see also Fig.~\ref{fig:1}). 
A possible decay path with two emitted photons is shown. 
%
%
%
%
}
\label{fig:6}
\end{figure}
%
%
%
%
%

To analyze the photon emission in the system, we need to 
consider the states spanning the relevant Hilbert space with $S=0, S_z=0$.
Beyond the state $\ket{S_{++}}$  where  both DQDs  are excited, we need to include  
the other states where one DQD is in an excited state and the other in the ground state and those were 
both DQD are in the ground states, as shown schematically in Fig.~\ref{fig:6}. 
For the intermediate states, in which only one DQD is excited whereas the other has relaxed, we 
have 
\begin{subequations}
\begin{align}
\ket{g_{R1}} &\equiv \hat{d}^{\dagger}_{L,+,\downarrow }  \hat{d}^{\dagger}_{R,-,\uparrow}\ket{0}\, , \label{eq:g_R1}\\
\ket{g_{L2}} &\equiv \hat{d}^{\dagger}_{L,-,\uparrow }\hat{d}^{\dagger}_{R,+,\downarrow}\ket{0}, \label{eq:g_L2}
\end{align}
\end{subequations}
with excitation energy $\delta  +\Delta\varepsilon_m $. 
The other intermediate states corresponding to the excitation energy $\delta  - \Delta\varepsilon_m$ are
\begin{subequations}
\begin{align}
\ket{g_{R2}} &\equiv \hat{d}^{\dagger}_{L,+,\uparrow }\hat{d}^{\dagger}_{R,-,\downarrow} \ket{0} \, , \label{eq:g_R2} \\
\ket{g_{L1}} & \equiv \hat{d}^{\dagger}_{L,-,\downarrow }\hat{d}^{\dagger}_{R,+,\uparrow} \ket{0}. \label{eq:g_L1}
\end{align}
\end{subequations}
The states with both DQD in their ground state are denoted by
\begin{subequations}
\begin{align}
\ket{G_{1}}  & \equiv  \hat{d}^{\dagger}_{L,-,\downarrow }\hat{d}^\dagger_{R,-,\uparrow}\ket{0} \label{eq:G_1} \\
\ket{G_{2}}   & \equiv \hat{d}^{\dagger}_{L,-,\uparrow }\hat{d}^\dagger_{R,-,\downarrow}\ket{0} \label{eq:G_2}
\end{align}
\end{subequations}
In addition, the condition $\Gamma_S \ll |\delta|$ has to be fulfilled to avoid 
that the low energy level $\ket{A_{-}} $ becomes resonant with $\ket{g_{R1}}$ and $\ket{g_{L2}}$ which would imply coherent coupling  between these two states.

%
%
%
%
%
%
\subsection{Interaction with the photons of the transmission line}
We have developed a mesoscopic model based on quantum network theory \cite{Yurke:1984,Devoret:1997,Vool:2017,Clerk:2010} 
in appendix \ref{app:A}.

The electron-photon interaction between the electrons in each DQD with the corresponding transmission line is given by
%
%
%
%
%
%
%
%
%
%
%
\begin{align}
\hat{H}_{int} 
&
=
i \sum_{k_L} g_{k_L} \left(  \hat{a}_{k_L,L}^{\phantom{\dagger}} - \hat{a}_{k_L,L}^{\dagger}  \right) \hat{n}_{L,2} \nonumber\\
&
+
i \sum_{k_R}  g_{k_R} \left(   \hat{a}_{k_R,R}^{\phantom{\dagger}} -  \hat{a}_{k_R,R}^{\dagger} \right) \hat{n}_{R,2}
  \, ,
\label{eq:4_H_int}
\end{align}
%
%
%
%
%
%
%
%
%
%
where $ \hat{a}_{k_L,L}^{\phantom{\dagger}}$ and $\hat{a}_{k_R,R}^{\phantom{\dagger}}$ are the bosonic annihilation operators for a photon in the left and right transmission lines, respectivelly.
The fermionic occupation operators are $\hat{n}_{L,2}=\sum_\sigma \hat{d}^\dagger_{L,2,\sigma}\hat{d}_{L,2,\sigma} $ and $\hat{n}_{R,2}=\sum_\sigma \hat{d}^\dagger_{R,2,\sigma}\hat{d}_{R,2,\sigma} $ for the QDs $L2$ and $R2$.
The coefficients $g_{k_L}$ and $g_{k_R}$ are given in Appendix \ref{app:A}.

The form of the interaction Hamiltonian Eq.~(\ref{eq:4_H_int}) is valid under the assumption that  the mutual capacitance between the two QDs in a DQD is sufficiently small  $C_m\ll C_{g}$
(see appendix \ref{app:A} for details).  
This condition ensures a sizeable electric dipole interaction between the transmission line and the DQD.

The electron-photon interaction Eq.~(\ref{eq:4_H_int}) also 
yields, in principle, a photon emission due to the relaxation between the two Andreev bound states (see Fig.~\ref{fig:2}) 
with a single photon of frequency $2\Gamma_S$ emitted in the left or in the right transmission line,
see appendices  \ref{app:C} and  \ref{app:D}, 
\cite{Cottet:2012s,Cottet_Kontos_Yeyati:2012}.
However, if the DQD states are weakly hybrized, i.e. large detuning $|t_{12}|\ll   |\delta|$, 
the rate for such photon emission 
is much weaker 
than the rate of the two correlated/entangled photon emission which is the process that we aim to realize (see Fig.~\ref{fig:1}) 
(see appendices \ref{app:D} for details). 
%

%
%
%
%
%
%
%
\subsection{Frequency-entangled photon pair}
\label{subsec:calculations}
For weak electron-photon coupling, we use the rotating-wave approximation (RWA)  to compute the photon-emission rate.
We assume the system to be initially  in the excited subspace of the Andreev doublets with no photons in the transmission lines (zero temperature limit).
Since RWA preserves the number of excitations, we can make the following Ansatz for the  total state of the system:
\begin{align}
& 
\ket{\Psi(t)} 
= 
c_{0+}(t) \ket{A_+} \otimes \ket{0,0} +c_{0-}(t)  \ket{A_-}  \otimes \ket{0,0}  \nonumber \\
&
+
\sum_{k_L} \big(  c_{k_L,L1}(t) \ket{g_{L1}} \otimes \ket{{k_L},0} + c_{k_L,L2}(t) \ket{g_{L2}} \otimes \ket{{k_L},0}  \big)  \nonumber \\
&
+
\sum_{k_R} \big(  c_{k_R,R1}(t) \ket{g_{R1}} \otimes \ket{0,{k_R}} + c_{k_R,R2}(t) \ket{g_{R2}} \otimes \ket{0,{k_R}}  \big)  \nonumber \\
&
+
\sum_{k_L,k_R}   c_{k_L,k_R,1}(t) \ket{G_{1}}  \otimes \ket{{k_L},{k_R}} \nonumber \\
&
+ 
\sum_{k_L,k_R}   c_{k_L,k_R,2}(t)  \ket{G_{2}} \otimes  \ket{{k_L},{k_R}}   \, .
\label{eq:7_Ansatz}
\end{align}
The Ansatz $\ket{\Psi(t)}$ includes the states, $\ket{A_\pm}$,  with the whole system in one of the two excited states and with no photons $\ket{0,0}$, the states, $\ket{g_{L1(2)}}$ and $\ket{g_{R1(2)}}$, and with one excited DQD and one photon in the system, $\ket{{k_L},0}$ or $\ket{0,{k_R}}$, and the states, $\ket{G_{1(2)}}$ with no excitations in DQDs and with two photons, with one photon in each line 
$\ket{{k_L},{k_R}}$.

By solving the system of equations for the coefficients, in the Markovian limit (Wigner-Weisskopf approximation) and applying a secular approximation,
(see appendices \ref{app:B} and  \ref{app:E}), 
one can show that all the coefficients vanish in the limit $t\rightarrow \infty$ except  \addGR{for $c_{k_L,k_R,n}$ ($n=1,2$)}. This implies that  
the system evolves towards one of the two ground states $\ket{G_{1(2)}}$  with two photons emitted.

Switching in the interaction picture, \addGR{
$c_{k_L,k_R,n}(t) = e^{-i\left( \omega_{k_L}+\omega_{k_R} \right) t}  \tilde{c}_{k_L,k_R,n} (t)$},  
and, for example, assuming as initial state $\ket{\Psi(t=0)} = \ket{A_{+}}$, 
we have for $t\rightarrow \infty$ 
%
%
%
%
%
%
%
%
%
%
\begin{widetext}
\begin{align}
\addGR{ \tilde{c}_{k_L,k_R,1}^{(+)} }
&=
\frac{g_{k_L}g_{k_R} }{2\hbar^2} \, \frac{t_{12}^2}{\hbar^2 \omega_{+} \omega_{-}} 
\left(\frac{1}{\omega_{k_L}-\omega_{+}  + \frac{i}{2}  \Gamma_{+}  }+\frac{1}{\omega_{k_R}-\omega_{-}  + \frac{i}{2}   \Gamma_{-}  }\right)
\left(
\frac{1}{\omega_{k_L}+\omega_{k_R}-\omega_{+} -\omega_{-}- \frac{\Gamma_S}{\hbar}   +\frac{i}{2}  \gamma_{+}  }
\right) 
\, ,
\label{eq:8_wavepacket_1}\\
\addGR{  \tilde{c}_{k_L,k_R,2}^{(+)} }
&=
-
\frac{g_{k_L}g_{k_R} }{2\hbar^2} \,
\frac{t_{12}^2}{\hbar^2 \omega_{+} \omega_{-}} 
\left(\frac{1}{\omega_{k_L}-\omega_{-}  + \frac{i}{2}  \Gamma_{-}  }+\frac{1}{\omega_{k_R}-\omega_{+}  + \frac{i}{2}   \Gamma_{+}  }\right)
\left(
\frac{1}{\omega_{k_L}+\omega_{k_R}-\omega_{+} -\omega_{-}- \frac{\Gamma_S}{\hbar}   +\frac{i}{2}  \gamma_{+}  } 
\right)
\, ,
\label{eq:9_wavepacket_2} 
\end{align}
\end{widetext}
where $\hbar\omega_{\pm}=\delta\pm\Delta\varepsilon_m$. 
The long time limit $(t \rightarrow \infty)$ corresponds to times much larger than the linewidths.

Equations~(\ref{eq:8_wavepacket_1}) and (\ref{eq:9_wavepacket_2}) are the main theoretical result of this work: they describe a two-photon frequency-entangled state.
When the final state of the two DQDs corresponds to $\ket{G_1}$ then  
the entagled state is described by the wavepacket 
Eq.~(\ref{eq:8_wavepacket_1}).
Otherwise, when the final state of the two DQDs corresponds to $\ket{G_2}$, the entagled state is described by the wavepacket 
Eq.~(\ref{eq:9_wavepacket_2}).
To better elucidate this result, we analyze, for example,  
Eq.~(\ref{eq:8_wavepacket_1}).
The first term in the first bracket in $\tilde{c}_{k_L,k_R,1}^{(+)}$ 
constrains the frequency $\omega_{k_L}$ of the left photon to be peaked at 
$\omega_{+}$. This implies that the frequency of the right photon $\omega_{k_R}$ 
is peaked at $\omega_{-} + \Gamma_S/\hbar$ 
due to the term in the second bracket.  
Conversely, the second  term in the first bracket in $\tilde{c}_{k_L,k_R,1}^{(+)}$ 
constrains the frequency $\omega_{k_R}$ of the right photon to be peaked at 
$\omega_{-}$, and the frequency of the left photon 
to be peaked at $\omega_{+}+\Gamma_S/\hbar$ 
due to the term in the last bracket.  
Note that when $\Gamma_S=0$, the state is not entangled, as each photon can be uniquely associated with a specific frequency.

\addGR{
Similarly, if the decay originates from the low-energy nonlocal Andreev bound state  
$\ket{A_{-}}$, 
a similar solution is obtained for the coefficients 
$\tilde{c}_{k_L,k_R,1}^{(-)}$ and  $\tilde{c}_{k_L,k_R,2}^{(-)}$
with $\Gamma_S$  replaced by  $-\Gamma_S$ 
and 
$\gamma_+$ replaced by  $\gamma_-$  
in Eq.~(\ref{eq:8_wavepacket_1}) Eq.~(\ref{eq:9_wavepacket_2}).
 }

%
%
%
%
%
%
%
%

\begin{figure}[t!]
    \centering
   (a)\phantom{\hspace{0.85\columnwidth}}\\[-12pt]
    \includegraphics[width=0.85\columnwidth]{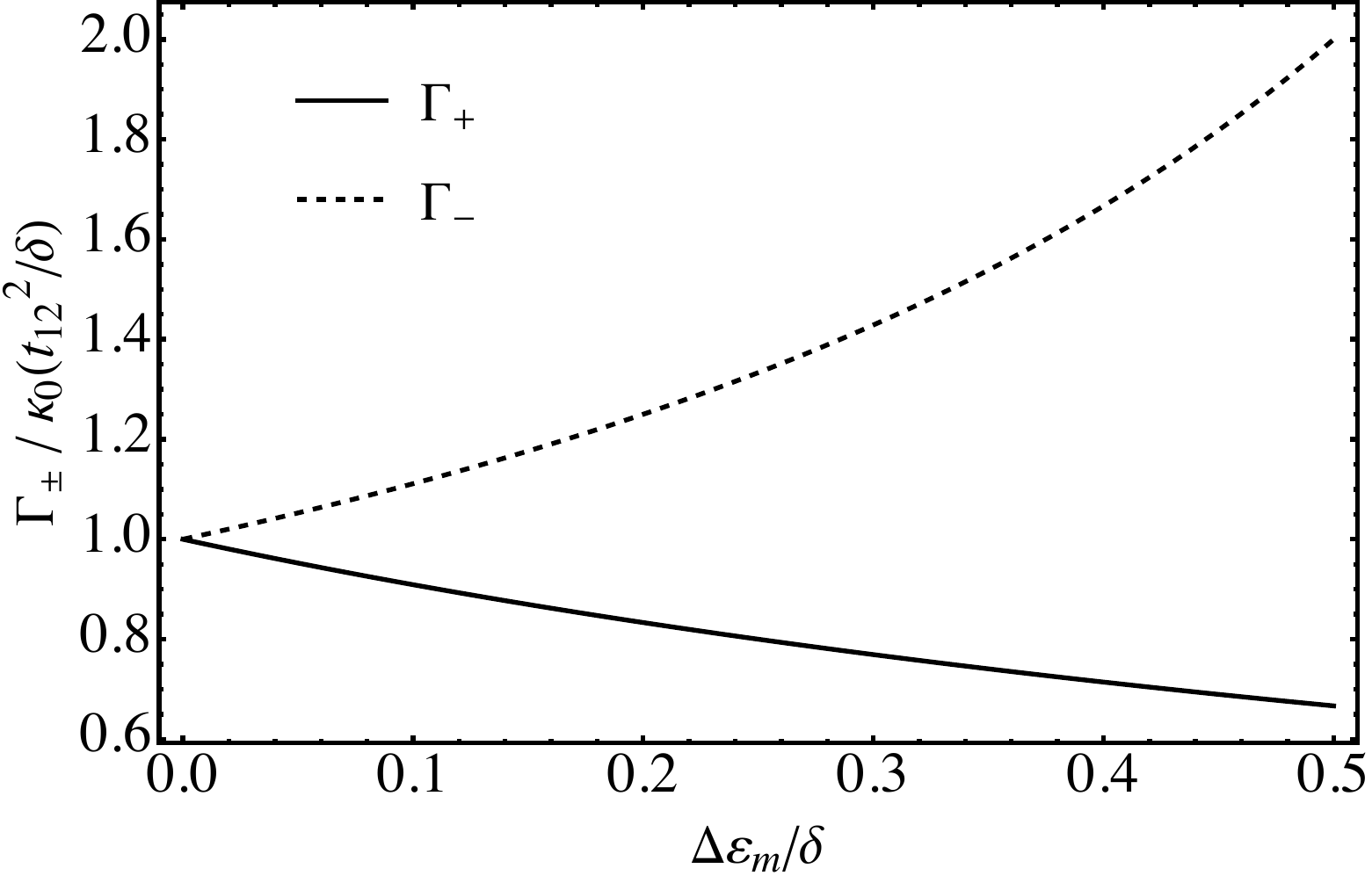}\\[12pt]
   (b)\phantom{\hspace{0.85\columnwidth}}\\[-12pt]
    \includegraphics[width=0.85\columnwidth]{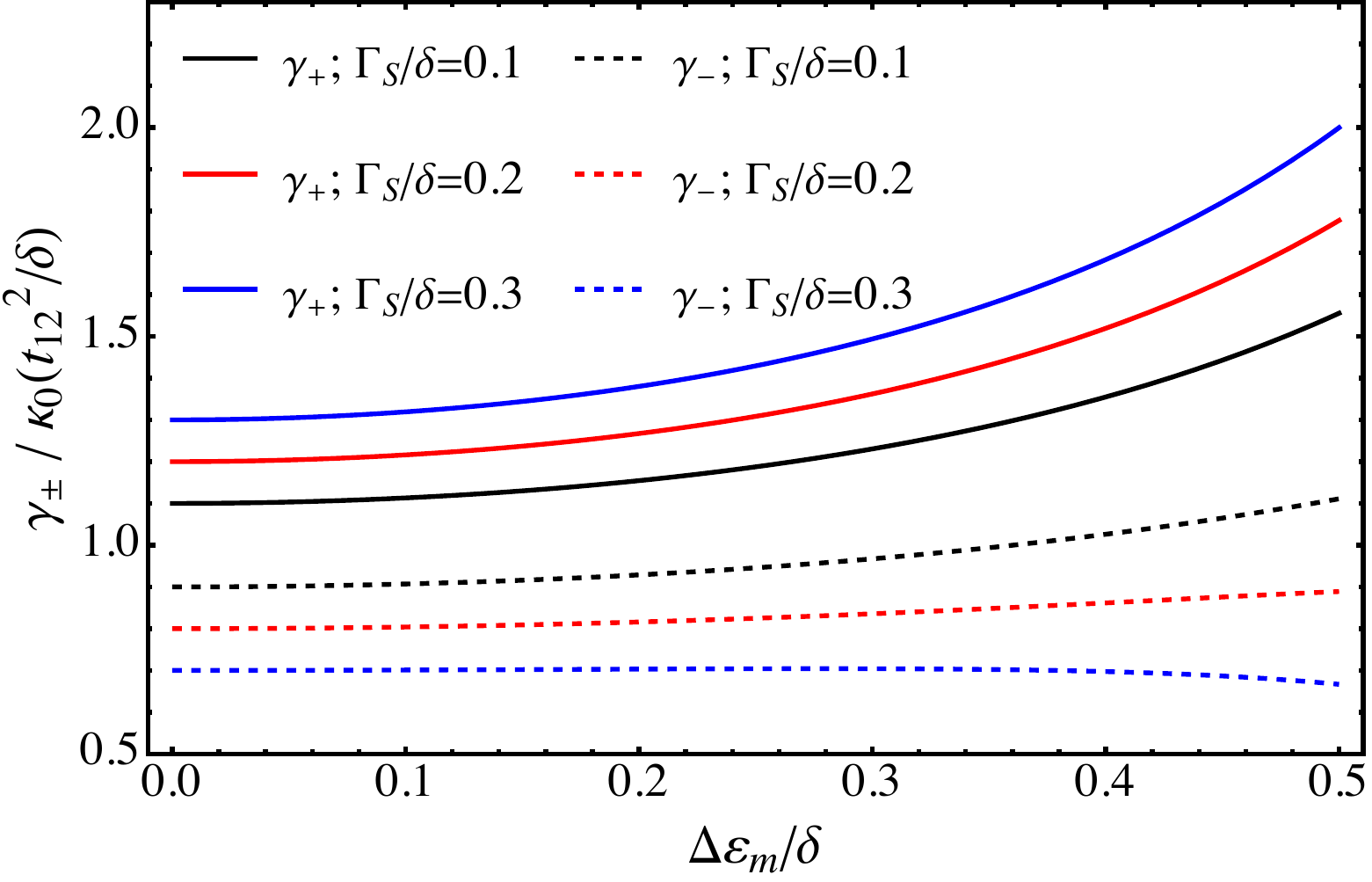}\\[3pt]
\caption{Linewidths of the emitted photon pair as a function of $\Delta\varepsilon_m/\delta$. The linewidths  
$\Gamma_{\pm}$ and $\gamma_{\pm}$ are scaled with 
$\kappa_0\left[ t_{12}^2/\delta \right]$.
}
\label{fig:7}
\end{figure}

The linewidths of the two emitted photons are given by 
%
%
%
%
%
%
%
%
\begin{align}
\Gamma_{\pm} 
&=   
{\left( \frac{t_{12}}{\hbar\omega_{\pm}}\right)}^2
\kappa_0 \big[ \delta \pm \Delta\varepsilon_m  \big]  
\, , \\
\gamma_{\pm}
&
= 
\frac{1}{2}
\left[ 
{\left( \frac{t_{12}}{\hbar\omega_{+}} \right)}^2
 \kappa_0 \left[\delta + \Delta\varepsilon_m \pm \Gamma_S \right]
 \right.
\nonumber\\
&
+
\left. 
{\left( \frac{t_{12}}{\hbar\omega_{-}} \right)}^2
 \kappa_0\left[\delta - \Delta\varepsilon_m \pm \Gamma_S \right] 
 \right]
 \, .
\end{align}
%
%
%
%
%
%
%
%
%
where the function $\kappa_0[\Delta E]$ is defined as 
%
%
%
%
%
%
%
%
%
\begin{equation}
\frac{ \kappa_0
  \left[ \Delta E \right] }{2\pi}
=
 {\left( \frac{C_x}{C_{eff}}  \right)}^2 
\frac{Z_T}{  R_K}    \,\left(  \frac{\Delta E}{\hbar} \right) \, ,
\label{eq:6_k_0}
\end{equation}
%
%
%
%
%
%
%
%
%
with $R_K=h/e^2$, $Z_T$ the characteristic impedance of the line 
and $ \Delta E$ the energy of the emitted photon.
The expression for $C_{eff}$ is given in appendix \ref{app:A}.
In the limit $|\delta| \gg \Delta\varepsilon_m \gg \Gamma_S$ one can obtain a simple estimate of the decay rates 
$
\sim {\left( t_{12}/\delta \right) }^2  \kappa_0\left[\delta \right] 
= 
\kappa_0\left[ t_{12}^2/\delta \right] $. This sets the typical scale for the linewidth.
In Fig.~\ref{fig:7} we plot the linewidths 
scaled with $\kappa_0\left[ t_{12}^2/\delta \right]$
as a function of the ratio $\Delta \varepsilon_m/\delta$.

%
%
%
%
%
%
%
%
%
%

The Eqs.~(\ref{eq:8_wavepacket_1}),(\ref{eq:9_wavepacket_2})  have been obtained assuming that the linewidths  are much smaller than the frequencies of the emitted photons.
This is valid under the condition that the frequency $\hbar\omega_{-}=\delta-\Delta \varepsilon_m$, appearing in the denominator of  $\Gamma_{-} $ 
and $\gamma_{\pm} $, satisfies the condition 
%
%
%
%
%
%
%
%
%
\begin{equation}
\delta-\Delta \varepsilon_m
\gg 
|t_{12}| 
 {\left( \frac{C_x}{C_{eff}}  \right)}
 \sqrt{ \frac{Z_T}{  R_K}  } \, .
\end{equation}
In other words, $\Delta \varepsilon_m$ must not be close to $\delta$, namely the difference between the Zeeman splittings is a small fraction of the detuning.

\addGR{
To summarize, if the system decays from  the nonlocal Andreev bound states $\ket{A_{+}}$ or $\ket{A_{-}}$ ,  
the total wavefunction of the electron and photon system at large times  is given by 
}
%
%
%
%
%
%
%
%
%
\begin{equation}
\addGR{
\ket{ \Psi_{\infty}^{(\pm)}}  
=
\ket{\Psi_{\pm,1}} \otimes  \ket{G_{1}}
+ 
\ket{\Psi_{\pm,2}}   \otimes  \ket{G_{2}}  
}
\end{equation}
%
%
%
%
%
%
%
%
%
%
\addGR{
with the entangled two-photon states 
$ 
\ket{\Psi_{\pm, n}} 
= 
\sum_{k_L,k_R}   
 \tilde{c}_{k_L,k_R,n}^{(\pm)} e^{-i\left( \omega_{k_L}+\omega_{k_R} \right) t}    \ket{{k_L},{k_R}}   
$ 
for  $n=1,2$. 
}

\addGR{
A priori, one can measure the final electronic spin state, $\ket{G_{1}}$ or $\ket{G_{2}}$, 
thereby obtaining a defined photon entangled state. 
This would require an additional spin-sensitive detection in the device. 
Otherwise, if the electrons are not measured, one can trace out the electronic states, yielding:}
\addGR{
\begin{equation}
\hat{\rho}_{\pm}
=
\ket{\Psi_{\pm,1}}\bra{\Psi_{\pm,1}}
+ 
\ket{\Psi_{\pm,2}}\bra{\Psi_{\pm,2}}  \, . 
\label{eq:rho_plus}
\end{equation}
}
We have also verified that the wavepackets Eqs.~(\ref{eq:8_wavepacket_1}) and (\ref{eq:9_wavepacket_2}) are normalized 
$\sum_{k_{L},k_R}\left| \addGR{ \tilde{c}_{k_{L},k_R,1}^{(\pm)} }  \right|^2 =  \sum_{k_{L},k_R}\left|  \addGR{ \tilde{c}_{k_{L},k_R,2}^{(\pm)}  }  \right|^2 =1/2$ 
(see appendix \ref{app:E}),
\addGR{such that $\mbox{Tr}\left[ \hat{\rho}_{\pm} \right]=1$. 
}

\addGR{
As discussed in detail in Appendix \ref{app:H},  
two-photon density matrices of the form in Eq.~(\ref{eq:rho_plus}) are inseparable and, indeed, contain entanglement. 
The crucial point is that the four possible states $\ket{\Psi_{+,1}},\ket{\Psi_{-,1}},\ket{\Psi_{+,2}},\ket{\Psi_{-,2}}$
do not have in common any two-frequency photon states, see Fig.~\ref{fig:NEW}.
In Appendix \ref{app:H}, we quantify this entanglement by calculating the logarithmic negativity, which corresponds to
an upper bound for the distillable entanglement.
In general, entanglement is hidden in the statistical mixture and can be extracted (via distillation) a priori using LOCC (Local Operations and Classical Communication).
In Sec.~\ref{subsec:distillation}, we present a straightforward method for achieving entanglement distillation in our proposal.
}

%
%
%
%
%
%
%
\subsection{\addGR{Entanglement distillation by frequency filtering}}
\label{subsec:distillation}

\addGR{In simple terms,} 
a consequence of the entanglement is the perfect correlations of the measured frequencies in the two waveguides. 
\addGR{If the electronic system starts from $\ket{A_+}$, }
the only possible frequency outcomes for the photonare
$(\omega_{+}+\Gamma_S/\hbar; \, \omega_{-})$ 
and 
$(\omega_{+} ; \, \omega_{-}+\Gamma_S/\hbar)$.
The two entangled photon states,$\ket{\Psi_{+, 1}} $ and 
$\ket{\Psi_{+, 2}} $, 
differ only in the directions of the two photons. 
Then, one detects in the left transmission line a given frequency 
of the two possible ones, with the corresponding correlated frequency 
on the right line.

\addGR{
Similarly, assuming that the system starts from $\ket{A_-}$, 
 if the electrons are not measured, one can trace out the electronic states, yielding
$
\hat{\rho}_{-}
=
\ket{\Psi_{-,1}}\bra{\Psi_{-,1}}
+ 
\ket{\Psi_{-,2}}\bra{\Psi_{-,2}}
$. The two entangled photon states,  $\ket{\Psi_{-, 1}} $ and 
$\ket{\Psi_{-, 2}} $,  differ only in the directions of the two photons. 
Again, entanglement manifests itself in the perfect frequency correlation in the detected photon pairs with frequencies $(\omega_{+}-\Gamma_S/\hbar; \, \omega_{-})$
and $(\omega_{+} ; \, \omega_{-}-\Gamma_S/\hbar)$. 
}

\addGR{These observations suggest that one can reveal the entanglement by using correlated frequency filters.}

\addGR{A more general approach can be adopted by assuming the absence of prior knowledge regarding the initial excited state of the CPS.
Then, in the subspace of two emitted photons in which we fix the possible frequency outcomes, 
we introduce the general density matrix
\begin{equation}
\hat{\rho} =  \,\, p_+ \,\, \hat{\rho}_+ \, + \, p_- \,\, \hat{\rho}_- 
\label{eq:rho_tot}
\end{equation}
where $p_{\pm}$ are the probabilities associated to the two states $\ket{A_{\pm}}$, with $p_{+}+p_{-}=1$.
An analysis of the entanglement contained in Eq.~(\ref{eq:rho_tot}) is presented in Appendix \ref{app:H}.
}

\begin{figure}[tbp]
\begin{flushleft}
\includegraphics[scale=0.24]{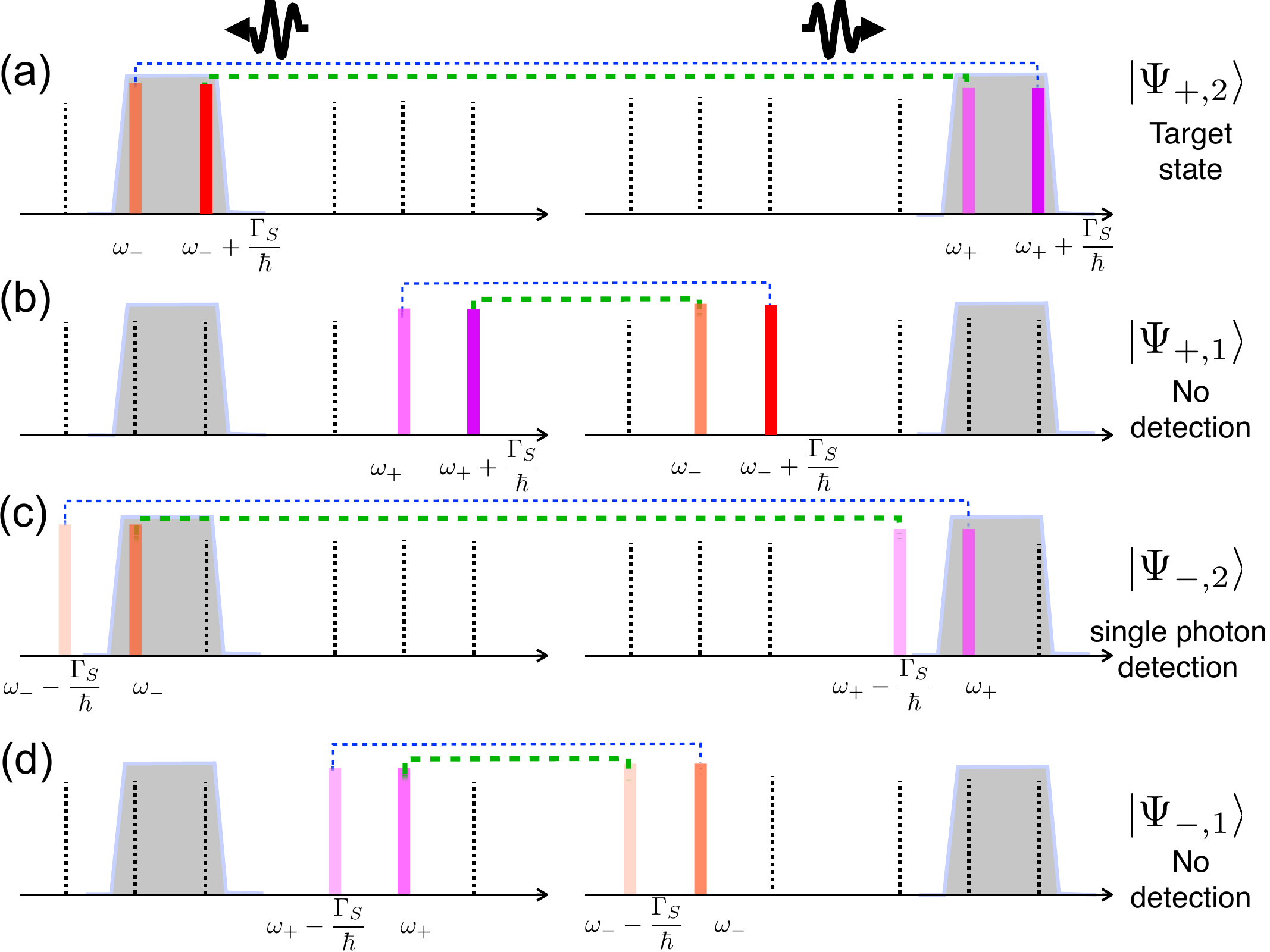} 
\end{flushleft}
\vspace{-2.5mm}
\caption{
\addGR{
Frequency filtering for entanglement distillation. 
In this example, the target entangled state is $\ket{\Psi_{+,2}}$, shown in {\bf (a)}, 
for which we have 
the frequency pairs $\left( \omega_L = \omega_{-}; \omega_R= \omega_{+}+\Gamma_S/\hbar \right)$ and 
$\left( \omega_L = \omega_{-} +\Gamma_S/\hbar; \omega_R= \omega_{+}\right)$.
The dashed horizontal lines (green and blue) connect the two frequencies that occur together in a frequency pair.
}
}
\label{fig:NEW}
\end{figure}

\addGR{
In our problem, a straightforward approach to distill the entanglement involves introducing frequency filters in the left and right transmission lines. 
These filters, with a specific bandwidth, allow for the extraction of the desired entangled state. 
In Fig.~\ref{fig:NEW}, we show an example for the target state $\ket{\Psi_{+,2}}$. 
While it is evident that the states $\ket{\Psi_{+,1}}$ and $\ket{\Psi_{-,1}}$ are not detected as the frequencies of the emitted photons are outside the detection windows, 
we also note that, in the ideal scheme, the simultaneous detection of two photons with frequencies $\omega_L = \omega_{-}$ and $\omega_R = \omega_{+}$ - appearing in Fig.~\ref{fig:NEW}c - 
is not possible, as this state does not appear in the density matrix given in Eq.~(\ref{eq:rho_tot}).
In the latter case one can detect only a single photon on the left or right transmission line.
}

\addGR{
Eventually, if frequency dephasing or other spurious effects affect the measurements, post-selection can be employed 
to exclude data corresponding to events that were not theoretically expected.
 }

%
%
%
%
%
%
%
%
%

%
%
%
%
%
%
%
\section{Reduction of photons emission in presence of phonon-induced charge DQD relaxation}
\label{sec:non-radiative}
\subsection{\addGR{Model and rate equation}}
The energy relaxation in a DQD can have different origins.
For example energy can be released by emitting phonons in the substrate
or by exchanging energy with the electromagnetic environment.
The phonon induced relaxation mechanism for DQD charge qubits has been studied 
in Si/SiGe  \cite{Wang:2013} 
and
in GaAs/AlGaAs \cite{Hofmann:2020}.
In particular, 
the decay rates associated to the phonon-assisted incoherent 
tunneling have been measured in a DQD 
as a function of the detuning $\delta$ and 
in the limit of large detuning $|\delta| \gg  |t_{12}|$.
The experimental results have been also compared with a theoretical model that describes the interaction between the electrons in the DQD  
and the phonons of the substrate \cite{Hofmann:2020}.

For $|t_{12}| \ll |\delta|$, the excited and ground states of the DQD charge qubit 
are strongly localized and 
they essentially corresponds to the states with electron on the QD1  
and on the QD2.
Using Fermi's golden rule, one can write the transition rate 
for the energy decay as \cite{Hofmann:2020} 
%
%
%
%
%
%
%
%
%
%
\begin{equation}
\Gamma_{rel}^{(phon)} \left( \delta \right) 
= 
{\left( \frac{ t_{12} }{ \delta } \right)}^2 
J_{phon} \left( \delta \right)   
\,
,
\label{eq:Gamma_rel_1}
\end{equation}
%
%
%
%
%
%
%
%
%
%
%
where  $J_{phon} (\varepsilon)$  is the so-called interaction spectral density (see appendix \ref{app:G}).
Another possible mechanism for the energy relaxation is based on the dipole interaction between the DQD and the local fluctuating electric field due to charge noise.  
Similarly to Eq.~(\ref{eq:Gamma_rel_1}), 
using Fermi-Golden rule in the regime $|t_{12}| \ll |\delta|$, 
the transition rate for this energy-decay channel is 
(see appendix \ref{app:G}) 
%
%
%
%
%
%
%
%
%
%
\begin{equation}
\Gamma_{rel}^{(ch)}
 \left( \delta \right) 
 = 
 {\left( \frac{ t_{12} }{ \delta } \right)}^2 
 J_{ch} \left( \delta \right) 
\,
,%
\label{eq:Gamma_rel_2}
\end{equation}
%
%
%
%
%
%
%
%
%
%
%
with 
$J_{ch} \left( \varepsilon \right)  =  
\lambda^2_{d} S_{ch} \left( \varepsilon \right) $ 
the interaction spectral density associated to the   local fluctuating electric field.
$S_{ch} \left( \omega \right)$
is 
the noise power spectrum of the electric field \cite{Clerk:2010} 
and $\lambda_d$ is defined in appendix \ref{app:G}. 
Notice that in our notation, $J_{phon}(\varepsilon)$   as well as   $ J_{ch} \left( \varepsilon \right) $ have dimensions of a rate.
Equations~(\ref{eq:Gamma_rel_1}) and (\ref{eq:Gamma_rel_2}) have the same prefactor ${( t_{12}/ \delta )}^2$  
which  
arises from 
the dipole matrix element between 
the two DQD charge states.

In general the electric charge noise is generated 
by an ensemble of two-level fluctuators capacitively coupled 
to the DQD, which generates a 
$J_{ch} \left( \hbar\omega \right) \propto 1/\omega$ 
\cite{Paladino2014}. This type of interaction spectral density is usually not relevant for the energy relaxation.

To estimate the efficiency of the photon emission in the CPS coupled to the two transmission lines, 
we therefore focus on the phonon induced relaxation as the main competing mechanism.
We extract take 
$\Gamma_{rel}^{(phon)} \left( \delta \right)$ 
from the experiment Ref.~\onlinecite{Hofmann:2020}, 
in order to set an upper bound on the phonon 
induced relaxation rate.
Infact this work deals with DQD in GaAs which is a piezoelectric substrate and therefore exhibits a strong coupling with piezoelectric phonons.

Then, by 
inverting Eq.~(\ref{eq:Gamma_rel_1}) 
one can directly 
construct 
the interaction spectral density 
$J_{phon}\left(\delta\right)$.
For the rates associated to energy relaxation processes involving only one DQD in the excited state,
we use the formula Eq.~(\ref{eq:Gamma_rel_1}) replacing 
$\delta$ with $\delta \pm \Delta \varepsilon_m$ 
which takes the Zeeman splitting into account 
[see Eqs.~(\ref{eq:w_rate_1}),(\ref{eq:w_rate_2}),(\ref{eq:w_rate_3}),(\ref{eq:w_rate_4}) in Appendix \ref{app:G}].
On the other hand, 
the energy relaxation rates for transitions starting from the nonlocal Andreev bound state have an extra prefactor $(1/4)$ which stems from the singlet 
and from the coherent superposition of the empty and the singlet state
(see Eqs.~(\ref{eq:w_rate_5},(\ref{eq:w_rate_6}),(\ref{eq:w_rate_7}),(\ref{eq:w_rate_8}) in appendix \ref{app:G}).

%
%
%
%
\begin{figure}[t]
\begin{flushleft}
\includegraphics[scale=0.24]{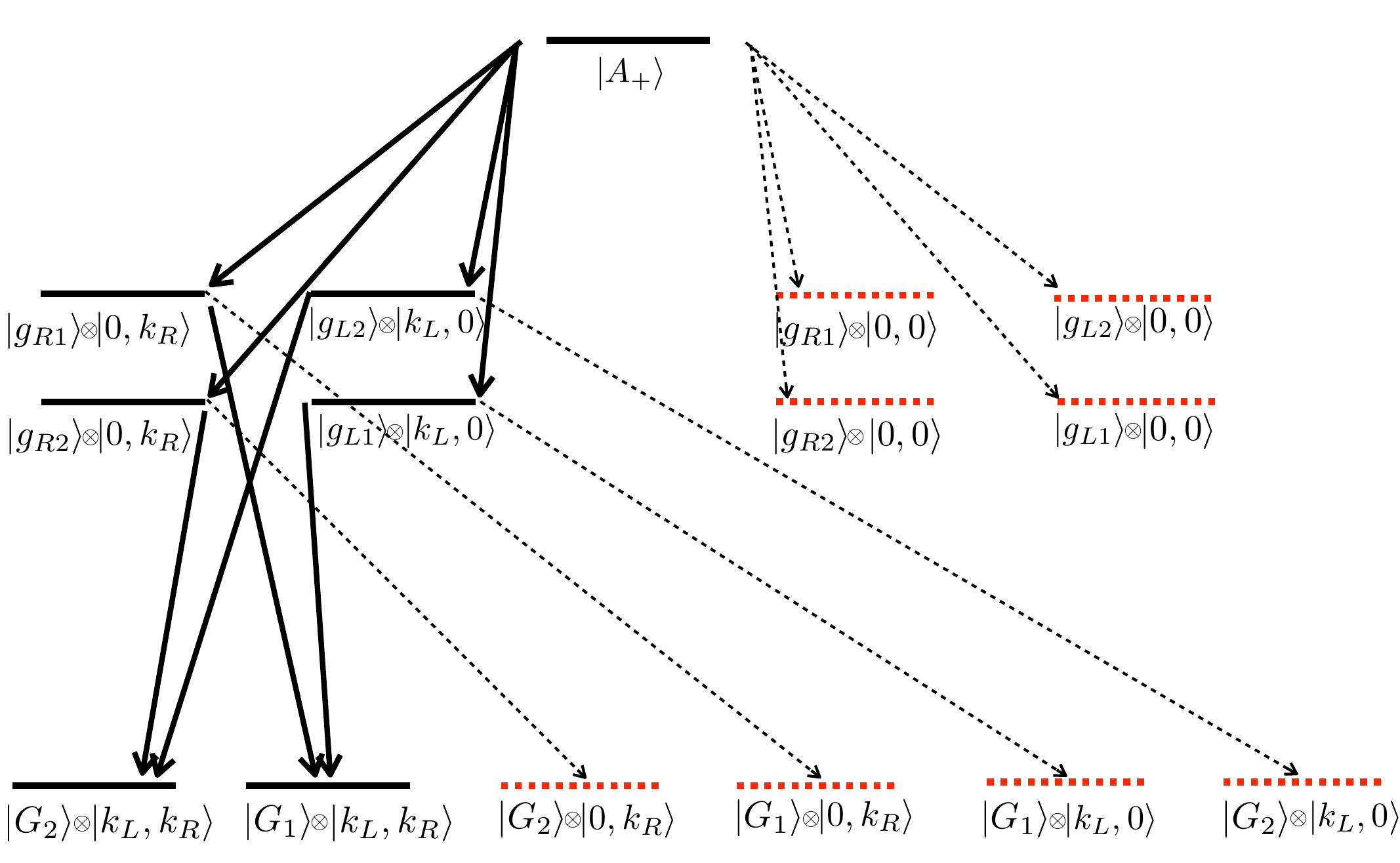} 
\end{flushleft}
\vspace{-2.5mm}
\caption{
Diagram for the transitions between the electronic levels in presence of phonon-assisted processes 
with the initial state 
$\ket{A_{+}}$ 
(see text for the notations and Appendix \ref{app:F}). 
The solid arrows represent processes with one photon emitted, while the dashed arrows indicate processes with one phonon emitted in the substrate of the device. Horizontal red-dotted lines denote final states in which one phonon has been emitted.
}
\label{fig:8}
\end{figure}
%
%
%
%
%

We introduce a rate equation that contains the photon-
as well as the phonon-assisted decay rates.
Fig.~\ref{fig:8}  shows the decay processes contained in the rate equation \addGR{that are relevant to compute the probability of two photons emitted $P_{2-ph} $}. 
We refer to Appendix \ref{app:F} for more details. 
The electronic states associated to the cascade 
pair photon emission have been defined in the previous section, see Fig.~\ref{fig:6}, 
and they are also reported for clarity in Fig.~\ref{fig:8} 
as solid line arrows.
\addGR{
The two-photon emitted states are shown as solid horizontal lines at the bottom of Fig.~\ref{fig:8}, each having two wavevectors, while the single-photon and single-phonon emitted states, represented by horizontal dashed lines at the bottom of Fig.~\ref{fig:8}, have only one wavevector for the photon.
}

We solve the rate equation and 
compute 
the probability of two photons emitted $P_{2-ph} $,
see  Eq.~(\ref{eq:P2ph}) in appendix \ref{app:F}.
A crucial observation is that all the rates 
scale 
as $\propto t_{12}^2$. 
Being $P_{2-ph} $ a function involving the ratio between  photon emission and  phonon assisted rates, 
the resulting $P_{2-ph}$  is independent of $t_{12}$.

\subsection{ \addGR{Results and discussion for the probability of two photon emission}}

\begin{figure}[t]
    \centering
    (a)\phantom{\hspace{0.85\columnwidth}}\\[-12pt]
    \includegraphics[width=0.85\columnwidth]{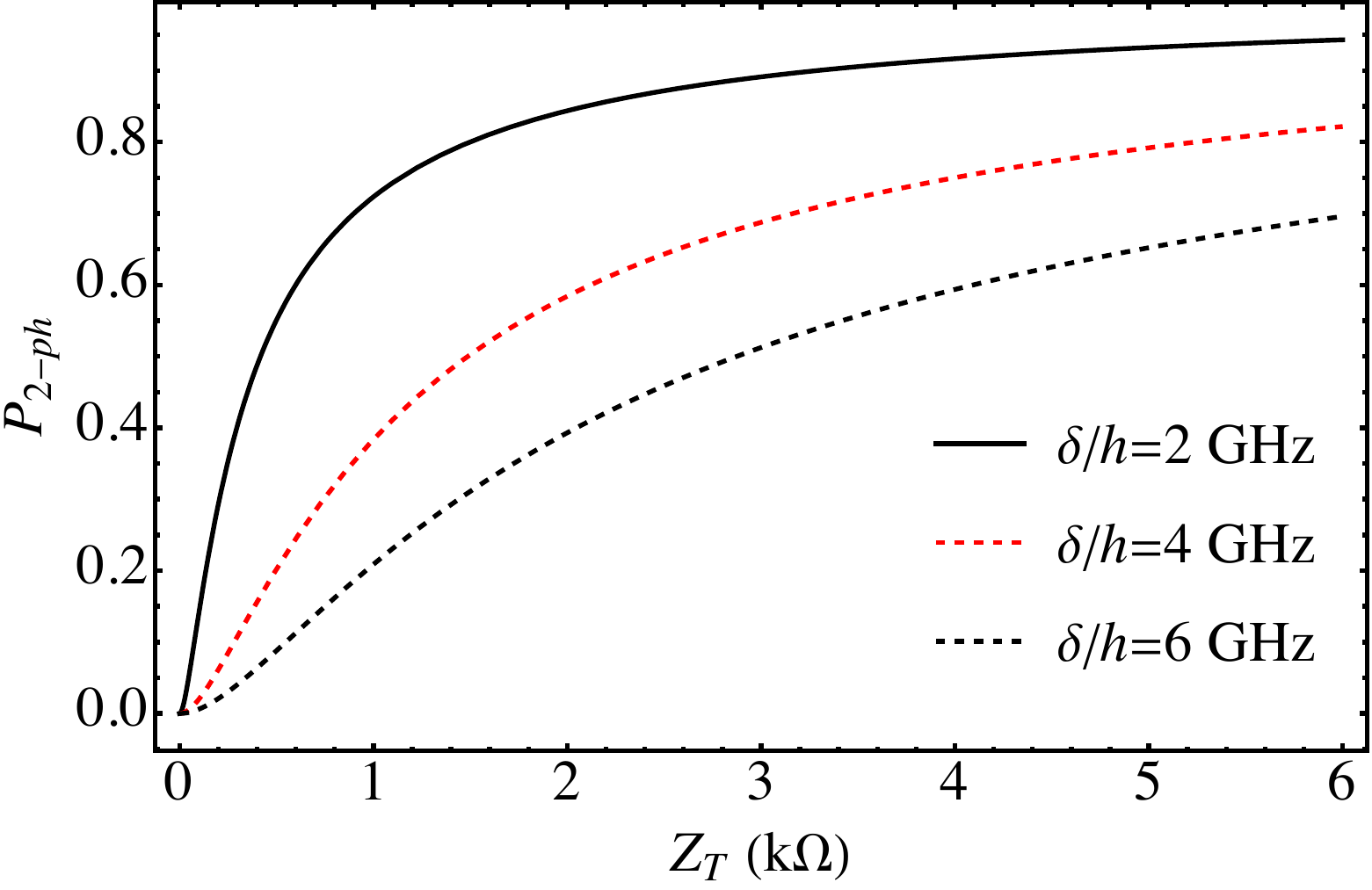}\\[12pt]
    (b)\phantom{\hspace{0.85\columnwidth}}\\[-12pt]
    \includegraphics[width=0.85\columnwidth]{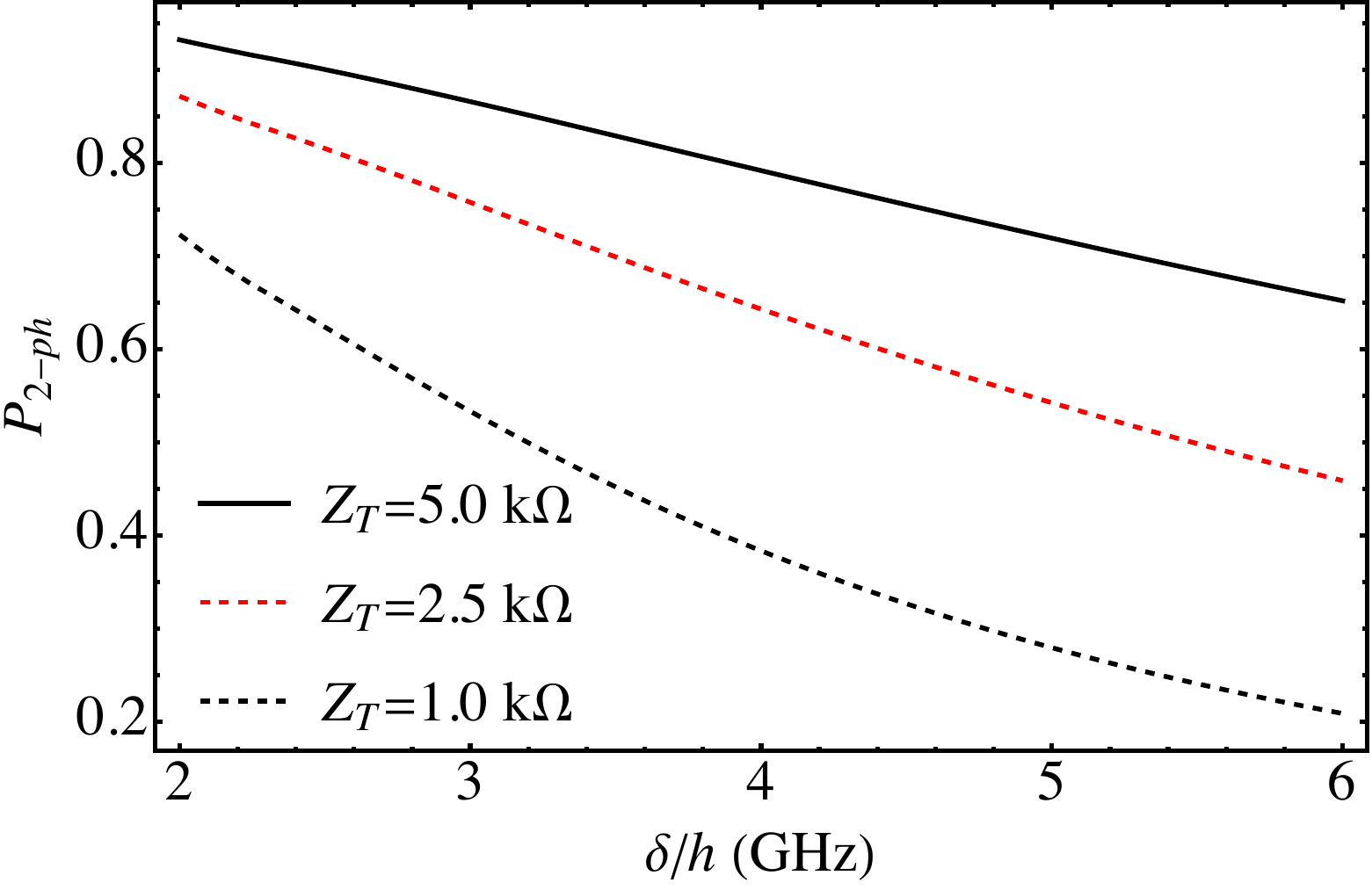}\\[3pt]
\caption{
Propability $P_{2-ph} $ for two photon emission 
in presence of competing phonon assisted decay processes 
in the CPS formed by two QDQs as a function of 
the impedance $Z_T$ (a)
and
the detuning $\delta$ (b).
Parameters: $C_x/C_{eff}=0.5$, 
$\Delta\varepsilon_m/h=0.5$ GHz, $\Gamma_s/h=0.25$ GHz.
}
\label{fig:9}
\end{figure}

Examples of results are reported in Fig.~\ref{fig:9}.
As expected, in Fig.~\ref{fig:9}(a), 
by increasing the characteristic impedance 
of the transmission line $Z_T$, 
we increase the coupling strength 
between the DQD 
and the photons 
such that the photon emission rates are enhanced, 
see Eq.~(\ref{eq:6_k_0}).
By contrast, 
Fig.~\ref{fig:9}(b) shows that increasing the detuning reduces the probabilty of two photon emission.
Infact, 
the photon emission rates 
are proportional to the energy of 
the emitted photon (see Eq.~(\ref{eq:6_k_0})), 
whereas
the phonon assisted relaxation rates are 
proportional 
to $\propto J_{phon}\left(\hbar\omega\right)$ 
which increases faster  
at large detuning. 
The numerical plot of the interaction spectral density $J_{phon}\left(\hbar\omega\right)$,  
as extracted from the experimental data of Ref.\onlinecite{Hofmann:2020} for GaAs/AlGaAs, 
is shown in Appendix \ref{app:G}.

Observing Fig.~\ref{fig:9}, 
we conclude that there is a wide range of 
parameters in which high-efficiency two-photon emission is achieved.
\addGR{Below, we offer a simple and intuitive explanation for this high efficiency.}

\addGR{
For this purpose, it is instructive to consider a semiconductor-based single DQD (a charge qubit with no transport) coupled to a high-impedance waveguide.
The isolated DQD hosts an effective extra charge which can tunnel between the two QDs.
With appropriate detuning, the DQD can be prepared in an excited state, from which it relaxes to the ground state.
In general, relaxation occurs through different channels:
by emitting a photon into the transmission line with rate  $\Gamma_{phot}$, by photon leakage into other circuit elements with rate $\Gamma_{lk}$, or by non-radiative decay with rate 
$\Gamma_{nr}$.
The efficiency of photon emission from a single quantum emitter is defined as follows ~\cite{Peng:2016,Gasparinetti:2017,FornDiaz:2017,Lu:2021} 
\begin{equation}
\eta = \frac{\Gamma_{phot}}{\Gamma_{phot} + \Gamma_{lk} +  \Gamma_{nr}} \, .
\end{equation}
While experiments with DQD charge qubits directly coupled to a transmission line are feasible, they have not yet been reported.
However, efficiency close to one has been experimentally demonstrated in a similar waveguide setup with superconducting qubits \cite{Peng:2016,Gasparinetti:2017,FornDiaz:2017,Zhou:2020,Lu:2021}.
To achieve this goal in this type of emission experiments, the qubit does not need to be long-lived, as long as it is strongly coupled to the waveguide, 
namely $\Gamma_{phot} \gg  \left( \Gamma_{lk},  \Gamma_{nr}\right)$.
In experiments \cite{Peng:2016,Gasparinetti:2017,FornDiaz:2017,Zhou:2020,Lu:2021} using standard $50 \, \Omega$ waveguides, reported linewidths were $\Gamma_{phot} \sim  10 \,\, \mbox{MHz}$
with efficiencies ranging from 60\% to 90\%.
}

\addGR{
In our modeling of a single DQD, 
we assume negligible photon leakage 
$(\Gamma_{lk} =0)$  
and consider only the rates associated with photon emission into the transmission line $\Gamma_{phot}$ and phonon emission into the substrate (non-radiative decay), with  $\Gamma_{nr}^{(phon)}$.
In a line with a characteristic impedance of a few $\mbox{K}\Omega$,  
the typical photon emission rate can then reach approximately 
$\Gamma_{phot} \sim 10 \, \mbox{MHz}$ (see Sec.~\ref{sec:feasibiltiy}), 
which is significantly higher than the typical phonon emission rate, in the range of 
$\Gamma_{nr}^{(phon)} \sim 10-100 \,  \mbox{Hz}$ for a DQD detuning $\delta < 10 \, \mbox{GHz}$ \cite{Hofmann:2020}.
}

\addGR{
Another crucial point is that once the first photon is emitted, leaving the second DQD excited, the likelihood of a second photon emission remains high as long as photon emission dominates over the phonon-induced decay channels. Indeed, coherence between the two photons emitted in the cascade decay by a superconducting emitter coupled to a waveguide was demonstrated in Ref.~\cite{Gasparinetti:2017}. 
}

%
%
%
%
%
%
%
%
%

\section{Effect of DQD charge dephasing}
\label{sec:charge-dephasing}
In this section we discuss the role of charge 
dephasing in DQD \cite{Paladino2014,Scarlino2019} and how it affects the performance of the CPS. 

Recently, multiple experimental studies have achieved longer charge-dephasing times for 
electrons and holes confined in electrostatically defined DQDs \cite{Scarlino2019,DePalma2023,Scarlino2023}.
In GaAs, Refs.~\onlinecite{Scarlino2019,Scarlino2023}  have demonstrated charge linewidth of the order of a few MHz around the  DQD sweet spot ($\delta \sim 0$), extracting a charge noise of 
$\sigma_{\delta} \approx 0.2 \mu$eV.
Assuming that similar linewidths are achieved for the DQDs in our proposal, this  
implies that, for the chosen operational point 
$|\delta| \gg |t_{12}|$, one can estimate a charge dephasing 
linewidth around $\approx 50$ MHz.

Charge noise dephasing can be described as a classical noise source that causes fluctuations in the detuning energy of the two DQDs. 
In this situation entanglement is not destroyed but it is actually hidden due to our lack of classical knowledge on the
system \cite{DARRIGO2014211}. 

However, charge noise is determined by slow fluctuations of the electromagnetic environment and therefore it does not impact directly the single-photon emission processes 
which take place on a much faster time scale.
This charge-noise induced frequency fluctuations must be taken into account in the design of 
the photon detection process, 
namely the photodetector must have a bandwidth 
comparable to the effective charge-dephasing linewidth.

%
%
%
%
%
%
%
\section{Effects of spin-relaxation and dephasing}
\label{sec:spin-relaxation}

In this section  we focus on spin relaxation and dephasing processes \cite{Hanson:2007,Burkard:2023}. 
Both of these phenomena adversely affect the entanglement of the spin-singlet electrons in the CPS. 

As an illustrative scenario, consider the extreme case where spin decoherence occurs immediately after the Cooper pairs are split and injected into the QDs of the CPS.
Under these circumstances, the electron system's initial state transitions from an entangled state to a statistical mixture of two states, each characterized by opposite spins located in the left and right part of the CPS \addGR{(see Fig.~\ref{fig:2}c).}
Nevertheless, despite this decoherence, the two DQDs remain in a doubly excited state. 
Consequently, the system is still capable of emitting correlated photons. 
Such photon correlation serves as an indirect marker of the initial spin-entangled state.

Accurate studies of the spin relaxation processes in largely detuned DQDs 
are lacking since 
this operational regime is atypical for electrically-defined semiconductor spin qubits.

In contrast, extensive research has been conducted on spin qubits defined in single QDs \cite{Hanson:2007}. 
For instance, studies on electron spins confined within gallium arsenide (GaAs) \cite{Scarlino2014,Camenzind2018} and silicon QDs \cite{Zwanenburg2013}, as well as hole spins in germanium QDs \cite{Lawrie2020}, reveal that typical spin relaxation rates are lower than 1 kHz. 
These rates are significantly slower (by three to four orders of magnitude) compared to our estimated rates of photons emission.

In addition, state-of-the-art spin qubits in  germanium \cite{Hendrickx2021} and in 
isotopically purified  
silicon \cite{Veldhorst2014} exhibit spin-dephasing times ranging from 
$1 \mu\mbox{s} < T_2 < 100 \mu\mbox{s}$ corresponding to dephasing rates of  
$1$ MHz and $10$ KHz \cite{Stano2022}.
These rates are also smaller than our estimated rates of photons emission.

%
%
%
\section{Experimental feasibility}
\label{sec:feasibiltiy}

\addGR{
We outline here only the conceptual approach, without providing a detailed blueprint for constructing and characterizing the device.
Clearly, further adjustments are required to bring the device into the correct operational regime,
such as tuning tunnel couplings, assessing the proximity effect, and constructing a charge stability diagram, among other tasks.
We briefly discuss these issues in Appendix \ref{app:I}.
}

In the following we summarize the energy scales involved in the proposed setup in order to realize on demand emission of two frequency-entangled photons. 
Our proposal for yielding two frequency 
entangled photons is based on 
 the regime $\delta > \Delta\varepsilon_m$, 
 $\delta \gg \Gamma_s$ and  $\delta \gg t_{12}$. 
Typical values of the parameters are given in Table \ref{tab:table01}.

We are considering a high impedance 
transmission lines   $Z_T \approx 0.2-0.4 \, R_K = 5-10 \, \mbox{K}\Omega$ 
with flat transmission 
in the vicinity of the frequency of the emitted photons, 
e.g. within GHz bandwidth.

To give a simple estimation, we set  $C_x/C_{eff}=1/2$, $Z_T/R_K=1/5$, 
and the energies $\Delta \varepsilon_m / h = 0.5$ GHz, $\Gamma_S/h=0.25$ GHz, 
$t_{12}/h=0.3$ GHz.
In this way, 
one obtains   the photon linewidths - defined in Sec.~\ref{subsec:calculations}  - 
$\Gamma_{+}/(2\pi) \approx 8$ MHz, $\Gamma_{-}/(2\pi) \approx 13$ MHz,
$\gamma_{+}/(2\pi) \approx 12$ MHz, $\gamma_{+}/(2\pi) \approx 9$ MHz 
at $\delta/h \, =2$ GHz.
Whereas, at $\delta/h \, =8$ GHz,  one has 
$\Gamma_{\pm}/(2\pi) \approx \gamma_{\pm}/(2\pi) \approx 2$ MHz.
In this regime of parameters, the linewidths of the individual photon emissions can be larger than the spin dephasing rate (see discussion in 
section \ref{sec:spin-relaxation}).

\renewcommand{\arraystretch}{2.5} 
\begin{table}[tbp]
\centering
  \begin{tabular}{|p{0.8cm}|p{1.1cm}|p{1.1cm}|c|p{1.1cm}|p{1.1cm}|} 
  \hline
$ 1/h$ & $\quad \,\, \delta$  & $\quad \,\, t_{12}$       & $\Delta\varepsilon_1$, $\Delta\varepsilon_2$ & $\quad \,\, \Delta\varepsilon_m$ &   $\quad \,\, \Gamma_S$       \\
\hline
GHz    & $2 - 8$   & $\,\, \lesssim 0.3$ &$0.25 - 0.75$                                 & $\,\, \lesssim 0.5 $       & $\,\, \lesssim 0.25$  \\
\hline
   \end{tabular}
    \caption{Ranges of the  relevant parameters of the CPS analyzed in this work.}
      \label{tab:table01} 
\end{table}

In an ideal noiseless device,  
we require 
the following conditions 
for the linewidths  
in order to be able to properly resolve the frequency 
of the emitted photons: 
(i) Large enough inhomogeneous Zeeman splitting  
$\Delta \varepsilon_m \gg \left( \gamma_{\pm}, \Gamma_{\pm} \right)$ 
and 
(ii) Large enough tunnel coupling with the 
superconductor $\Gamma_S \gg \gamma_{\pm}$.
The second condition is necessary to achieve photon entangled pairs 
resolvable in frequency.
Eventually, the effects of charge dephasing, as discussed in Section \ref{sec:charge-dephasing}, must be taken into account when collecting multiple single-pair photon measurements.

%
%
%
%
%
%
%
\subsection{
Implementation of a CPS in a 
superconductor-semiconductor hybrid platform}

The extraction of spin-entangled electron pairs in transport has been successfully demonstrated (with the exception of entanglement proof) in 1D semiconducting devices built with either carbon nanotubes\cite{Herrmann:2010,Schindele:2012,Schindele:2014},  or semiconducting nanowires\cite{Hofstetter:2009,Hofstetter:2011,Das:2012,Fulop:2014,Fulop:2015}, predominantly based on III-V semiconductors, particularly InAs.

Recently, the proximitized semiconducting platform has been extended to two-dimensional hetero\-structure, also based on III-V materials \cite{Shabani:2016,Suominen:2017,Wang:2023}. Here, a thin layer of InAs defines a conducting quantum well located close to the surface of the wafer which is, after the growth of the hetero\-structure stack, covered with a thin layer of superconducting Al without braking the vacuum. This 2D platform is much more versatile in particular if more complex structure are considered where also on chip transmission lines are incorporated. All fabrication can be done top-down. In contrast, when starting with nano\-wires, these wires need to be placed with micro\-manipulators. This works in principle reliably for a single device, but it is hardly scalable. Hence, we 
argue that a 2D quantum-well device 
can be a good candidate for realizing our experimental proposal.

On the other hand, a disadvantage of working with III-V materials is their large nuclear moment, which results in a strongly fluctuating magnetic field due to the hyperfine interaction. 
In this respect, 
great success has already been demonstrated with Si/Ge 2D hetero\-structures, also with near-surface Ge quantum wells that can be proximitized by Al as a superconductor in much the same way as in the III-V platform~\cite{Hendrickx2018,Scappucci2021}.
There are two key advantages with the type-IV platform: (i) Si and Ge can be grown free of isotopes carrying nuclear moments, and (ii) one can tune the Fermi energy into the hole band where large $g$-factors can be obtained and where, due to the absence of s-orbital character, the coupling with remaining nuclear spins is further suppressed. Hence, it is expected that much larger spin coherence times, e.g. 
$> 1$\,$\mu$s are feasible than in similar experiments with III-V semiconductors. 
 
\addGR{
To conclude this section, we mention that thin layers of Al, with a thickness of a few to $10 \, \mbox{nm}$, can be highly resilient to magnetic fields when the field is applied parallel to the Al films \cite{Krause:2022}.
Combining this with the fact that both III-V materials, e.g., InAs and InSb, as well as hole states in Ge, can exhibit large g-factors, ensures the feasibility of achieving sufficient Zeeman splitting without destroying superconductivity.
Finally, very recently, it has been demonstrated that a two-dimensional electron gas in indium arsenide can be proximitized even with a thin layer of niobium, which is known to be resilient to much higher magnetic fields \cite{Telkamp:2025}.
}

%
%
%
%
%
%
%
\subsection{Implementation of high impedance transmission line}

A microwave transmission line can be modelled by a network of interconnected lumped element LC-oscillators. 
Its properties are characterized by the inductance per unit length ($L_0$) and capacitance per unit length ($C_0$). 
The phase velocity and characteristic impedance of the electromagnetic field within the transmission line are determined by $v = 1/ \sqrt{L_0 C_0}$ and 
\addGR{$Z_0 = \sqrt{L_0 / C_0}$},  respectively \cite{Pozar}.

Typically, superconducting transmission lines, constructed using geometric inductance and capacitance, exhibit a characteristic impedance of approximately $Z_0 \sim 50 \Omega$, which is significantly lower than the quantum resistance ($R_K\sim 26 \, \mbox{k}\Omega$). 
However, recent studies suggest the feasibility of achieving transmission-line impedances comparable to or higher than the quantum resistance. 
This advancement can be realized by employing Josephson inductance of arrays of Josephson junctions \cite{Masluk2012}, or kinetic inductance of  disorder superconducting alloys \cite{Samkharadze2016}.  
Granular aluminum (GrAl)  has emerged as an alternative material due to its high quality and simplified fabrication process, allowing for the highest achievable kinetic inductance, approximately $L_{\square} \sim nH/ \square$ \cite{Grnhaupt2019}.

High impedance Josephson junction arrays have been extensively explored to construct superinductors \cite{Masluk2012}, a fundamental building block for realizing fluxonium qubits, and to implement multimode high-impedance photonic environments \cite{Kuzmin2019,PuertasMartnez2019} for exploring many-body problems using microwave photons \cite{Mehta2023}. 
Additionally, high-impedance resonators are beneficial for increasing light-matter interactions \cite{Stockklauser:2017}, facilitating the achievement of strong and 
ultrastrong coupling regimes \cite{Scarlino2023} 
by enhancing coupling to vacuum fluctuations. 
Currently, impedances around $5-6 \, \mbox{k}\Omega$ are readily accessible in experiments.

For the realization of spin qubits, applying a finite, non-negligible magnetic field to the device is necessary. In this context, implementing a high-impedance transmission line using high kinetic inductance materials, such as NbN \cite{Niepce2019}, TiN \cite{Shearrow2018}, NbTiN \cite{Samkharadze2016}, or GrAl \cite{Grnhaupt2019}, is advantageous, as these materials have demonstrated resilience to magnetic fields up to Tesla magnetic field strengths.

%
%
%
%
%
%
%
\subsection{Single photon detection in the microwave  regime}
The detection of single photons has emerged as a pivotal ingredient for the development of the field of quantum optics.
The endeavor to transpose this technique into the microwave spectrum encounters notable challenges, primarily attributed to the substantial reduction in photon energy—approximately five orders of magnitude lower than its optical counterpart. 
Nevertheless, the field has witnessed considerable progress, particularly through the development of single microwave photon detectors (SMPD), leveraging the potential of superconducting qubits technologies \cite{Besse2018,Wang2023}. 
An innovative SMPD has been recently introduces in Ref.\cite{Wang2023,Balembois2023}, that capitalizes on a transmon qubit's transition to an excited state upon photon interaction, employing a four-wave mixing process to ensure the irreversible transfer of photon energy.

In addressing the inefficiencies of microwave photon detection, an alternative paradigm has been explored based on semiconducting nanowire DQDs \cite{Khan2021}. 
These components are resonantly coupled to cavities, enabling the efficient and continuous conversion of microwave photons to electrical current. 
This novel photodiode device marks a departure from conventional detection methods, highlighting the potential for continuous, time-correlated photon counting in the microwave domain as  a crucial step forward in quantum information technologies and solid-state system readout applications.

In addition, 
an innovative approach has been the introduction of a microwave photon-to-electron converter, utilizing the unique properties of granular aluminuma high kinetic inductance superconductors, to significantly enhance 
light-matter interactions \cite{Stanisavljevi2023}.
This converter showcases a remarkable quantum efficiency and low dark current, marking a significant leap towards the ideal detection of microwave frequency photons. 
The converter's success is attributed to its innovative use of photoassisted tunneling in a superconducting tunnel junction, closely integrated with a high impedance microwave resonator, thus overcoming the natural limitations of semiconductor-based detectors in capturing microwave frequency photons.

The amalgamation of these cutting-edge developments underscores a concerted effort to address the inherent challenges of microwave photon detection. 
By enhancing light-matter interaction and utilizing innovative semiconducting materials, these advancements present a promising pathway for achieving efficient, continuous microwave single photon detection, thereby facilitating the realization of the proposed device.

%
%
%
%
%
%
%
%
%
\section{Conclusions}
\label{sec:conclusions}

We have proposed a setup for transferring entanglement from electrons in a superconducting nanodevice to microwave photons in transmission lines in a  waveguide circuit QED architecture. 

Specifically, we have investigated a method to generate pairs of frequency-entangled propagating photons starting from a many-body quantum condensate, i.e. a BCS superconductor composed of spin-singlet Cooper pairs.

Beyond the fundamental interest in extracting entanglement from quantum many-body systems, 
having a deterministic on-demand source of frequency-entangled photon pairs 
is crucial for applications in quantum information based on photonic technologies.

In quantum optics (THz frequency range), frequency- (or color-) entanglement between two photons was demonstrated in Ref.~\onlinecite{Ramelow:2009} and verified through nonclassical two-photon interference. Time-resolved detection has also enabled Bell-test measurements \cite{Guo:17,Lingaraju:2022}. 
Generally, advances in integrated photonics are now making frequency-bin systems a promising approach for quantum information processing \cite{Lu:2023apt}.
Similarly, by using a phase modulator operating in the microwave regime to generate frequency sidebands, the photon-frequency entanglement in our proposal can be directly measured via nonclassical two-photon interference and quantum tomography reconstruction \cite{Imany:2018}.

In quantum microwave systems, the quantum superposition of a single microwave photon at two different frequencies 
has been achieved \cite{Zakka:2022}, and theoretical proposals for generating frequency-entangled photon pairs have been put forward \cite{Stolyarov:2022}. 

In conclusion, our all-on-chip scheme represents a promising advancement for quantum-information processing in the microwave domain.

%
%
%
%
%
%
%
%
\begin{acknowledgments}
{\sl Acknowledgments.}---
G.R. acknowledges financial support from the Provincia Autonoma di Trento (PAT) and from the National Quantum Science and Technology Institute through the PNRR MUR Project under Grant PE0000023-NQSTI.
P.S. acknowledges support from the Swiss National Science Foundation through the Grant Ref. No. 200021-200418, and the Grant Ref. No. UeM019-16 – 215928, and from the SERI through grant 101042765 SEFRI MB22.00081.
G.R. thanks Iacopo Carusotto, Daniele De Bernardis, Simone Felicetti, Andr{\'a}s P{\'a}lyi, Stefano Azzini, Giuseppe Falci, Marco Liscidini and Philipp Hauke  
for  stimulating discussions.
\end{acknowledgments}

%
%
%
%
%
%
%
%
\appendix

%
%
%
%
\section{Derivation of the interaction between the DQD and the transmission line}
\label{app:A}
In this appendix, we derive the model Hamiltonian for a DQD  coupled to a microwave transmission line using quantum circuit theory 
\cite{Yurke:1984,Devoret:1997,Clerk:2010,Vool:2017} 
and an electrostatic model for capacitively coupled QDs 
with voltage gates \cite{vanderWiel:2003}.

\begin{figure}[t!]
\begin{center}
\includegraphics[scale=0.26]{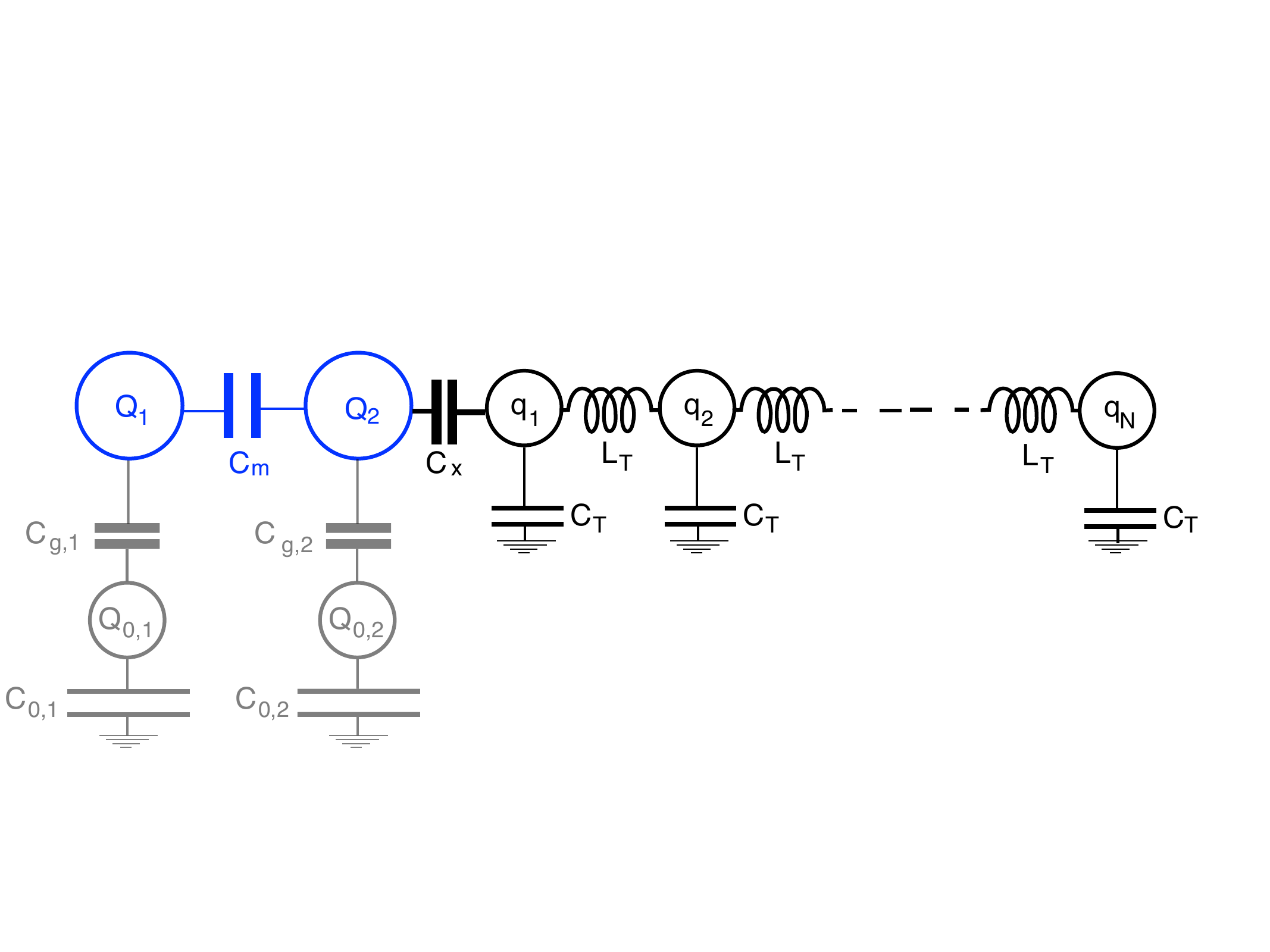}
\end{center}
\caption{A double dot is characterized by charges in the two sites $Q_i$  ($i=1,2$), with an interdot capacitance $C_m$.
Each dot is connected to an island (or node) with charge $Q_{0,i}$ via a gate capacitance $C_{g,i}$.
The nodes with charges $Q_{0,i}$ are also connected to the ground via the capacitance $C_{0,i}$.
The dot 2 is connected via the capacitance $C_x$ to a finite-size transmission line formed by the inductances $L_T$ and 
the ground capacitances $C_T$. Each island of the chain has charge $q_n$ ($n=1,2,\dots,N$).
}
\label{fig:SUP-1}
\end{figure}

We consider two QDs as shown in Fig.~\ref{fig:SUP-1}. 
Each QD has charge $Q_i$ and voltage $V_i$, with $i=1,2$.
In addition to the tunnel coupling (not shown in the figure), the two QDs are capacitively coupled through the interdot capacitance $C_m$.
Each QD is also capacitively coupled to an island (or node) with charge $Q_{0,i}$ via the capacitances $C_{g,i}$.
The nodes with $Q_{0,i}$  are also connected to the ground via the capacitances $C_{0,i}$.
Assuming the limit $Q_{0,i} \rightarrow \infty$  and $C_{0,i} \rightarrow \infty$ and keeping fixed the ratio $Q_{0,i} /C_{0,i}  = V_{g,i}$, 
the two islands model the gate voltages applied to the QDs.
QD 2 is also coupled to the first island of the chain via the capacitance $C_x$.
The chain is formed by discrete lumped elements 
corresponding to inductances $L_T$ and capacitances $C_T$, as shown in Fig.~\ref{fig:SUP-1}, having 
$Z_T=\sqrt{L_T/C_T}$ as characteristic impedance.
The charges and the voltages on the islands of the chain are denoted, respectively, by $q_n$ and $V_{T,n}$  $(n=1,2,\dots,N)$.
In the limit $N \rightarrow \infty$, 
the chain models the semi-infinite transmission line  through which the  microwave photons propagate.

%
%
%
%
%
%
%
%
%

The total electrostatic energy of the system is  
\begin{equation}
E_{el.} =
\frac{1}{2} 
\sum_{i=1}^{2}
\left(
Q_{0,i} V_{g,i} 
+
Q_i V_i
\right)
+
\frac{1}{2} 
\sum_{n=1}^{N} q_n V_{T,n}
\, . 
\label{eq:E_tot}
\end{equation}
From the equations connecting  the charges and the voltages at the nodes with charges $Q_{0,i}$, 
\begin{equation}
Q_{0,i} = C_{g,i} \left(V_{g,i} -  V_i  \right)  + C_{0,i} V_{g,i}   	
\, ,
\end{equation}
we express the gate voltages $V_{g,i}$  as a function of the charge $Q_{0,i}$  
\begin{equation}
V_{g,i} = \frac{Q_{0,i} + C_{g,i} V_i}{C_{0,i} + C_{g,i} }  
\, .
\end{equation}
We insert $V_{g,i} $ into Eq.~(\ref{eq:E_tot}) and 
take the limit $Q_{0,i} \rightarrow \infty$, $C_{0,i} \rightarrow \infty$ while  keeping the fixed ratio $Q_{0,i},/C_{0,i}=V_{g,i}$ such that these nodes represent ideal voltage generators.
We obtain the following expression for the electrostatic energy (omitting constant terms) 
\begin{equation}
E_{el.}
=
 \frac{1}{2} 
\sum_{i=1}^{2}
\left( Q_i + C_{g,i} V_{g,i} \right) V_i
+
 \frac{1}{2} 
\sum_{n=1}^{N} q_n V_{T,n}
\, . 
\end{equation}
The equations connecting the charges and the voltages for the two QDs and the node $1$ of the chain are
\begin{align}
Q_1 &= C_{g,1} \left( V_1- V_{g,1}  \right)  + C_m \left( V_1- V_2  \right)   \,, \\
Q_2 &= C_{g,2} \left( V_2- V_{g,2}  \right)   + C_m \left( V_2- V_1  \right)  + C_x \left( V_2- V_{T,1}  \right) \,, \\
q_1 &= C_T V_{T,1} + C_x \left( V_{T,1} - V_2 \right)  \,.
\end{align}
We introduce the vector of the charges as 
$\vec{Q}_{\mathrm{v}}^{T} \equiv ( Q_1 + C_{g,1} V_{g,1},\,  Q_2 + C_{g,2} V_{g,2}, \,   q_1)$, in which the applied gate voltages appear, 
and 
the voltage vector $\vec{V}_{\mathrm{v}}^{T}= (V_1,\, V_2,\, V_{T,1})$.
Next, we introduce the matrix representation and define the symmetric capacitance matrix $\Bar{\bar{C}} $ as 
\begin{equation}
\Bar{\bar{C}} 
=
\left(
\begin{array}{ccc}
C_{g,1}+C_m 	& -C_m				 & 0 \\
-C_m 		& C_{g,2} + C_m + C_x 	& - C_x \\
0			& - C_x			& C_T +C_x
\end{array}
\right)
\end{equation}
such that $\vec{Q}_{\mathrm{v}} = \Bar{\bar{C}}  \vec{V}_{\mathrm{v}}$.
The electrostatic energy can be expressed as  
\begin{equation}
E_{el.} 
=
\frac{1}{2} \vec{Q}_{\mathrm{v}}^{T} \Bar{\bar{C}}^{-1}  \vec{Q}_{\mathrm{v}}
+
\frac{1}{2} \sum_{n=2}^{N} \frac{q_n^2}{C_T} \,.
\end{equation}
We must now invert the capacitance matrix $\Bar{\bar{C}}$.
This can be done  analytically.  
For the diagonal elements $[1,1]$ and $[2,2]$ 
and the off-diagonal element $[1,2]$, we have the following results
\begin{align}
{\left[ \Bar{\bar{C}}^{-1}  \right]}_{11}\!
& =\!\!
\frac{1}{C_{1,DQD}}
\left[
1-
\frac{
\left( 
\frac{C_m^2}{\left( C_{g,1}+C_m\right)\left( C_{g,2}+C_m\right)}
\right)
\frac{C_x}{C_{2,DQD}}
}
{1+ C_x\left( \frac{1}{C_T} + \frac{1}{C_{2,DQD}} \right)}
\right]
 ,
\\
{\left[ \Bar{\bar{C}}^{-1}  \right]}_{22}
&
=
\frac{1}{C_{2,DQD}}
\left[
1-
\frac{
\frac{C_x}{C_{2,DQD}}
}
{1+ C_x\left( \frac{1}{C_T} + \frac{1}{C_{2,DQD}} \right)}
\right]
\, ,
\\
{\left[ \Bar{\bar{C}}^{-1}  \right]}_{12}
&
=
\frac{1}{C_{12,DQD}}
\left[
1-
\frac{
\frac{C_x}{C_{2,DQD}}
}
{1+ C_x\left( \frac{1}{C_T} + \frac{1}{C_{2,DQD}} \right)}
\right]
\, ,
\end{align}
where we have set 
\begin{align}
\frac{1}{C_{1,DQD} } 
&
=
\frac{C_{g,2}+C_m}{C_{g,1}C_{g,2}+C_m\left( C_{g,1} + C_{g,2} \right)}
\, ,
\\
\frac{1}{C_{2,DQD} } 
&
=
\frac{C_{g,1}+C_m}{C_{g,1}C_{g,2}+C_m\left( C_{g,1} + C_{g,2} \right)}
\, ,
\\
\frac{1}{C_{12,DQD} } 
&
=
\frac{C_m}{C_{g,1}C_{g,2}+C_m\left( C_{g,1} + C_{g,2} \right)}
\, .
\end{align}
For $C_x=0$, we recover the electrostatic energy for the voltage-biased DQD  in terms of the displaced charges, which contains  $C_{1,DQD}$  and $C_{2,DQD}$ 
for the diagonal components and $C_{12,DQD}$ for the off-diagonal component    \cite{vanderWiel:2003}.
The finite coupling $C_x\neq0$ leads to a 
renormalization of the charging energies of the two DQDs.

Then, we look at the elements $[2,3]$ and 
$[1,3]$ that are related to the interactions between the charges of the DQD with the charge at the node $1$ of the chain
\begin{align}
{\left[ \Bar{\bar{C}}^{-1}  \right]}_{23}
&
=
\left( \frac{C_x}{C_T C_{2,DQD} } \right)
\frac{
1
}
{1+ C_x\left( \frac{1}{C_T} + \frac{1}{C_{2,DQD}} \right)}
 \, ,
 \\
{\left[ \Bar{\bar{C}}^{-1}  \right]}_{13}
&=
\frac{C_m}{C_{g,1}+C_m}\,\,
{\left[ \Bar{\bar{C}}^{-1}  \right]}_{23}
\, .
\end{align}
Notice that, even if only QD $2$ is directly coupled to the chain, 
an effective electrostatic interaction exists between the charge of the QD $1$ and the charge of the first node of the chain.
The interaction term is given by $E_{el}^{(int)} =E_{el,1}^{(int)}+E_{el,2}^{(int)}$ with
\begin{align}
E_{el,1}^{(int)}
&
= 
\frac{q_1}{2} 
{\left[ \Bar{\bar{C}}^{-1}  \right]}_{13}
 \left(Q_1+ C_{g,1} V_{g,1} \right) 
\, ,
\\
\label{eq:EL-int}
E_{el,2}^{(int)}
&
= 
\frac{q_1}{2} 
{\left[ \Bar{\bar{C}}^{-1}  \right]}_{23}
 \left( Q_2+ C_{g,2} V_{g,2}  \right) 
\, .
\end{align}
In the limit $ C_m\ll C_{g,1}$, we can
neglect the electrostatic interaction 
between the QD $1$  and the first node of the chain
\begin{equation}
E_{el}^{(int)} 
\simeq E_{el,2}^{(int)}
=
\frac{1}{2} q_1
{\left[ \Bar{\bar{C}}^{-1}  \right]}_{23}
{\big( -|e| n_2 + C_{g,2} V_{g,2} \big) }
\, ,
\label{eq:EL-int-2}
\end{equation}
in which we have introduced the electron number $n_2$ on the dot $2$ ($|e|$ is  the  absolute value of the electron charge).\\

The next step is to diagonalize the transmission line in terms of the eigenmodes.
We start from the quantum Hamiltonian of the line of finite size,
in which we have the electrostatic energy in term of the charge operators $\hat{q}_n$ and the inductive energy as a function of the flux node operators $\hat{\Phi}_n$
with the canonical commutation relation $[\hat{\Phi}_n,\hat{q}_m]= i\hbar \delta_{n,m}$ 
\cite{Yurke:1984,Devoret:1997,Clerk:2010,Vool:2017} 
\begin{equation}
\hat{H}_{T} 
=  
\frac{1}{2} 
{\left[ \Bar{\bar{C}}^{-1}  \right]}_{33} \hat{q}_1^2
+
\sum_{n=2}^{N} \frac{\hat{q}_n^2}{2 C_T} + \sum_{n=1}^{N-1}\frac{ {\left( \hat{\Phi}_{n+1} - \hat{\Phi}_n \right) }^2 }{2 L_T} \, .
\end{equation}
Notice that, due to the coupling with the DQD via $C_x$,  
the capacitance of the first node is renormalized and, instead of $C_T$, the diagonal element reads
\begin{equation}
{\left[ \Bar{\bar{C}}^{-1}  \right]}_{33}
=
\frac{1}{C_T} 
\left( 
\frac{1}{1 +  \frac{ C_x/C_T }{1 + C_x/C_{2,DQD}}   } 
\right) 
 \, .
\end{equation}
In order to simplify the discussion and to have a simple model for the transmission line, we assume $C_T\gg C_x$ such that 
we can neglect the renormalization of the capacitance of the first node. 
This does not affect our main result.
We use the canonical transformation from the local coordinates to the orthonormal eigenmodes with open boundary conditions
\begin{equation}
\left(
\begin{array}{c}
\hat{\Phi}_n \\
\hat{q}_n 
\end{array}
\right)
=  
\sum_{k=0}^{N-1} 
\sqrt{ \frac{2-\delta_{k,0}}{N} } \cos\left( \frac{\pi}{N} k \left[ n-\frac{1}{2}\right] \right)
\left(
\begin{array}{c}
 \hat{\Phi}_k \\
 \hat{q}_k
\end{array}
\right)
\, , 
\end{equation}
with $\left[ \hat{\Phi}_{k_1},\hat{q}_{k_2} \right]= i\hbar \delta_{k_1,k_2}$.
The Hamiltonian of the line reads
\begin{equation}
\hat{H}_{T} =  \sum_{k=1}^{N-1} \left(  \frac{\hat{q}_k^2}{2 C_T} +  \frac{1}{2} C_T \omega_k^2  \hat{\Phi}_k^2 \right)
\, , 
\end{equation}
with $\omega_T= 2/\sqrt{L_T C_T} $ and $\omega_k = \omega_T \sin[ \pi k/(2N) ]$.
The mode with $k=0$ $(\omega_k=0)$ corresponds to a uniform charge distribution that we can disregard.
Introducing the bosonic creation and annihilation operators 
\begin{align}
\hat{\Phi}_k &= \sqrt{ \frac{\hbar}{2\omega_kC_T} } \left( \hat{a}_k^{\phantom{\dagger}} + \hat{a}_k^{\dagger} \right)
\, ,
\\
\hat{q}_k &
=  - i \sqrt{ \frac{\hbar \omega_k C_T}{2} } \left( \hat{a}_k^{\phantom{\dagger}} - \hat{a}_k^{\dagger} \right)
\, ,
\end{align}
and the Hamiltonian of the line takes the standard diagonal form
\begin{equation}
\hat{H}_T =\sum_{k=1}^{N-1}  \hbar\omega_k  \hat{a}_k^{\dagger}  \hat{a}_k^{\phantom{\dagger}} 
\, .
\end{equation}
The terms proportional to $C_{g,2} V_{g,2}$ in Eq.~(\ref{eq:EL-int-2}), that corresponds physically to a voltage applied at the beginning of the line,  
can be removed by applying a canonical transformation that shifts the bosonic operators 
$ \hat{a}_k^{\phantom{\dagger} } \rightarrow  \hat{a}_k^{\phantom{\dagger} } + \bar{\alpha}_k$ 
and therefore we can ignore it.
Finally the interaction terms between the occupation number of QD $2$ and the modes of the line reads 
\begin{equation}
\hat{H}_{int}
 = 
 - \frac{|e|}{2} 
{\left[ \Bar{\bar{C}}^{-1}  \right]}_{23}
 \hat{n}_2 
 \hat{q}_1
=
i \hat{n}_2  \sum_{k=1}^{N-1} g_k  \left( \hat{a}_k^{\phantom{\dagger}} - \hat{a}_k^{\dagger} \right)
\end{equation}
with the coupling strength
\begin{equation}
g_k 
\!\!
=
\!\!
\sqrt{ \frac{\pi  }{2  } } 
\left( \frac{C_x}{ C_{eff} } \right)
\sqrt{ \frac{  Z_T}{  R_K} } \frac{1}{\sqrt{N}} \sqrt{ \frac{\omega_k}{\omega_T} } \sqrt{1 -  {\left( \frac{\omega_k}{\omega_T} \right)}^2  } \hbar \omega_T \, , 
\label{eq:g_k_N}
\end{equation}
with $R_K=h/e^2$ and 
\begin{equation}
\frac{1}{ C_{eff} }
=
\frac{1}{ C_{2,DQD} } 
\left(  \frac{
1
}
{1+ \frac{C_x}{C_{2,DQD}} }
\right)
\, .
\end{equation}
%
%
%
%
%
%

%
%
%
%
\section{Derivation of the function $\kappa_0\left[ \Delta E \right]$}
\label{app:B}
In the next Appendices, we solve the equations for the electronic system, i.e. two DQDs withins a CPS, 
interacting with the two transmission lines.
In the weak coupling regime, we can safely use the Rotating Wave Approximation (R.W.A.).
For this purpose, the relevant quantity is the following response function  
\begin{equation}
K_{\Delta E}(t) 
= 
\frac{1}{\hbar^2}
 \sum_{k=1}^{N-1} 
 g_k^2 \,\, e^{-i \left( \omega_k  - \frac{\Delta E}{\hbar} \right) \,  t}    \,.
\label{eq:Kernel}
\end{equation}
Using the density of the states of the lines
\begin{equation}
\rho(\omega) 
=  
\lim_{N\rightarrow\infty} \frac{1}{N} \sum_{k=1}^{N-1} \delta\left(\omega-\omega_k \right)
=
\frac{2}{\pi \omega_T} 
\,\,  
\frac{\theta(\omega) 
\,\, \theta(\omega_T-\omega)}{\sqrt{1 -   {\left( \frac{\omega}{\omega_T} \right)}^2   }}
\, ,
\end{equation}
assuming the limit $N\rightarrow \infty$ and transforming the  discrete sum over the frequencies in an integral,
we cast the function $K_{\Delta E}(t) $ as 
\begin{equation}
K_{\Delta E}(t) 
\!
=  \!\!
{\left( \frac{C_x}{ C_{eff} } \right)}^2
\!
\frac{Z_T}{  R_K}   
\!
\int^{\omega_T}_{0}  \!\!\!\!\!\! d\omega
\omega 
\sqrt{ 1 -  {\left( \frac{\omega}{\omega_T} \right)}^2   }
 e^{-i \left( \omega  - \frac{\Delta E}{\hbar} \right)  t}   .
\end{equation}
In the calculations of the photons emitted by the electronic transitions, 
we need to compute convolutions of the form 
\begin{align*}
  \int_{t_0}^{t}  \!\!\!\! dt' 
K_{\Delta E}\left( t-t' \right)  c(t').   
\end{align*}
The time scale for the decay of $K_{\Delta E}(t-t') $  is $1/\omega_T$, i.e. the high-frequency cut-off of the line.
If the characteristic time scale for the evolution of the function $c(t)$ is much smaller than $1/\omega_T$, 
the system is in the Markovian regime and we can write
\begin{equation}
\int_{t_0}^{t}  \!\!\!\! dt' 
K_{\Delta E}\left( t-t' \right)  c(t')  
\simeq
c(t)
 \int_{-\infty}^{t} \!\!\!\!\!\! dt'  K_{\Delta E}\left( t-t' \right) 
\label{eq:Markov}  \, ,
\end{equation}
with
\begin{equation}
\int^{\infty}_{0}  \!\!\!\!\!\!  d\tau K_{\Delta E}\left( \tau \right)
\!
=
\!
 \mbox{Re}\left[ \int^{\infty}_{0}  \!\!\!\!\!\!  d\tau  K_{\Delta E}\left( \tau \right)     \right] 
 + 
i 
 \mbox{Im}\left[  \int^{\infty}_{0}  \!\!\!\!\!\!  d\tau K_{\Delta E}\left( \tau \right)     \right] 
\end{equation}
in which the real part of the integral is equal to the function 
\begin{align}
\frac{ \kappa_0\left[ \Delta E \right] }{2}
&
=
\mbox{Re}
\left[ 
\int^{\infty}_{0}   \!\!\!\!\!\!  d\tau K_{\Delta E}\left( \tau \right)
\right] 
\nonumber 
\\
&
=
\pi  
{\left( \frac{C_x}{ C_{eff} } \right)}^2
\frac{Z_T}{  R_K}   
\,\,  
 \left(\frac{\Delta E}{\hbar} \right) 
\sqrt{ 1 -  {\left(    \frac{\Delta E}{\hbar \omega_T}   \right)}^2   } \nonumber
 \\
& 
\simeq  
\pi 
\,
{\left( \frac{C_x}{ C_{eff} } \right)}^2
\,
\frac{Z_T}{  R_K}     \left(\frac{\Delta E}{\hbar} \right)     \, .
\label{eq:Kappa_0}
\end{align}
where in the last equality of Eq.~(\ref{eq:Kappa_0}) we have assumed $\Delta E \ll\hbar \omega_T$.
The function $\kappa_0\left[ \Delta E \right]$ is related to the photon emission rate, as reported in the main text and as explained in the following Appendices.
As expected, in waveguide QED with QDs coupled to a 1D transmission line via a capacitance (charge-charge interaction), the rate is directly proportional to the energy of the emitted photon, namely 
$\propto \Delta \omega^{d}$ with $d=1$ the space dimension.

In summary, we have demonstrated that the Markovian limit consists of replacing  
\begin{equation}
\frac{1}{\hbar^2} \sum_{k=1}^{N-1} g_k^2 \,\, e^{-i \left( \omega_k  - \frac{\Delta E}{\hbar}  \right) \, t}
\longrightarrow 
\frac{\kappa_0 \left[ \Delta E \right]}{2}
 \,\, \delta\left(   t \right)  \, .
\label{eq:Markov_approx}  
\end{equation}
Furthermore, integrating  in time the left and right side of Eq.~(\ref{eq:Markov_approx})   from $0^-$ to $\infty$, we find the relation
\begin{equation}
\frac{\pi}{\hbar^2} 
\sum_{k=1}^{N-1} g_k^2 \,\, \delta\left( \omega_k  -  \frac{\Delta E}{\hbar}  \right)
=
\pi 
{\left( \frac{C_x}{ C_{eff} } \right)}^2
\frac{Z_T}{  R_K}    \,   \left( \frac{\Delta E}{\hbar} \right)   \,. 
\label{eq:kappa_0_def}
\end{equation}
Before finishing this Appendix, we analyze the normalization in frequency space.
We evaluate the following sum in the Markovian approximation and in the large high-frequency cut-off limit
$\omega_T\gg\Delta E/\hbar$ 
\begin{align}
&
\lim_{N\rightarrow\infty} 
\sum_{k=1}^{N-1}  
\frac{ g_k^2  / \hbar^2}{ {\left(\omega_k - \frac{\Delta E}{\hbar} \right)}^2  + {\left[ \frac{\kappa_0[\Delta E]}{2} \right]}^2 }
=
\nonumber \\
&
=
{\left( \frac{C_x}{ C_{eff} } \right)}^2  \frac{Z_T}{ R_K}  
\int^{\omega_T}_{0}\!\!\!\!\!\! d\omega \,\, 
\frac{\omega \sqrt{1 - \frac{\omega^2}{\omega_T^2}} }{ {\left(\omega -  \frac{\Delta E}{\hbar} \right)}^2  + {\left[ \frac{\kappa_0[\Delta E]}{2} \right]}^2 }
\nonumber \\
&
\simeq 
\int^{+\infty}_{-\infty}\!\!\!\!\!\! d\omega \,\, 
\frac{\hbar\omega}{\Delta E}
\underbrace{ \frac{\frac{1}{\pi}   \frac{\kappa_0[\Delta E]}{2}  }{ {\left(\omega - \frac{\Delta E}{\hbar} \right)}^2  + {\left[ \frac{\kappa_0[\Delta E]}{2} \right]}^2 }}_{\approx \delta(\omega-\Delta E/\hbar)}
= 1 \, .
\end{align}

%
%
%
%
\section{The model Hamiltonian of the CPS with two DQDs}
\label{app:C}
In this appendix we discuss the electronic system formed by two DQDs and the superconducting contact forming the CPS.
As discussed in the main text, 
by opportunely setting the controlling gate voltages applied to a single DQD, 
one can achieve a situation in which the relevant states $(N_1,N_2)$  are   $(1,0)$, $(0,1)$ and $(0,0)$.
To simplify the notation, we assume the symmetric case with equal parameters for  the left and the right DQD.

Then the Hamiltonian of a single DQD  
is $(\nu=L,R)$ 
\begin{align}
&
\hat{H}_{DD,\nu} 
=  
 \left( \bar{\varepsilon} + \frac{\delta}{2}  \right) \hat{n}_{\nu,1}
+ \left( \bar{\varepsilon} - \frac{\delta}{2}  \right) \hat{n}_{\nu,2}
\nonumber \\
&
+ \Delta\varepsilon_1 
\left( \hat{n}_{\nu,1,\uparrow} - \hat{n}_{\nu,1,\downarrow} \right) 
+ 
\Delta\varepsilon_2  
\left( \hat{n}_{\nu,2,\uparrow} - \hat{n}_{\nu,2,\downarrow} \right) 
\nonumber \\
\nonumber
&
+ U_{\text{intra}} \hat{n}_{\nu,1,\downarrow} \hat{n}_{\nu,2,\downarrow}
+U_{\text{inter}}\left( \hat{n}_{\nu,1,\downarrow} \hat{n}_{\nu,1,\downarrow}+
\hat{n}_{\nu,2,\downarrow} \hat{n}_{\nu,2,\downarrow}\right)
\\
&
 +\sum_{\sigma=\uparrow,\downarrow} \left[ \frac{ t_{12}}{2}  \,  \hat{d}_{\nu,1,\sigma}^{\dagger}  \hat{d}_{\nu,2,\sigma}^{\phantom{\dagger}} +\mbox{h.c.} \right] 
\,,  
\label{eq:H_dd_nu}
\end{align}
where the last term corresponds to the spin-conserving tunneling between  QD$1$ and  QD$2$.
We assume that both the intra-dot, $U_{\text{intra}}$, and the inter-dot, $U_{\text{inter}}$, Coulomb repulsion energies are large and therefore the double occupation of the DQD is suppressed.
Setting $\Delta \varepsilon_m = (\Delta \varepsilon_2 - \Delta \varepsilon_1)/2$, 
with $\delta > \Delta \varepsilon_m > 0$, 
the eigenstates of the DQD with one electron are delocalized 
and they have the following energy levels
\begin{align}
E_{\pm,\uparrow}
&=  
\bar{\varepsilon} 
+
\frac{\Delta\varepsilon_1+ \Delta\varepsilon_2}{4} 
\pm
\sqrt{ {\left(  \frac{\delta -  \Delta\varepsilon_m  }{2} \right)}^2 + 
{\left|  \frac{t_{12}}{2}  \right|}^2 } \, ,
\\ 
E_{\pm,\downarrow}
&=  
\bar{\varepsilon} 
-
\frac{\Delta\varepsilon_1+ \Delta\varepsilon_2}{4} 
\pm
\sqrt{ {\left(  \frac{\delta +  \Delta\varepsilon_m  }{2} \right)}^2 + 
{\left|  \frac{t_{12}}{2}  \right|}^2 }
\, .
\end{align}
Hereafter we assume the limit $|t_{12}| \ll  \delta \pm  \Delta\varepsilon_m$.
Then the energy differences of the levels for each spin are simply 
\begin{align}
E_{+,\uparrow}  -E_{-,\uparrow}     & \simeq \delta - \Delta \varepsilon_m \, , \\
E_{+,\downarrow}-E_{-,\downarrow}   & \simeq \delta + \Delta \varepsilon_m \, . 
\end{align}
We set the mixing angles
\begin{equation}
\sin\left(\theta_{\uparrow}\right) 
= \frac{ t_{12} }{  \sqrt{ {\left(  \delta - \Delta \varepsilon_m  \right)}^2 +{\left|  t_{12}  \right|}^2 } }
\simeq
\frac{t_{12}}
{  \delta  - \Delta \varepsilon_m  } 
\, ,
\end{equation}
\begin{equation}
\sin\left(\theta_{\downarrow}\right) 
= \frac{ t_{12} }{  \sqrt{ {\left(  \delta + \Delta \varepsilon_m  \right)}^2 +{\left|  t_{12}  \right|}^2 } }
\simeq
\frac{t_{12}}
{  \delta  + \Delta \varepsilon_m  } 
\, ,
\end{equation}
and the notation
\begin{equation}
s_{\sigma} = \sin\left(\frac{ \theta_{\sigma} }{2} \right) 
\,,
\quad
c_{\sigma} =  \cos\left(\frac{ \theta_{\sigma} }{2} \right) 
\, .
\end{equation}
For each DQD ($\nu=L,R$) the eigenstates can be expressed as 
\begin{equation}
\ket{\nu, +,\sigma}
= 
c_{\sigma} \ket{\nu,1,\sigma} -  s_{\sigma}  \ket{\nu,2,\sigma} 
\approx 
 \ket{\nu,1,\sigma}  
\, ,   
\end{equation}
\begin{equation}
\ket{\nu,-,\sigma}
=  
s_{\sigma}  \ket{\nu,1,\sigma} +  c_{\sigma}  \ket{\nu,2,\sigma} 
\approx   
 \ket{\nu,2,\sigma}  \, . 
\end{equation}
We now consider the two DQDs as a whole system.
For the next discussion, we only need  the level with total spin component $S_z=0$ in each subspace. 

 It is useful to decompose the energy spectrum in four subspaces, with label $s,s'=\pm$  for the two DQDs.
These subspaces correspond to the cases in which 
(i) both DQDs are in the excited state $(+,+)$, 
(ii) the left DQD in the excited and the right one in the ground state $(+,-)$ and (iii) viceversa $(-,+)$,
and finally (iv) both DQDs are in the ground state $(-,-)$.
We denote the energies of these states as $E_{ss'}^{(\uparrow\downarrow)}$  for the left DQD $(s)$  with spin $\uparrow$ and right  DQD $(s')$ with spin 
$\downarrow$.
Similar definition holds for $E_{ss'}^{(\downarrow\uparrow)}$.

In the subspace $(+,+)$ we have two degenerate  energy levels  corresponding to two electrons with opposite spins
\begin{equation}
E_{++}^{(\uparrow\downarrow)}=
 E_{++}^{(\downarrow\uparrow)}
=  E_{+,\uparrow} + E_{+,\downarrow} 
\simeq 2  \bar{\varepsilon} + \delta    
\,.
\end{equation}
In presence of an inhomogeneous Zeeman splitting,  
the subspaces $(+,-)$ and $(-,+)$ are spanned by two nondegenerate 
energy levels 
\begin{align}
E_{+-}^{(\uparrow\downarrow)} =  E_{-+}^{(\downarrow\uparrow)}  &
= E_{+,\uparrow}  + E_{-,\downarrow}
\simeq 
2  \bar{\varepsilon} -  \Delta \varepsilon_m \, , \\
E_{-+}^{(\uparrow\downarrow)} =  E_{+-}^{(\downarrow\uparrow)}   &= E_{+,\downarrow}  + E_{-,\uparrow}
\simeq 
2  \bar{\varepsilon} +  \Delta \varepsilon_m   
\, .
\end{align}
Finally, in the subspaces $(-,-)$   we have two degenerate  energy levels  corresponding to two electrons with opposite spins
\begin{equation}
E_{--}^{(\uparrow\downarrow)}= 
E_{--}^{(\downarrow\uparrow)}
= E_{-,\uparrow}  + E_{-,\downarrow}
\simeq 2  \bar{\varepsilon}  - \delta 
\, .
\end{equation}

We now analyze the case when the QDs $1$ of each DQD is tunnel-coupled to a superconducting nanocontact. 

Since the double occupation of a single DQD is forbidden, 
then the  relevant states are the empty state and the singly-occupied states of each DQD
with a single electron delocalized on two different QDs.

As the superconductor is a $s$-wave BCS with singlet states for the Cooper pairs, the subspace of the triplet states $S=1$ is decoupled from the singlet and the empty states.
In the large superconducting gap limit, 
the effective Hamiltonian $ \hat{H}_{CPS} $ connects only the states $\ket{0}$ (empty state) and $\ket{S_{11}}$, the delocalized singlet in the QDs $1$
\cite{Rozhkov2000,Recher:2001,Meng2009,Eldridge:2010,Trocha:2015}
\begin{equation}
\hat{H}_{CPS} 
=
-  \Gamma_S 
\Big(
\ket{0}\bra{S_{11}} + \ket{S_{11}}\bra{0} 
\Big)
\, .
\end{equation}
One can express the delocalized singlet $\ket{S_{11}}$ in terms of the delocalized eigenstates of the two DQDs 
- see Eq.~(\ref{eq:singles_ss}) 
\begin{align}
&
\ket{S_{11}}
= \nonumber \\
&
=
c_{\uparrow} c_{\downarrow}   \ket{S_{++}}
+
c_{\uparrow}  s_{\downarrow} \ket{S_{+-}}
+
c_{\downarrow}  s_{\uparrow}  \ket{S_{-+}}
+
s_{\uparrow} s_{\downarrow} \ket{S_{--}} 
\, .
\end{align}
We consider the following case
\begin{itemize}
\item the hybridization angle is small 
$ \sin\theta_{\sigma} \simeq \theta_{\sigma} \ll 1 $; 
\item the coupling with the superconducting contact is smaller than the typical energy spacing  beetween the subspaces $(++),(+-),(-+),(--)$, namely 
$\Gamma_S \ll \delta$;
\item the singlet level $\ket{S_{++}}$ is resonant with the empty state of the two DQDs $\ket{0}$, namely $E_{+,+} \simeq  E_{0}$ .
\end{itemize}
Hence, we can approximate the Hamiltonian $\hat{H}_{CPS}$ as
\begin{equation}
\hat{H}_{CPS} 
\simeq
- 
\Gamma_S 
\Big(
\ket{0}\bra{S_{++}} + \ket{S_{++}}\bra{0} 
\Big)
\, .
\end{equation}
The electron-photon interaction is spin-indipendent and therefore the electronic transitions conserve the angular momentum.
Therefore we can  continue to disregard the triplet states 
as well as  
the states with $S_z=\pm 1$ in the subspaces $(+,-)$,  $(-,+)$ and $(-,-)$.

To analyze the electronic system, it is useful to use a mixed representation in which we consider the singlet $\ket{S_{++}}$ $(S=S_z=0)$ 
for the subspace  $(++)$ whereas 
we use the factorized basis $\ket{\uparrow\downarrow},\ket{\downarrow\uparrow}$ 
for the subspaces $(+-),(-+),(--)$.
Since the state $\ket{0}$ and $\ket{S_{++}}$ are resonant, they hybridize and form two coherent states, Andreev bound states doublet
\begin{equation}
\ket{A_{\pm}} = \frac{1}{\sqrt{2}} \left[ \ket{S_{++}} \mp \ket{0}  \right] \, ,
\end{equation}
with energies
\begin{equation}
E_{A_{\pm}}=    E_{++} \pm  \Gamma_S 
\, .
\end{equation}

In conclusion, in the regime and approximations discussed above, 
the Hamiltonian of the two DQDs in the CPS in the diagonal form 
reads 
\begin{equation}
\hat{H}_{el}^{(S_z=0)}
= 
\hat{H}_{el} ^{(++)}  + \hat{H}_{el} ^{(+-)} + \hat{H}_{el} ^{(-+)} + \hat{H}_{el} ^{(--)} \label{eq:H_el_eff} \\
\end{equation}
with
\begin{align}
\hat{H}_{el} ^{(++)} 
 &=
E_{A_{+}} \hat{\Pi}_{A_{+}}
+
E_{A_{-}} \ \hat{\Pi}_{A_{-}}  \, ,
\\
\hat{H}_{el} ^{(+-)} &= E_{+-}^{\uparrow\downarrow} 
\hat{\Pi}_{ \left(L, +,\uparrow ; R,-,\downarrow \right) } 
+
E_{+-}^{\downarrow\uparrow} \hat{\Pi}_{\left( L,+,\downarrow ; R,-,\uparrow \right) }  \, ,\\
 \hat{H}_{el} ^{(-+)}  &=
E_{-+}^{\uparrow\downarrow} 
\hat{\Pi}_{\left( L,-,\uparrow ;   R,+,\downarrow \right) }
+
E_{-+}^{\downarrow\uparrow} 
\hat{\Pi}_{\left(L, -,\downarrow ; R, +,\uparrow \right) } \, ,  \\
 \hat{H}_{el} ^{(--)} & =
E_{--}\Big(  \hat{\Pi}_{\left( L,-,\uparrow; R, -,\downarrow\right) }
+
\hat{\Pi}_{\left( L,-,\downarrow ; R, -,\uparrow \right) } 
\Big)
  \, .
\end{align}
We have introduced the projection operators as
\begin{equation}
\hat{\Pi}_{\left( X \right)} \equiv \ket{X} \bra{X} \,
\end{equation}
with $\ket{X}$ eigenstates of the system.

%
%
%
%
\section{The effective electron-photon interaction of the CPS with two DQDs}
\label{app:D}
The Hamiltonian of the photons in the two transmission lines is 
\begin{equation}
\hat{H}_{ph} 
=
\sum_{k_L} \hbar \omega_{k_L}  \hat{a}_{k_L}^{\dagger}  \hat{a}_{k_L}^{\phantom{\dagger}}
+
\sum_{k_R} \hbar \omega_{k_R}  \hat{a}_{k_R}^{\dagger}  \hat{a}_{k_R}^{\phantom{\dagger}} \, .
\end{equation}
After eliminating the constant terms proportional to the applied gate voltages, 
the electron-photon interaction between the electron states in each DQD with the corresponding half transmission line is given by
\begin{align}
\hat{H}_{int} 
&
=
i \sum_{k_L} g_{k_L} \left(  \hat{a}_{k_L}^{\phantom{\dagger}} - \hat{a}_{k_L}^{\dagger}  \right) \hat{n}_{2,L}
\nonumber \\
&
+
i \sum_{k_R}  g_{k_R} \left(   \hat{a}_{k_R}^{\phantom{\dagger}} -  \hat{a}_{k_R}^{\dagger} \right) \hat{n}_{2,R}
  \, .
\end{align}

We discuss the interaction between a single DQD and a single transmission line.
Using the projection operators, we consider the delocalized eigenstates of the DQD, namely the electronic states  of the separated system 
corresponding to the two DQD  for  $\Gamma_S=0$
\begin{equation}
\hat{n}_{2,L} = \hat{n}_{2,L}^{(diag)}  +\hat{n}_{2,L}^{(off)} 
\label{eq:n2L_expanded} 
\end{equation}
with 
\begin{align}
\hat{n}_{2,L}^{(diag)}
& =
s_{\uparrow}^2 \hat{\Pi}_{(L+\uparrow)} + s_{\downarrow}^2 \hat{\Pi}_{(L+\downarrow)} +
c_{\uparrow}^2 \hat{\Pi}_{(L-\uparrow)} + c_{\downarrow}^2 \hat{\Pi}_{(L-\downarrow)} \, ,
\label{eq:n2L_expanded_diag} 
\\
\hat{n}_{2,L}^{(off)} 
=
& 
-  c_{\uparrow} s_{\uparrow} \left(  \hat{\Pi}_{(L-\uparrow)}^{(L+\uparrow) } + \mbox{h.c.}  \right)
-  c_{\downarrow} s_{\downarrow} \left(  \hat{\Pi}_{(L-\downarrow)}^{(L+\downarrow) } + \mbox{h.c.}  \right) 
\label{eq:n2L_expanded_off}   \,, 
\end{align}
where we have introduced the projection operators (off-diagonal) between two states  $\ket{x}$,$\ket{x'}$ of a single DQD as
\begin{equation}
\hat{\Pi}_{\left( x'\right) }^{\left(x \right)} = \ket{x} \bra{x'} \,.
\end{equation}
The diagonal elements in Eq.~(\ref{eq:n2L_expanded}) can be removed from the interaction hamiltonian using the following unitary transformation
\begin{equation}
\hat{U}_{D,L}  
= 
\prod_{k_L}   e^{  i 
\frac{g_{k_L}}{\hbar\omega_{k_L}} \left( \hat{a}_{k_L,L}^{\phantom{\dagger} } - \hat{a}_{k_L,L}^{\dagger} \right)  
\hat{n}_{2,L}^{(diag)}
}
\label{eq:U_D}
  \, .
\end{equation}
This  unitary transformation  leads to a energy renormalization of the individual electronic levels 
(Lamb-shift) and to higher order interaction terms that scale as $g_k/\left(  \hbar \omega_k \right) \sim g_k / \left( E_{x} - E_{x'} \right) \ll 1$ 
(weak coupling regime) where $x$ and $x'$ are the electronic states involved in the  emission (or absorption) of a photon.

We point out that the operator Eq.~(\ref{eq:U_D}) commutes with the electronic Hamiltonian without the superconducting contact 
but it does not commute with $\hat{H}_{CPS}$ for which we have 
\begin{align}
&
\hat{U}_{D,L}^{\dagger} \hat{H}_{CPS} \hat{U}_{D,L} 
\simeq
 \nonumber \\
& - \frac{i}{\sqrt{2}} \Gamma_s \left( s^2_{\uparrow}  - s^2_{\downarrow}  \right) 
  \sum_{k_L} \frac{ g_{k_L} }{ \hbar \omega_{k_L} }  \left( \hat{a}_{k_L,L}^{\phantom{\dagger} } + \hat{a}_{k_L,L}^{\dagger} \right)
  \hat{\Pi}^{(S_{++})}_{(00)} 
  \nonumber\\
 &+ \mbox{(other terms)} \, .
\label{eq:D_exp}
\end{align}
After the unitary transformation, the CPS Hamiltonian contains an electron-photon interaction that 
leads to single photon emission associated to the injection of the Cooper pair from the superconducting contact to the two DQDs.
When the two states $\ket{S_{++}}$ and  $\ket{0}$ are resonant, this interaction leads to a relaxation decay 
between the two delocalized Andreev bound states $\ket{A_{\pm}}$ 
with a photon emitted with frequency $\Gamma_S$.
We thus conclude that,
in general, during the assisted photon emission tunneling of the Cooper pair, a single photon can be produced in the left or right transmission line.
This is an extra process 
that it is not relevant to produce entangled photon pairs and to witness the entanglement of the Cooper pair.

From Eq.~(\ref{eq:D_exp}) we infer that such process scales  
\begin{equation}
 \Gamma_S  s^2_{\sigma} \frac{g_k}{\hbar\omega_k} \sim   s^2_{\sigma} g_k \ll g_k \, ,
\end{equation}
namely, as 
the dots' states are weakly hybrized,  $s^2_{\sigma} = \sin^2\left(  \theta_{\sigma} / 2 \right) \ll 1$,
the coupling strength is much smaller than the coupling $g_k$ and therefore we neglect it hereafter.
This is the reason why we have designed a setup in which the transmission line is coupled to the DQD through the QD $2$ whereas the QD $1$ is tunnel-coupled to the superconducting contact.
In the following we neglect this process.

Let's now discuss the off-diagonal elements Eq.~(\ref{eq:n2L_expanded_off}) in the Hilbert space given by the two double dots and   write the operators  as
\begin{align}
\hat{\Pi}_{\left( L-\uparrow \right)}^{\left(L +\uparrow \right) }  
& =
\underbrace{ \hat{\Pi}_{(L -\uparrow,;R+\uparrow)}^{ ( L+\uparrow ;R+\uparrow) }  }_{\mbox{(subspace $S=1$, $S_z=1$)}}
+
\underbrace{    \hat{\Pi}_{ (L-\uparrow;R+\downarrow)}^{ (L+\uparrow;R+\downarrow) }   }_{\mbox{(subspace  $S_z=0$)}} \nonumber \\
&
+
\underbrace{   \hat{\Pi}_{ (L-\uparrow;R-\uparrow)}^{ (L+\uparrow;R-\uparrow) }   }_{\mbox{(subspace $S=1$, $S_z=1$)}}
+
 \underbrace{     \hat{\Pi}_{ (L-\uparrow;R-\downarrow)}^{ (L+\uparrow;R-\downarrow) }    }_{\mbox{(subspace  $S_z=0$)}}
  \, 
\end{align}
where we have defined the two-electron states 
\begin{align}
    \ket{L,s,\sigma;R,s',\sigma'}=\hat{d}^{\dagger}_{L,s,\sigma}\hat{d}^{\dagger}_{R,s',\sigma'}\ket{0}.
\end{align}
As we have a CPS, only singlet states are injected in the DQDs and therefore we can disregard the triplet states with $S_z=\pm 1$.
The part of the operator $ \hat{n}_{2,L} $ relevant for the interaction reduces to 
\begin{align}
& {\left( \hat{n}_{2,L} \right)}^{(off)}_{(S_z=0)} = \nonumber \\
&=
-  c_{\uparrow} s_{\uparrow} 
\left(  
\hat{\Pi}_{ (L-\uparrow;R+\downarrow)}^{ (L+\uparrow;R+\downarrow) }   +   \hat{\Pi}_{ (L-\uparrow;R-\downarrow)}^{ (L+\uparrow;R-\downarrow) } 
+ \mbox{h.c.}  
\right)
\nonumber\\
&
\!\!-  c_{\downarrow} s_{\downarrow} 
\left(  
\hat{\Pi}_{ (L-\downarrow;R+\uparrow)}^{ (L+\downarrow;R+\uparrow) }   +   \hat{\Pi}_{ (L-\downarrow;R-\uparrow)}^{ (L+\downarrow;R-\uparrow) } 
\right) 
= \nonumber \\
&
\!
=\!\!-  c_{\uparrow} s_{\uparrow} 
\left[
\frac{1}{\sqrt{2}} 
\left( 
\hat{\Pi}_{ (L-\uparrow;R+\downarrow)}^{ (T_{++}^{(0)}) }  
+
\hat{\Pi}_{ (L-\uparrow;R+\downarrow)}^{ (S_{++} ) }  
\right)
+   
\hat{\Pi}_{ (L-\uparrow;R-\downarrow)}^{ (L+\uparrow;R-\downarrow) } 
\right]  
\nonumber\\
&
-  c_{\downarrow} s_{\downarrow}
 \left[  
 \frac{1}{\sqrt{2}}
 \left(
\hat{\Pi}_{ (L-\downarrow;R+\uparrow)}^{ (T_{++}^{(0)}) }  
- 
\hat{\Pi}_{ (L-\downarrow;R+\uparrow)}^{ (S_{++} ) }   
\right) 
+ 
\hat{\Pi}_{ (L-\downarrow;R-\uparrow)}^{ (L+\downarrow;R-\uparrow) } 
\right] \nonumber\\
&
+ \mbox{H.c.}  
  \, ,
\end{align}
in the last line we have used 
\begin{align}
\ket{L,+,\uparrow;R,+,\downarrow}
&=
\frac{1}{\sqrt{2}} \left( \ket{T_{++}}^{ (0) } +  \ket{S_{++}}  
\right)
\label{eqLuRd} \, ,  \\
\ket{L,+,\downarrow;R,+,\uparrow}
&=
\frac{1}{\sqrt{2}} \left( \ket{T_{++}}^{ (0) } -  \ket{S_{++}}   \label{eqLdRu}  \right) \, ,
\end{align}
where 
\begin{align}
    \ket{T_{++}}^{ (0) }=\frac{1}{\sqrt{2}} 
\Big( \hat{d}^\dagger_{L,+,\uparrow} \hat{d}^\dagger_{R,+,\downarrow}  + \hat{d}^\dagger_{L,+,\downarrow}  \hat{d}^{\dagger}_{R,+,\uparrow}  \Big)\ket{0}.
\end{align}
Again, since the triplet state with $S_z=0$ is decoupled in the dynamics of the system, we can take only the singlet state in the subspace $(++)$:
\begin{align}
{\left( \hat{n}_{2,L} \right)}_{eff} 
& =
-  c_{\uparrow} s_{\uparrow}
\left[ 
\frac{1}{\sqrt{2}}  
\hat{\Pi}_{ (L-\uparrow;R+\downarrow)}^{ (S_{++} ) }  
+ 
\hat{\Pi}_{ (L-\uparrow;R-\downarrow)}^{ (L+\uparrow;R-\downarrow) } 
+ 
\mbox{H.c.}  
\right]
\nonumber\\
&- 
c_{\downarrow} s_{\downarrow}
\left[
- \frac{1}{\sqrt{2}}
\hat{\Pi}_{ (L-\downarrow;R+\uparrow)}^{ (S_{++} ) }   
+   \hat{\Pi}_{ (L-\downarrow;R-\uparrow)}^{ (L+\downarrow;R-\uparrow) } 
+ \mbox{H.c.} 
 \right]   \label{eq:n2_eff_1}.
\end{align}

\begin{figure}[t!]
\begin{center}
\includegraphics[scale=0.235]{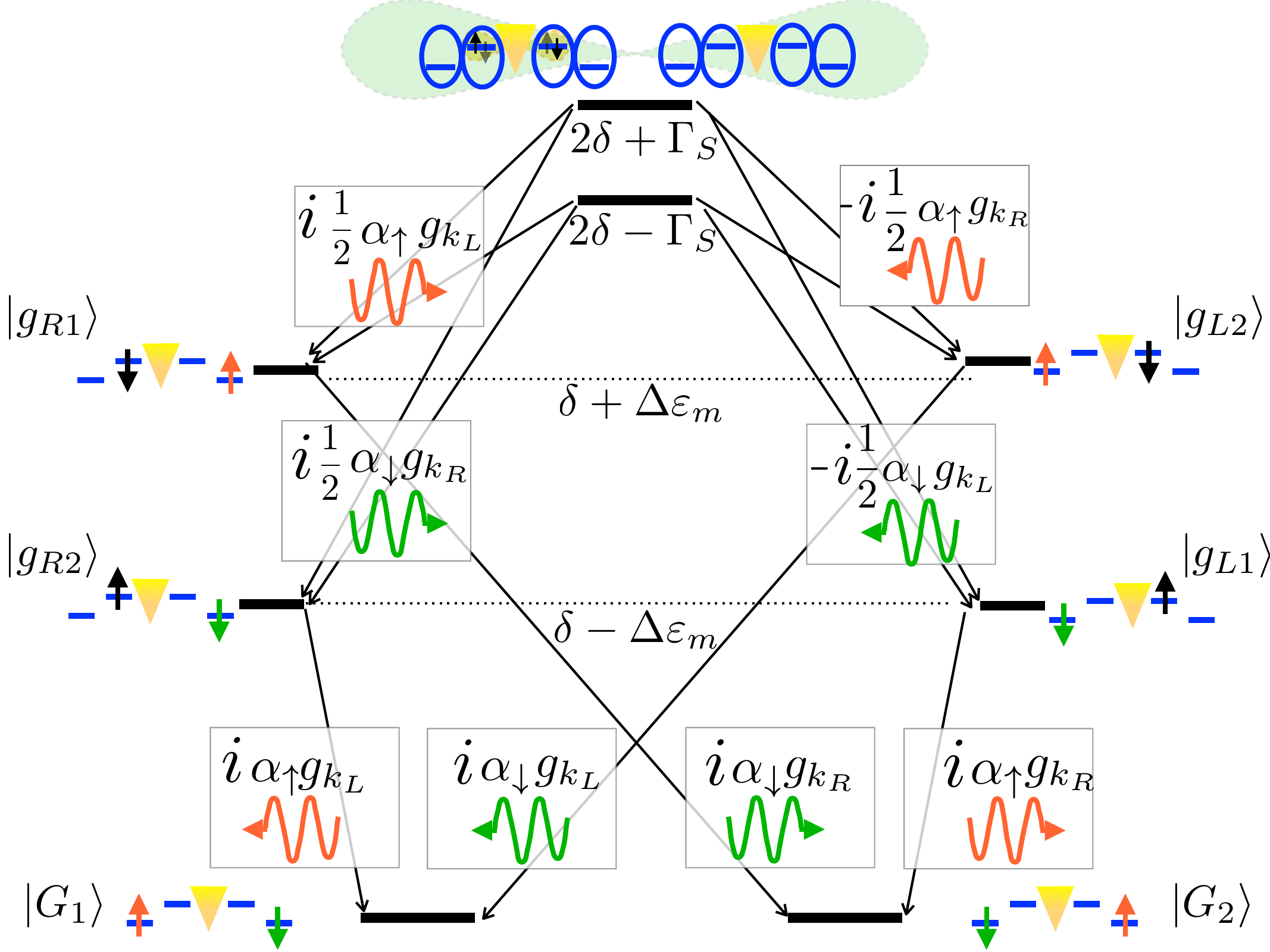} 
\end{center}
%
\caption{
 Diagram for the relaxation transitions with photon emission.
Each spin refers to a single electron in each DQD.
}
\label{fig:Sup-5}
\end{figure}

We know that the singlet state $\ket{S_{++}}$ is coherently coupled to the empty state $\ket{00}$, 
forming a doublet of Andreev bound states $\ket{A}_{+}$ and $\ket{A}_{-}$.
Therefore we need to express the operator ${\left( \hat{n}_{2,L} \right)}_{eff} $ in term of the energy eigenstates of the full electron Hamiltonian including $\hat{H}_{CPS}$.
This can be done by expressing the singlet state in terms of the Andreev bound states  
\begin{equation}
\ket{S_{++}} = \frac{1}{\sqrt{2}}  \left( \ket{A_+} + \ket{A_-} \right) \, .
\end{equation}
Thus we obtain 
\begin{align}
{\left( \hat{n}_{2,L} \right)}_{eff} 
&
=
-  
\frac{ c_{\uparrow} s_{\uparrow} }{2}
 \left(
 \hat{\Pi}_{ (L-\uparrow;R+\downarrow)}^{ (A_{+} ) }   + \hat{\Pi}_{ (L-\uparrow;R+\downarrow)}^{ (A_{-} ) }  
 \right) 
 \nonumber \\
 & +
 \frac{ c_{\downarrow} s_{\downarrow} }{2}
\left( 
\hat{\Pi}_{ (L-\downarrow;R+\uparrow)}^{ (A_{+} ) } 
+
\hat{\Pi}_{ (L-\downarrow;R+\uparrow)}^{ (A_{-} )  } 
  \right) 
\nonumber \\  
&
 - 
 c_{\uparrow} s_{\uparrow}    
 \hat{\Pi}_{ (L-\uparrow;R-\downarrow)}^{ (L+\uparrow;R-\downarrow) } 
-
c_{\downarrow} s_{\downarrow} 
 \hat{\Pi}_{ (L-\downarrow;R-\uparrow)}^{ (L+\downarrow;R-\uparrow) } 
+ \mbox{H.c.}  
  \, .
\end{align}
By repeating all the same steps for the operator $ \hat{n}_{2,R}$ of the right double dot, one obtains a similar result
\begin{align}
 {\left( \hat{n}_{2,R} \right)}_{eff}  &= 
\frac{ c_{\uparrow} s_{\uparrow} }{2}
 \left(
 \hat{\Pi}_{ (L+\downarrow;R-\uparrow)}^{ (A_{+} ) }   + \hat{\Pi}_{ (L+\downarrow;R-\uparrow)}^{ (A_{-} ) }  
 \right) 
 \nonumber \\
 &
 -
 \frac{ c_{\downarrow} s_{\downarrow} }{2}
\left( 
\hat{\Pi}_{ (L+\uparrow;R-\downarrow)}^{ (A_{+} ) } 
+
\hat{\Pi}_{ (L+\uparrow;R-\downarrow)}^{ (A_{-} )  } 
  \right) 
 \nonumber \\ 
&
 - 
 c_{\uparrow} s_{\uparrow}    
 \hat{\Pi}^{ (L-\downarrow;R+\uparrow)}_{ (L-\downarrow;R-\uparrow) } 
-
c_{\downarrow} s_{\downarrow} 
 \hat{\Pi}_{ (L-\uparrow;R-\downarrow)}^{ (L-\uparrow;R+\downarrow) } 
+ \mbox{H.c.}  
  \, . 
\end{align}
The effective operators ${\left( \hat{n}_{2,L} \right)}_{eff}$  and ${\left( \hat{n}_{2,R} \right)}_{eff}$ are symmetric by exchanging $L\leftrightarrow R$; the different sign for the first two terms originates  in the different sign appearing in front of the singlet in 
Eqs.~(\ref{eqLdRu}) and (\ref{eqLuRd}).
Setting  the coefficients 
\begin{align}
\alpha_{\uparrow} 
&=  c_{\uparrow} s_{\uparrow} 
 \simeq
 s_{\uparrow}
 \simeq \frac{t_{12}}{\delta - \Delta\varepsilon_m}
 \, , \\
\alpha_{\downarrow} 
&=  c_{\downarrow} s_{\downarrow} 
 \simeq 
s_{\downarrow} 
 \simeq \frac{t_{12}}{\delta + \Delta\varepsilon_m}
\, , 
\end{align}
in the regime of weak coupling strength between electrons and photons, and applying the rotating wave approximation (RWA) 
we write  
\begin{align}
&
\hat{H}_{int,L}^{(RWA)}
=
i \sum_{k_L} 
 g_{k_L}
\hat{a}_{k_L}^{\dagger} 
\Big[
\alpha_{\downarrow} 
 \hat{\Pi}_{ (L-\downarrow;R-\uparrow)}^{ (L+\downarrow;R-\uparrow) } 
 + 
\alpha_{\uparrow}  
 \hat{\Pi}_{ (L-\uparrow;R-\downarrow)}^{ (L+\uparrow;R-\downarrow) } 
\nonumber \\
&
+ \frac{  \alpha_{\uparrow}  }{2} 
\sum_{p=\pm}
 \hat{\Pi}_{ (L-\uparrow;R+\downarrow)}^{ (A_{p} ) } 
 -
 \frac{ \alpha_{\downarrow} }{2}
\sum_{p=\pm}
\hat{\Pi}_{ (L-\downarrow;R+\uparrow)}^{ (A_{p} ) } 
\Big]
+ \mbox{H.c.}   
\label{eq:Hint_eff_L}
 \, ,
\end{align}
and
\begin{align}
& \hat{H}_{int,R}^{(RWA)}
=
i \sum_{k_R} g_{k_R}
\hat{a}_{k_R}^{\dagger} 
\Big[
 \alpha_{\downarrow} 
 \hat{\Pi}_{ (L-\uparrow;R-\downarrow)}^{ (L-\uparrow;R+\downarrow) } 
 + 
 \alpha_{\uparrow}
 \hat{\Pi}^{ (L-\downarrow;R+\uparrow)}_{ (L-\downarrow;R-\uparrow) } 
\nonumber \\
&
+
 \frac{  \alpha_{\downarrow} }{2}
\sum_{p=\pm}
\hat{\Pi}_{ (L+\uparrow;R-\downarrow)}^{ (A_{p} ) } 
-
\frac{ \alpha_{\uparrow}  }{2}
\sum_{p=\pm}
 \hat{\Pi}_{ (L+\downarrow;R-\uparrow)}^{ (A_{p} ) }   
\Big]
+ \mbox{H.c.}
\, . 
\label{eq:Hint_eff_R}
\end{align}
In Fig.~\ref{fig:4} we give an 
example of the possible electronic relaxation processes with photons emission 
as described by the effective Hamiltonian Eq.~(\ref{eq:Hint_eff_L}) and Eq.~(\ref{eq:Hint_eff_R}). 
%
%
The energy level diagram, along with the interaction coupling strength, is   presented in Fig.~\ref{fig:Sup-5}, using the notation for the states from the main text.

%
%
%
%
%
%
%
%
%
%
%
%
%

%
%
%
%
\section{Solution of the two photon emission dynamics}
\label{app:E}
We solve the electron-photon model considering the Hamiltonian $\hat{H}_{el}^{(S_z=0)}$ of Eq.~(\ref{eq:H_el_eff}), 
for the electronic systems and the Hamiltonian $\hat{H}_{int,L}^{(RWA)}$ of  Eq.~(\ref{eq:Hint_eff_L}) and $\hat{H}_{int,R}^{(RWA)}$ of 
Eq.~(\ref{eq:Hint_eff_R}).
To proceed it is useful to exploit the notations introduced in the maintext for the electronic states according to the scheme reported in 
Fig.~\ref{fig:Sup-5} and the total effective Hamiltonian reads  
\begin{align}
& \hat{H}_{tot}
 = \nonumber\\
& =  
\sum_{X} \, E_X \, \ket{X}\bra{X}  
+
\sum_{k_L} \hbar \omega_{k_L} \hat{a}^{\dagger}_{k_L}  \hat{a}^{ \phantom{\dagger} }_{k_L} 
+
\sum_{k_R} \hbar \omega_{k_R} \hat{a}^{\dagger}_{k_R}  \hat{a}^{ \phantom{\dagger} }_{k_R} 
\nonumber \\
& +  \hat{H}_{int,L}^{(RWA)} +  \hat{H}_{int,R}^{(RWA)} \, , 
\end{align}
 $\ket{X} =  \left\{ \ket{A_+},\ket{A_-},\ket{g_{L1}},\ket{g_{L2}},\ket{g_{R1}},\ket{g_{R2}},\ket{G_1},\ket{G_2} \right\} $   
and the interaction Hamiltonians are
\begin{align}
 \hat{H}_{int,L}^{(RWA)} 
 &= i 
 \sum_{k_L} g_{k_L} \hat{a}^{ \dagger }_{k_L} 
 \big[ 
\alpha_{\downarrow} \ket{G_1}\bra{g_{R1}} 
+ 
\alpha_{\uparrow} \ket{G_2}\bra{g_{R2}} 
\nonumber \\
&
\frac{\alpha_{\uparrow}}{2}  \sum_{s=\pm} \ket{g_{L2}}\bra{A_s}  
- 
\frac{\alpha_{\downarrow}}{2}  \sum_{s=\pm}  \ket{g_{L1}}\bra{A_s} 
\big]
+ \mbox{H.c.}  
\, ,
\nonumber\\
 \hat{H}_{int,R}^{(RWA)} 
 &=i
\sum_{k_R} g_{k_R} \hat{a}^{ \dagger }_{k_R,R} 
 \big[ 
 \alpha_{\uparrow} \ket{G_1}\bra{g_{L1}} 
 +
 \alpha_{\downarrow} \ket{G_2}\bra{g_{L2}} 
 \nonumber \\
 &
\frac{\alpha_{\downarrow}}{2}  \sum_{s=\pm} \ket{g_{R2}}\bra{A_s}
 - 
\frac{\alpha_{\uparrow}}{2}  \sum_{s=\pm} \ket{g_{R1}}\bra{A_s} 
\big]
+ \mbox{H.c.} 
\end{align}

We assume the system is initially excited in the subspace of the Andreev doublets with no photons in the transmission lines (zero temperature limit).
Since RWA preserves the number of excitations, we can make the following Ansatz for the  total state of the system:
\begin{align}
& 
\ket{\Psi(t)} 
= 
c_{0+}(t) \ket{A_+} \otimes \ket{0,0} +c_{0-}(t)  \ket{A_-}  \otimes \ket{0,0}  \nonumber \\
&
+
\sum_{k_L} \big(  c_{k_L,L1}(t) \ket{g_{L1}} \otimes \ket{{k_L},0} + c_{k_L,L2}(t) \ket{g_{L2}} \otimes \ket{{k_L},0}  \big)  \nonumber \\
&
+
\sum_{k_R} \big(  c_{k_R,R1}(t) \ket{g_{R1}} \otimes \ket{0,{k_R}} + c_{k_R,R2}(t) \ket{g_{R2}} \otimes \ket{0,{k_R}}  \big)  \nonumber \\
&
+
\sum_{k_L,k_R}   c_{k_L,k_R,1}(t) \ket{G_{1}}  \otimes \ket{{k_L},{k_R}} \nonumber \\
&
+ 
\sum_{k_L,k_R}   c_{k_L,k_R,2}(t)  \ket{G_{2}} \otimes  \ket{{k_L},{k_R}}   \, .
\label{eq:7_Ansatz}
\end{align}

The coefficients satisfy the following equations 
%
%
%
%
\begin{widetext}
\begin{align}
i \hbar 
\frac{\partial }{\partial t} c_{0,\pm}(t)
&
= 
\left(2 \delta \pm \Gamma_S \right)  c_{0,\pm}(t)
+
\frac{i}{2}  
\sum_{k_L}  g_{k_L}   \big[  \alpha_{\downarrow}   c_{k_L,L1}(t)  - \alpha_{\uparrow}   c_{k_L,L2}(t)   \big]
+
\frac{i}{2} 
 \sum_{k_R} g_{k_R}  \big[ \alpha_{\uparrow}  c_{k_R,R1}(t) -\alpha_{\downarrow}   c_{k_R,R2}(t) \big]   \, ,
\label{eq:eq,_C0pm}
\\
i \hbar 
\frac{\partial }{\partial t} c_{k_L,L1}(t) 
&=
\left( \hbar\omega_{k_L} + \delta  - \Delta \varepsilon_m \right)  c_{k_L,L1}(t) 
- i 
\frac{\alpha_{\downarrow}}{2} g_{k_L} \big( c_{0,+}(t) + c_{0,-}(t)  \big) 
-  i
\alpha_{\uparrow} \sum_{k_R} g_{k_R} c_{k_L,k_R,1}(t) 
\, , 
\label{eq:cL1} 
\\
i \hbar 
\frac{\partial }{\partial t} c_{k_R,R1}(t) 
&=
\left( \hbar\omega_{k_R} + \delta  + \Delta \varepsilon_m \right)  c_{k_R,R1}(t) 
-i
\frac{\alpha_{\uparrow}}{2} g_{k_R} \big( c_{0,+}(t) + c_{0,-}(t)  \big) 
- i
\alpha_{\downarrow} \sum_{k_L} g_{k_L} c_{k_L,k_R,1}(t)   
\, ,
\label{eq:cR1}
\\
i \hbar \frac{\partial }{\partial t} c_{k_L,L2}(t) 
&=
\left( \hbar\omega_{k_L} + \delta  + \Delta \varepsilon_m \right)   c_{k_L,L2}(t) 
+
i
\frac{\alpha_{\uparrow}}{2} g_{k_L} \big( c_{0,+}(t) + c_{0,-}(t)  \big) 
- 
i
\alpha_{\downarrow} \sum_{k_R} g_{k_R} c_{k_L,k_R,2}(t)  
\, ,\label{eq:cL2} \\
i \hbar 
\frac{\partial }{\partial t} c_{k_R,R2}(t) 
&=
\left( \hbar\omega_{k_R} + \delta  - \Delta \varepsilon_m \right) c_{k_R,R2}(t) 
+
i
\frac{\alpha_{\downarrow}}{2} g_{k_R} \big( c_{0,+}(t) + c_{0,-}(t)  \big) 
- 
i
\alpha_{\uparrow} \sum_{k_L} g_{k_L} c_{k_L,k_R,2}(t)    
\, ,\label{eq:cR2}
\\
i \hbar 
\frac{\partial }{\partial t} c_{k_L,k_R,1}(t) 
&=
\left(  \hbar\omega_{k_L} +  \hbar\omega_{k_R}  \right)  c_{k_L,k_R,1}(t)
+ i \alpha_{\downarrow} g_{k_L} c_{k_R,R1}(t) 
+ i \alpha_{\uparrow} g_{k_R} c_{k_L,L1}(t) 
\, ,
\label{eq:Cklkr1}
 \\
i \hbar 
\frac{\partial }{\partial t} c_{k_L,k_R,2}(t) 
&=
\left(  \hbar\omega_{k_L} +  \hbar\omega_{k_R} \right)  c_{k_L,k_R,2}(t)
+ i \alpha_{\uparrow} g_{k_L} c_{k_R,R2}(t) 
+ i \alpha_{\downarrow} g_{k_R} c_{k_L,L2}(t) 
\, . 
\label{eq:Cklkr2}
\end{align}
%
%
%
%
\end{widetext}

For the remainder of the calculation, we will use the interaction picture, setting the coefficients accordingly to the relations 
\begin{align}
 c_{k_L,L1}(t) & =  \tilde{c}_{k_L,L1}(t)  \,\,\, e^{-\frac{i}{\hbar} \left( \hbar\omega_{k_L} + \delta - \Delta\varepsilon_m \right)  t} \, , \label{eq:ckL1_tilde}	 \, , \\
 c_{k_R,R1}(t) & =  \tilde{c}_{k_R,R1}(t)  \,\,\, e^{-\frac{i}{\hbar} \left( \hbar\omega_{k_R} + \delta + \Delta\varepsilon_m \right)  t}   \label{eq:ckR1_tilde} \, , \\
c_{k_L,L2}(t) &=   \tilde{c}_{k_L,L2}(t)  \,\,\, e^{-\frac{i}{\hbar} \left( \hbar\omega_{k_L} + \delta + \Delta\varepsilon_m \right)  t}  \label{eq:ckL2_tilde}	\, , \\
c_{k_R,R2}(t) & =   \tilde{c}_{k_R,R2}(t)  \,\,\, e^{-\frac{i}{\hbar} \left( \hbar\omega_{k_R} + \delta - \Delta\varepsilon_m \right)  t}  \label{eq:ckR2_tilde} \, , \\
c_{0,\pm}(t) & = \tilde{c}_{0,\pm}(t)  e^{-\frac{i}{\hbar} \left( 2 \delta \pm \Gamma_S  \right) t} \, .
\end{align}
We start by solving  Eqs.~(\ref{eq:Cklkr1}), (\ref{eq:Cklkr2}) for the coefficients associated to the two photon states 
with boundary conditions $c_{k_L,k_R,1}(t_0)= c_{k_L,k_R,2}(t_0)=0$.

We insert the solution $c_{k_L,k_R,1}(t)$  of  Eq.~(\ref{eq:Cklkr1})
into Eq.~(\ref{eq:cL1}), (\ref{eq:cR1}), obtaining the following equation 
for $\tilde{c}_{k_L,L1}(t)$  
\begin{align}
&
\frac{\partial \tilde{c}_{k_L,L1}}{\partial t} 
=
- \frac{\alpha_{\downarrow}}{2\hbar} g_{k_L}
\big[ c_{0,+}(t)+  c_{0,-}(t) \big]
e^{\frac{i}{\hbar} \left( \hbar\omega_{k_L} + \delta - \Delta\varepsilon_m \right)  t} 
\nonumber \\
&
-
\int^{t}_{t_0} \!\!\! dt' 
\alpha_{\uparrow}^2 
\big[
\sum_{\tilde{k}_R} \frac{g_{\tilde{k}_R}^2}{\hbar^2}  e^{ -i \left( \omega_{\tilde{k}_R} - \frac{\delta-\Delta\varepsilon_m}{\hbar} \right)   \left( t-t' \right) }
\big]
\tilde{c}_{k_L,L1}(t')
\nonumber  \\
& 
-
\int^{t}_{t_0} \!\!\! dt' 
\alpha_{\uparrow}\alpha_{\downarrow} \frac{g_{k_L}}{\hbar^2}
\nonumber \\
&\times \big[
\sum_{\tilde{k}_R} g_{\tilde{k}_R} 
e^{ -i \left( \omega_{\tilde{k}_R} - \frac{\delta-\Delta\varepsilon_m}{\hbar} \right)  t }
e^{i \left( \omega_{k_L} - \frac{\delta+\Delta\varepsilon_m}{\hbar} \right)   t' }
\tilde{c}_{\tilde{k}_R,R1}(t')
\big] \, , \label{eq:Semifinal-1}
\end{align}
and for $\tilde{c}_{k_L,R1}(t)$
\begin{align}
&
\frac{\partial \tilde{c}_{k_R,R1} }{\partial t}
=
- \frac{\alpha_{\uparrow}}{2\hbar} g_{k_R}
\big[ c_{0,+}(t)+  c_{0,-}(t) \big]
e^{\frac{i}{\hbar} \left( \hbar\omega_{k_R} + \delta + \Delta\varepsilon_m \right)  t} 
\nonumber \\
&
-
\int^{t}_{t_0} \!\!\! dt' 
\alpha_{\downarrow}^2 
\big[
\sum_{\tilde{k}_L} \frac{g_{\tilde{k}_L}^2}{\hbar^2}  e^{ -i \left( \omega_{\tilde{k}_L} - \frac{\delta+\Delta\varepsilon_m}{\hbar} \right)   \left( t-t' \right) }
\big]
\tilde{c}_{k_R,R1}(t')
\nonumber  \\
& 
-
\int^{t}_{t_0} \!\!\! dt' 
\alpha_{\uparrow}\alpha_{\downarrow} \frac{g_{k_R}}{\hbar^2} \nonumber\\
&
\times
\big[
\sum_{\tilde{k}_L} g_{\tilde{k}_L} 
e^{ -i \left( \omega_{\tilde{k}_L} - \frac{\delta+\Delta\varepsilon_m}{\hbar} \right)  t }
e^{i \left( \omega_{k_R} - \frac{\delta-\Delta\varepsilon_m}{\hbar} \right)   t' }
\tilde{c}_{\tilde{k}_L,L1}(t')
\big] \, . 
\label{eq:Semifinal-2}
\end{align}
The last term on the right-side in Eq.~(\ref{eq:Semifinal-1}) and in Eq.~(\ref{eq:Semifinal-2}) does not depend on the time difference $t-t'$ but it depends on 
the time sum $t+t'$ and we neglect it assuming the secular approximation.  
For the function Eq.~(\ref{eq:Kernel})  we  use the Markov approximation [see Eqs.~(\ref{eq:Markov}), (\ref{eq:Markov_approx})] and we obtain
\begin{align}
\frac{\partial \tilde{c}_{k_L,L1}}{\partial t}
&=
- \frac{\Gamma_{-}}{2}
 \,\, \tilde{c}_{k_L,L1}(t) 
\nonumber \\
&
- \frac{\alpha_{\downarrow} g_{k_L}}{2\hbar} 
\big[ c_{0,+}(t)+  c_{0,-}(t) \big]
e^{\frac{i}{\hbar} \left( \hbar\omega_{k_L} + \delta - \Delta\varepsilon_m \right)  t} 
 \label{eq:ckL1}\\
\frac{\partial \tilde{c}_{k_R,R1} }{\partial t}
&=
- \frac{ \Gamma_{+}}{2}
  \,\, \ \tilde{c}_{k_R,R1}(t) 
 \nonumber\\
&
- \frac{\alpha_{\uparrow} g_{k_R}}{2\hbar} 
\big[ c_{0,+}(t)+  c_{0,-}(t) \big]
e^{\frac{i}{\hbar} \left( \hbar\omega_{k_R} + \delta + \Delta\varepsilon_m \right)  t} 
 \label{eq:ckR1} \, ,
\end{align}
in which we have introduced  
$\Gamma_{-} =  \alpha_{\downarrow}^2 \kappa_0 \left[ \delta- \Delta\varepsilon_m  \right] $
and
$\Gamma_{+} =  \alpha_{\uparrow}^2 \kappa_0 \left[ \delta+ \Delta\varepsilon_m  \right] $.

In a similar way, we insert these solutions $c_{k_L,k_R,2}(t)$
of Eq.~(\ref{eq:Cklkr2})
into Eqs.~(\ref{eq:cL2}), (\ref{eq:cR2}),  applying the Markovian and the secular approximations we find 
\begin{align}
\frac{\partial \tilde{c}_{k_L,L2}  }{  \partial t}
&=
- \frac{\Gamma_{+}}{2}
\,\, \tilde{c}_{k_L,L2}(t) 
\nonumber \\
&
+
 \frac{\alpha_{\uparrow}}{2\hbar} g_{k_L}
\big[ c_{0,+}(t)+  c_{0,-}(t) \big]
e^{\frac{i}{\hbar} \left( \hbar\omega_{k_L} + \delta + \Delta\varepsilon_m \right)  t} 
\,,    \label{eq:ckL2} \\
\frac{\partial \tilde{c}_{k_R,R2} }{\partial t}
&=
- \frac{\Gamma_{-}}{2} \,\, \ \tilde{c}_{k_R,R2}(t)
\nonumber \\
&
 + \frac{\alpha_{\downarrow}}{2\hbar} g_{k_R}
\big[ c_{0,+}(t)+  c_{0,-}(t) \big]
e^{\frac{i}{\hbar} \left( \hbar\omega_{k_R} +  \delta-  \Delta\varepsilon_m \right)  t} 
\,.  \label{eq:ckR2}
\end{align}
We insert the solutions of the Eqs.~(\ref{eq:ckL1}), (\ref{eq:ckR1}), (\ref{eq:ckL2}), (\ref{eq:ckR2}) with boundary conditions 
$\tilde{c}_{k_L,L1} (t_0)=\tilde{c}_{k_R,R1}(t_0)=\tilde{c}_{k_L,L2}(t_0) =\tilde{c}_{k_R,R2}(t_0)=0$ 
 in 
Eqs.~(\ref{eq:ckL1_tilde}),(\ref{eq:ckR1_tilde}) and (\ref{eq:ckL2_tilde}),(\ref{eq:ckR2_tilde}) and 
then we insert them into the equation Eq.~(\ref{eq:eq,_C0pm}).
In this way we obtain a set of close equations for the the coefficient $\tilde{c}_{0,+}(t)$ and $\tilde{c}_{0,-}(t)$
in which we have terms of the type  $\sim e^{\frac{i}{\hbar} \Gamma_s t }  e^{\frac{i}{\hbar} \Gamma_s t'} $.
The latter terms  do not depend only on the time difference $t-t'$ but  on 
the time sum $t+t'$ and we neglect them assuming the secular approximation.

Using the Markovian approximation we arrive at the following system of coupled differential equations:
\begin{align}
\frac{\partial \tilde{c}_{0,+} }{\partial t}& = - 
\frac{ \gamma_{+} }{2} \tilde{c}_{0,+}  \,, \label{eq:c_0+_tilde} \\
\frac{\partial \tilde{c}_{0,-} }{\partial t}& = - 
\frac{ \gamma_{-} }{2} 
\tilde{c}_{0,-}  \,, \label{eq:c_0-_tilde}
\end{align}
with $\gamma_{\pm}
= 
(\alpha_{\downarrow}^2/2) \kappa_0\left[\delta + \Delta\varepsilon_m  \pm \Gamma_s \right]
+
(\alpha_{\uparrow}^2/2) \kappa_0\left[ \delta - \Delta\varepsilon_m \pm \Gamma_s \right]$.

\addGR{Given the boundary conditions at $t=0$ for $c_{0,+}$ and $c_{0,-}$,
one can solve Eqs.~(\ref{eq:c_0+_tilde}, \ref{eq:c_0-_tilde}) and then 
}
 insert the solutions for \addGR{$\tilde{c}_{0,\pm}(t)$} in the equations for 
$\tilde{c}_{k_L,L1} (t)$, $ \tilde{c}_{k_R,R1}(t)$, $\tilde{c}_{k_L,L2}(t)$,  $\tilde{c}_{k_R,R2}(t)$, the coefficients associated to the states with one emitted photon,  
and solve them.
Finally, by the knowledge of the solutions for the coefficients $c_{k_L,L1}(t), \,c_{k_R,R1}(t), \,c_{k_L,L2}(t), \,c_{k_R,R2}(t)$ 
we can write the solutions for the coefficients associated to two emitted photons $c_{k_L,k_R,1}(t)$ and $c_{k_L,k_R,2}(t)$. 

\addGR{For example, if} we assume as initial time $t_0=0$ \addGR{the initial state is $\ket{A_+}$, this corresponds} to initial conditions $c_{0,-}(0)=0$, $c_{0,+}(0)=1$.
Assuming the limit $t\rightarrow \infty$, namely $\Gamma_{\pm} t\gg 1$ and $\gamma_{\pm} t\gg 1$,  we have the 
following solution
\begin{widetext}
\begin{align}
& \lim_{t\rightarrow\infty}
\addGR{ \tilde{c}_{k_L,k_R,1}^{(+)}(t)  }
=
\alpha_{\uparrow}\alpha_{\downarrow}g_{k_L}g_{k_R}/\left( 2 \hbar^2 \right) \times  \nonumber \\
&
\left\{
\frac{1}{\omega_{k_L}- \left( \omega_{+} + \frac{\Gamma_S}{\hbar} \right)  +
i \left( \gamma_+- \Gamma_{-} \right)/2}
\left[
\frac{1}{\omega_{k_R}-\omega_{-}  +i  \Gamma_{-}/2  }
-
\frac{1}{\omega_{k_L}+\omega_{k_R}-\omega_{+} -\omega_{-}- \frac{\Gamma_S}{\hbar}  
+i  \gamma_{+}/2 }
\right]
\right.
\nonumber\\
&
\left.
\frac{1}{\omega_{k_R}-\left( \omega_{-} + \frac{\Gamma_S}{\hbar} \right) + i \left(\gamma_+ - \Gamma_{+} \right)/2}
\left[
\frac{1}{\omega_{k_L}-\omega_{+}  +i  \Gamma_{+}/2 }
-
\frac{1}{\omega_{k_L}+\omega_{k_R}-\omega_{+} -\omega_{-}- \frac{\Gamma_S}{\hbar}   
+i  \gamma_{+}/2  }
\right]
\right\}
\, , 
\label{eq:final-coutdown}
\end{align}
\end{widetext}
and, after some algebra, it reduces to  Eq.~(\ref{eq:8_wavepacket_1}),of the main text.
A similar result holds for $\tilde{c}^{(+)}_{k_L,k_R,2}(t)$ Eq.~(\ref{eq:9_wavepacket_2}) 

\addGR{
Starting from the initial state $\ket{A_-}$, one obtains similar results for the coefficients  $\tilde{c}_{k_L,k_R,n}^{(-)}$   with $\Gamma_S$ replaced by $-\Gamma_S$ and 
$\gamma_{+}$ replaced by $\gamma_{-}$ in  Eq.~(\ref{eq:final-coutdown})
}

\addGR{
More generally, if the system is initially in an arbitrary linear combination of the two states $\ket{A_+}$ and $\ket{A_-}$,
the final solution is the linear combination of the two particular solutions  
$\tilde{c}_{k_{L},k_R,n}^{(+)}$ 
and 
$\tilde{c}_{k_{L},k_R,n}^{(-)}$.
This is a consequence of the linearity of the equations of motion presented here.
}

We now proceed to show that the two-photon wavefunction is normalised. 
We assume that we are in the limit of large times but omit writing $\lim_{t\rightarrow\infty}$ 
not to clutter the notation. 
We calculate 
\begin{widetext}
\begin{align}
\sum_{k_{L},k_R}\left| \addGR{ \tilde{c}_{k_{L},k_R,1}^{(+)}  } \right|^2 
&
=
\frac{\alpha_\uparrow^2 \alpha_\downarrow^2}{4\hbar^4} \sum_{k_{L},k_R} g_{k_L}^2 g_{k_R}^2
\left|
\frac{1}{\omega_{k_L}-\omega_{+}  +i  \frac{ \Gamma_{+}}{2} }+\frac{1}{\omega_{k_R}-\omega_{-}  +i  \frac{ \Gamma_{-}}{2}  }\right|^2
\frac{1}{\left[\omega_{k_L}+\omega_{k_R}-\omega_{+} -\omega_{-}- \frac{\Gamma_S}{\hbar}\right]^2   +  \frac{\gamma_{+}^2}{4}  }
\\
\label{eq:norm1-0}
\approx &
\frac{\alpha_\uparrow^2 \alpha_\downarrow^2}{4\hbar^4} \sum_{k_{L},k_R} g_{k_L}^2 g_{k_R}^2
\left[
\frac{1}{(\omega_{k_L}-\omega_{+})^2  +  \frac{\Gamma_{+}^2}{4}  }+\frac{1}{(\omega_{k_R}-\omega_{-})^2  +  \frac{\Gamma_{-}^2}{4}  }\right]
\frac{1}{\left[\omega_{k_L}+\omega_{k_R}-\omega_{+} -\omega_{-}- \frac{\Gamma_S}{\hbar}\right]^2   +  \frac{ \gamma_{+}^2}{4}  }
\,,
\end{align}
\end{widetext}
where the cross-term 
\begin{align}
2 \text{Re}\left[   \frac{1}{\omega_{k_L}-\omega_{+}  +i  \frac{\Gamma_{+}}{2}  } \times \frac{1}{\omega_{k_R}-\omega_{-}  -i  \frac{\Gamma_{-}}{2}  } \right]
\end{align}
is negligible in the weak-coupling limit, that is when $|\omega_+-\omega_- |=2\Delta\varepsilon_m \gg  \Gamma_{+},\Gamma_{-}$.
In the weak-coupling limit, the Lorentzian functions in ~(\ref{eq:norm1-0}) can be treated as $\delta$ functions and Eq.~(\ref{eq:norm1-0}) simplifies to 
\begin{align}
& \sum_{k_{L},k_R}\left|   \addGR{ \tilde{c}_{k_{L},k_R,1}^{(+)}  }      \right|^2
=
\frac{\alpha_\uparrow^2 \alpha_\downarrow^2}{2\hbar^4}\frac{\pi^2}{\gamma_{-}} \times 
 \nonumber \\
&
\left[ \sum_{k_{L},k_R} g_{k_L}^2 g_{k_R}^2
    \frac{2}{\Gamma_+}\delta(\omega_{k_L}-\omega_+)
    \delta\left(\omega_{k_R}-\omega_{-}=\frac{\Gamma_S}{\hbar}\right)
\right.     
    \nonumber \\
    &+
    \left.
    \sum_{k_{L},k_R} g_{k_L}^2 g_{k_R}^2
    \frac{2}{\Gamma_-}\delta(\omega_{k_R}-\omega_-)\delta\left(\omega_{k_L}-\omega_{+}-\frac{\Gamma_S}{\hbar}\right)
    \right].
\end{align} 
Finally using Eq.~(\ref{eq:kappa_0_def}) we find
\begin{equation}
\sum_{k_{L},k_R}\left|\addGR{ \tilde{c}_{k_{L},k_R,1}^{(+)}  }   \right|^2=\frac{1}{2} \, .
\end{equation}
Proceeding in a similar fashion, we also find that 
\begin{equation}
\sum_{k_{L},k_R}\left|  \addGR{ \tilde{c}_{k_{L},k_R,2}^{(+)}  }  \right|^2=\frac{1}{2} \, .
\end{equation}

\addGR{Clearly, the same result holds for the coefficient $\tilde{c}_{k_{L},k_R,n}^{(-)}$.}

%
%
%
%
\section{Rate equation in presence of phonon emission  processes}
\label{app:F}
In this Appendix we study the case in which the electronic system can relax via phonon emission. 
For the electronic levels, we use the notation of the main text.

To calculate the probability of two photons emitted $P_{2-ph}$, 
in addition to the photon emission rates, we also need the transition rates of the radiationless processes  
indicated by the dashed   arrows in Fig.~\ref{fig:8}: These are competing processes which 
give rise to the pathways 
where
the system does not emit two  photons  but instead it ends up in one of the other two possible states, with one emitted photon 
or with no emitted photon at all. This observation simplifies the notation for the rate equations. 
Indeed, to calculate $P_{2-ph}$, 
we need to write only the equations for the occupation probability 
for the states with no phonons emitted, represented by the 
horizontal black lines in Fig.~\ref{fig:8}.

The electronic states labelled $\ket{g_{L1}}\,, \ket{g_{R1}}$ and $\ket{g_{L2}}\,,\ket{g_{L2}}$
- Eqs.~(\ref{eq:g_R1}, \ref{eq:g_L2}, \ref{eq:g_R2}, \ref{eq:g_L1}) - 
correspond to states in which 
only one DQD is excited.

Referring to Fig.~\ref{fig:8},  the  states labeled by the tensor product 
$\ket{x}\otimes\ket{0,0}$, where $x=g_{R1}, g_{R2}, g_{L1}, g_{L2}$, 
indicate 
that the electronic system has relaxed to the state $\ket{x}$
with no emitted photon but with the emission of a phonon. 
On the other hand, the  states 
$\ket{g_{R1(R2)}}\otimes\ket{0,k_R}$,  and 
$\ket{g_{L1(L2)}}\otimes\ket{k_L,0}$,  indicate the same electronic 
state but with one emitted photon.
We can denote the occupation probabilities of the latter states 
as $p_{g_{R1}}, p_{g_{R2}}, p_{g_{L1}}, p_{g_{L2}}$,
using a single index 
since the state of the emitted photon can be inferred from the electronic state, see diagram Fig.~\ref{fig:8}.

The states labelled  $\ket{G_{1}}\,,\ket{G_{2}}$ - Eqs.~(\ref{eq:G_1},\ref{eq:G_2}) -
correspond to the two degenerate ground states of the whole two DQDs system.

Referring again to Fig.~\ref{fig:8}, the  states labeled by the tensor product 
$\ket{G_{1(2)}}\otimes\ket{k_L,0}$ and $\ket{G_{1(2)}}\otimes\ket{0,k_R}$, 
indicate 
that the electronic system has relaxed to one of the two ground states  
with one photon emitted in either the left $(k_L)$ or right $(k_R)$ transmission line, and one phonon also emitted.
Similarly, the state $\ket{G_{1(2)}}\otimes\ket{k_L,k_R}$ indicates the 
ground states with two emitted photons.
We   use the notation $p_{G_1}$ and $p_{G_2}$ with a single index for 
the occupation probabilities of the latter states 
with two emitted photons, Fig.~\ref{fig:8}.

We start by writing the equations for the occupation probability for the Andreev bound states $\ket{A_{\pm}}$
\begin{equation}
\frac{d p_{\pm}}{dt} 
= - \left( 
R_{\pm}^{(phot)} 
+ 
R_{\pm}^{(phon)} 
\right) 
p_{\pm} \label{eq:p_pm_decay}
\end{equation}
with 
\begin{align}
R_{\pm}^{(phot)} 
& = 
W_{\pm \rightarrow g_{R1}}^{(phot)} + 
W_{\pm \rightarrow g_{R2}}^{(phot)} + 
W_{\pm \rightarrow g_{L1}}^{(phot)} + 
W_{\pm \rightarrow g_{L2}}^{(phot)} 
\\
R_{\pm}^{(phon)}
& = 
W_{\pm \rightarrow g_{R1}}^{(phon)} + 
W_{\pm \rightarrow g_{R2}}^{(phon)} +  
W_{\pm \rightarrow g_{L1}}^{(phon)} + 
W_{\pm \rightarrow g_{L2}}^{(phon)} 
\end{align}
where the rates 
$W_{\pm \rightarrow x}^{(phot)} $ 
and
$W_{\pm \rightarrow x}^{(phon)} $ 
are the rates associated to the processes 
in which one of the two Andreev doublet $\ket{A_{\pm}}$ decays to 
the final electronic state $x$  (where $x = g_{R1}, g_{R2}, g_{L1}, g_{L2}$)   
with one photon $(phot)$  or one phonon $(phon)$ emitted, respectively
(see diagram Fig.~\ref{fig:8}).

Assuming the initial time $t_0=0$, the solutions of the Eq.~(\ref{eq:p_pm_decay}) are straightforward
\begin{equation}
p_{\pm}(t) = e^{-\left( R_{\pm}^{(phot)} + R_{\pm}^{(phon)}  \right)  t} p_{\pm} \left( 0 \right) \, . \label{eq:p_pm_sol}
\end{equation}

We now consider the rate equations for the occupation probability $p_{g_{R1}}, p_{g_{R2}}, p_{g_{L1}}, p_{g_{L2}}$ 
associated to the 
states 
$\ket{g_{R1(R2)}}\otimes\ket{0,k_R}$,  and 
$\ket{g_{L1(L2)}}\otimes\ket{k_L,0}$,  in which one photon has been emitted. 
These equations are given by
\begin{align}
\frac{d p_{g_{R1}} }{dt} \!&= \!
\sum_{s=\pm}  p_{s}  W_{s\rightarrow g_{R1}}^{(phot)}  
-  p_{g_{R1}} 
\left( 
W_{g_{R1}  \rightarrow G_1}^{(phot)} 
+ 
W_{g_{R1}  \rightarrow G_{1} }^{(phon)} 
\right)   , \label{eq:pg_R1} \\
\frac{d p_{g_{R2}} }{dt} \!&= \! 
\sum_{s=\pm}  p_{s}  W_{s\rightarrow g_{R2}}^{(phot)} 
- p_{g_{R2}} \left( 
W_{g_{R2} \rightarrow G_2}^{(phot)}  
+ 
W_{g_{R2} \rightarrow G_{2} }^{(phon)}   
\right)   , \label{eq:pg_R2} \\
\frac{d p_{g_{L1}}}{dt}  \!&= \!
\sum_{s=\pm}  p_{s}  W_{s\rightarrow g_{L1}}^{(phot)} 
- p_{g_{L1}}   \left( 
W_{g_{L1} \rightarrow G_1}^{(phot)}  
+ 
W_{g_{L1} \rightarrow G_{1} }^{(phon)}   \right)  , \label{eq:pg_L1} \\
\frac{d  p_{g_{L2}}  }{dt}  \!&= \!   
\sum_{s=\pm}  p_{s}  
W_{s\rightarrow g_{L2}}^{(phot)} 
-  p_{g_{L2}} \left(
W_{g_{L2} \rightarrow G_2}^{(phot)}  
+ 
W_{g_{L2}  \rightarrow G_{2} }^{(phon)}  \right)  \label{eq:pg_L2}  . 
\end{align}

The rates $W_{x \rightarrow G_{1(2)}}^{(phot)}$ 
 (where $x = g_{R1}, g_{R2}, g_{L1}, g_{L2}$) 
describe the photon emission processes  
from the initial states 
$\ket{x}\otimes \ket{k_{L},0}$ or $\ket{x}\otimes \ket{0,k_R}$
to the final states 
$\ket{G_{1(2)}}\otimes \ket{k_{L},k_R}$.
In this case, the initial states have only one of the two DQDs 
in the excited state - since   one photon has been already emitted - and the final states corresponding to one of the two possibles electronic ground states, see Fig.~\ref{fig:8}.

Similarly, the rates $W_{x \rightarrow G_{1(2)}}^{(phon)}$ 
describe the  radiationless processes with one emitted phonon 
from the initial states 
$\ket{x}\otimes \ket{k_{L},0}$ or $\ket{x}\otimes \ket{0,k_R}$
towards the final states 
$\ket{G_{1(2)}}\otimes \ket{0,k_R}$ or 
$\ket{G_{1(2)}}\otimes \ket{k_L,0}$.
In this case,  the final states correspond to one of the two possibles electronic ground states with only one emitted photon at the end, see Fig.~\ref{fig:8}.

With the boundary conditions $p_{g_{R1}}(0) = p_{g_{R2}}(0) =p_{g_{L1}}(0) = p_{g_{L2}}(0) =0$, 
the solutions of the Eqs.~(\ref{eq:pg_R1}), (\ref{eq:pg_R2}), (\ref{eq:pg_L1}), (\ref{eq:pg_L2}) are simply
\begin{align}
p_{g_{R1}} (t) &
= 
\!\!\! \int^t_0\!\!\!\! d t' 
e^{- R_{R1} \left(t-t' \right)}  \sum_{s=\pm} p_{s}(t')  W_{s\rightarrow g_{R1}}^{(phot)}    
\label{eq:pg_R1_sol}  \,,  \\
p_{g_{R2}} (t) &
= 
\!\!\! \int^t_0\!\!\!\! d t' 
e^{- R_{R2} \left(t-t' \right)}   \sum_{s=\pm} p_{s}(t')  W_{s\rightarrow g_{R2}}^{(phot)}   
\label{eq:pg_R2_sol}  \,,  \\
p_{g_{L1}} (t) &
= 
\!\!\! \int^t_0\!\!\!\! d t' 
e^{- R_{L1} \left(t-t' \right)}   \sum_{s=\pm} p_{s}(t')  W_{s\rightarrow g_{L1}}^{(phot)}    
\label{eq:pg_L1_sol}  \,,  \\
p_{g_{L2}} (t)  &
= 
\!\!\! \int^t_0\!\!\!\! d t' 
e^{- R_{L2} \left(t-t' \right)}  \sum_{s=\pm} p_{s}(t')  W_{s\rightarrow g_{L2}}^{(phot)}    
 \label{eq:pg_L2_sol}   \,, 
\end{align}
where we set 
\begin{align}
R_{R1} &= W_{g_{R1} \rightarrow G_1}^{(phot)}  
+ 
W_{g_{R1} \rightarrow G_{1R} }^{(phon)}   \,,  \\
R_{R2} &= W_{g_{R2} \rightarrow G_2}^{(phot)}  
+ 
W_{g_{R2} \rightarrow G_{2R} }^{(phon)}   \,,  \\
R_{L1} &= 
W_{g_{L1} \rightarrow G_1}^{(phot)}  + 
W_{g_{L1}  \rightarrow G_{1L} }^{(phon)}   \,,  \\
R_{L2} &= W_{g_{L2} \rightarrow G_2}^{(phot)}  
+ 
W_{g_{L2} \rightarrow G_{2L} }^{(phon)}    \,.
\end{align}

Finally, the rate equations for the occupation probabilities of the two ground states of the two DQDs with two emitted photons  are 
\begin{align}
\frac{d p_{G_1} }{dt}&=   p_{g_{R1}}  
W_{g_{R1} \rightarrow G_1}^{(phot)} 
+ p_{g_{L1}}  
W_{g_{L1} \rightarrow G_1}^{(phot)}  \, , \\
\frac{d p_{G_2} }{dt}&=   p_{g_{R2}}  
W_{g_{R2} \rightarrow G_2}^{(phot)} 
+ p_{g_{L2}}  
W_{g_{L2} \rightarrow G_2}^{(phot)}  \,. 
\end{align}

We note that  transitions from the states 
$\ket{x}\otimes\ket{0,0}$ ($x=g_{R1}, g_{R2}, g_{L1}, g_{L2}$),
to the electronic ground states with two photons emitted,
$\ket{G_{1(2)}}\otimes\ket{k_L,k_R}$, 
are energetically forbidden. 

We are interested in the final probability that the electronic states relaxes in one of the two ground states and two photons have been emitted, 
i.e. assuming the final time $t\rightarrow \infty$.
We denote this quantity $P_{2ph}$ and it reads
\begin{align}
P_{2-ph}  
&= \lim_{t\rightarrow\infty} \left( p_{G_1}(t)+ p_{G_2}(t)  \right) \nonumber \\
&
=
\!\!\! \int^{\infty}_0\!\!\!\! d t
\big[ 
 p_{g_{R1}}(t)  W_{g_{R1} \rightarrow G_1}^{(phot)} 
 + 
 p_{g_{L1}}(t)  W_{g_{L1} \rightarrow G_1}^{(phot)} 
 \nonumber \\
 &
 +
 p_{g_{R2}}(t) W_{g_{R2} \rightarrow G_2}^{(phot)} 
 + 
 p_{g_{L2}}(t)  W_{g_{L2} \rightarrow G_2}^{(phot)}
\big] 
\, . \label{eq:P_2_ph}
\end{align}
Inserting Eqs.~(\ref{eq:p_pm_sol}) into 
Eqs.~(\ref{eq:pg_R1_sol}), (\ref{eq:pg_R2_sol}), (\ref{eq:pg_L1_sol}),  (\ref{eq:pg_L2_sol}), 
one obtains the solutions for $p_{g_{R1}}(t), p_{g_{R2}}(t),  p_{g_{L1}}(t),  p_{g_{L2}}(t)$ 
and by integration of  Eq.~(\ref{eq:P_2_ph}) one get $P_{2-ph}$.
After some straightforward algebra, the final result reads
\begin{align}
P_{2-ph} 
&=  
\sum_{s=+,-}
p_{s}
\left( \frac{1}{R_s^{(phot)} + R_s^{(phon)}}  \right) \times \nonumber\\
&
\left[
W_{s\rightarrow g_{R1}}^{(phot)} 
\left(
\frac{  
W_{g_{R1} \rightarrow G_1}^{(phot)}   
}{
W_{g_{R1} \rightarrow G_1}^{(phot)}     
+ W^{(phon)}_{g_{R1} \rightarrow G_{1}} }
\right)
\right. 
\nonumber\\
&
+
W_{s\rightarrow g_{L1}}^{(phot)} 
\left(
\frac{  
W_{g_{L1} \rightarrow G_1}^{(phot)}   
}{
W_{g_{L1} \rightarrow G_1}^{(phot)}     
+ W^{(phon)}_{g_{L1} \rightarrow G_{1}} }
\right)
\nonumber\\
&
+
W_{s\rightarrow g_{R2}}^{(phot)} 
\left(
\frac{  
W_{g_{R2} \rightarrow G_2}^{(phot)}   
}{
W_{g_{R2} \rightarrow G_2}^{(phot)}     
+ W^{(phon)}_{g_{R2} \rightarrow G_{2}} }
\right)
\nonumber\\
&
\left.
+
W_{s\rightarrow g_{L2}}^{(phot)} 
\left(
\frac{  
W_{g_{L2} \rightarrow G_2}^{(phot)}   
}{
W_{g_{L2} \rightarrow G_2}^{(phot)}     
+ W^{(phon)}_{g_{L2} \rightarrow G_{2}} }
\right)
\right] \, .
\label{eq:P2ph}
\end{align}

The final formula has a simple interpretation.
The term $( R_s^{(phot)} + R^{(phon)}_s )$  is the sum of the two kind of possible processes in which, 
starting from the Andreev doublet, the electronic system can emit a single photon or can decay via a phonon emission,  
ending up in a state in which only one of the two DQDs is still in the excited state.\\
The terms of the type $\sim W_{x \rightarrow G_{1(2)}}^{(phot)} / (W_{x \rightarrow G_{1(2)}}^{(phot)}     
+ W^{(phon)}_{x \rightarrow G_{1(2)}}  )$
expresses the competition between the two possible processes in which, 
starting from the electronic state with only one DQD in an excited state (after having emitted one photon) 
the electronic system finally decays into one of the two ground states,  
by emitting the second photon or by emitting a phonon.

%
%
%
%
\section{Rates for the phonon emission processes}
\label{app:G}

The  model Hamiltonian for the DQD  interacting with phonons can be written as \cite{Hofmann:2020} 
\begin{equation}
\hat{H}_{DD}^{(ph)}
=
\sum_q \hbar \omega_q b_{q}^{\dagger }  b_{q}^{\phantom{\dagger}}
+
\sum_{q} \lambda_q \left( b_{q}^{\phantom{\dagger}} + b_{-q}^{\dagger }  \right)   \left( \hat{n}_1  + e^{i q d} \hat{n}_2    \right) 
\,,  
\end{equation}
where $ b_{q}^{\dagger }$ and $b_{q}^{\phantom{\dagger}}$ are the bosonic operators for the phonons with wavevectors $q$ and frequencies $\omega_q$,
$d$ is the average distance between the two QDs forming the DQD.
We consider a DQD in the limit in which the energy detuning is larger than the tunneling amplitude $|\delta| \gg |t_{12}|$.
Assuming $\delta  > 0 $,  the high energy state is weakly hybridized and it essentially corresponds to the state on the QD $1$ 
whereas the low energy state is weakly hybridized too and one can identify  it with the state on the second QD $2$.
In this regime, Fermi's golden rule  for the transition rate for the energy decay from the high energy state to the low energy state  yields 
\begin{equation}
\Gamma_{rel}^{(phon)} \left( \delta \right) 
= {\left( \frac{ t_{12} }{  \delta } \right)}^2 
J_{phon} \left(  \delta \right)   \, , 
\label{eq:Gamma_rel_1_app}
\end{equation}
corresponding to Eq.~(\ref{eq:Gamma_rel_1}) in the main text, 
with the interaction spectral density $J_{phon} (\varepsilon)$ given by
\begin{equation}
J_{phon} \left(  \varepsilon \right)   = \frac{\pi}{2\hbar^2} \sum_q  {\left| \lambda_q \right|}^2   {\left| 1 - e^{i q d } \right|}^2 \delta\left( \varepsilon - \hbar\omega_q \right) \,.
\end{equation}
Another possible model for the energy relaxation is based on the coupling between the DQD's dipole moment and the local fluctuating electric field due to   charge noise.
The corresponding Hamiltonian can be written as 
\begin{equation}
\hat{H}_{DD}^{(ch)}
=
\hat{H}_{(ch)} + \,  \lambda_{d}  \,\, \delta\hat{E} \,\, \left( \hat{n}_1  - \hat{n}_2  \right) 
\,.
\end{equation}
where $\hat{H}_{(ch)}$ is the (unknown) Hamiltonian of the fluctuating charges 
and $\lambda_{d} $ is proportional to the dipole momennt of the DQD. 
The transition rate for the energy decay from the high energy state  
to the low energy state  can be calculated using  Fermi-golden rule and, in the regime $|\delta| \gg |t_{12}|$, we obtain
\begin{equation}
\Gamma_{rel}^{(ch)} 
\left( \delta \right) = {\left( \frac{ t_{12}}{ \delta } \right)}^2 \, 
J_{ch} \left( \varepsilon \right) 
\label{eq:Gamma_rel_2_app}
\end{equation}
with 
$J_{ch} \left( \varepsilon \right)  =  
\lambda^2_{d} S_{ch} \left( \varepsilon \right) $ 
the interaction spectral density associated to the   local fluctuating electric field and
the noise power spectrum\cite{Clerk:2010}  of the operators $ \delta\hat{E}$
\begin{equation}
S_{ch} \left( \varepsilon \right)   
= 
\frac{\lambda^2_{d} }{\hbar^2}
\int^{+\infty}_{-\infty} \!\!\! dt'  
\,\, \left< \delta\hat{E}(t)  \delta\hat{E}(0)   \right> 
\,\, e^{i\frac{ \varepsilon }{\hbar} t} \,.
\end{equation}
Notice that, according our notation, $J_{ch} ( \varepsilon )   $ as well as  $J_{phon}(\varepsilon)$ have dimensions of a rate.
From comparison of Eq.~(\ref{eq:Gamma_rel_1}) with Eq.~(\ref{eq:Gamma_rel_2}), we see that the prefactor ${( t_{12}/\delta )}^2$ is the same: 
this factor arises from the matrix element between the two states of the DQD.
Hence, independent of the microscopic origin, one can expect that $\Gamma_{rel}(\varepsilon) \approx {( t_{12}/\varepsilon)}^2 J(\varepsilon)$ where 
$J(\omega)$ is a function that can be measured experimentally by studying the energy relaxation as a function of the DQD detuning.
In other words, $J(\varepsilon)$ is a place holder either for $J_{phon}(\varepsilon)$ or $J_{ch}(\varepsilon)$. 
Here we analyze the decay rates associated to the phonon emission as experimental data are available to obtain $J_{phon}(\varepsilon)$, Ref.~\onlinecite{Hofmann:2020}.

\begin{figure}[btph]
 \centering
   (a)\phantom{\hspace{0.85\columnwidth}}\\[-12pt]
 \includegraphics[width=0.8\columnwidth]{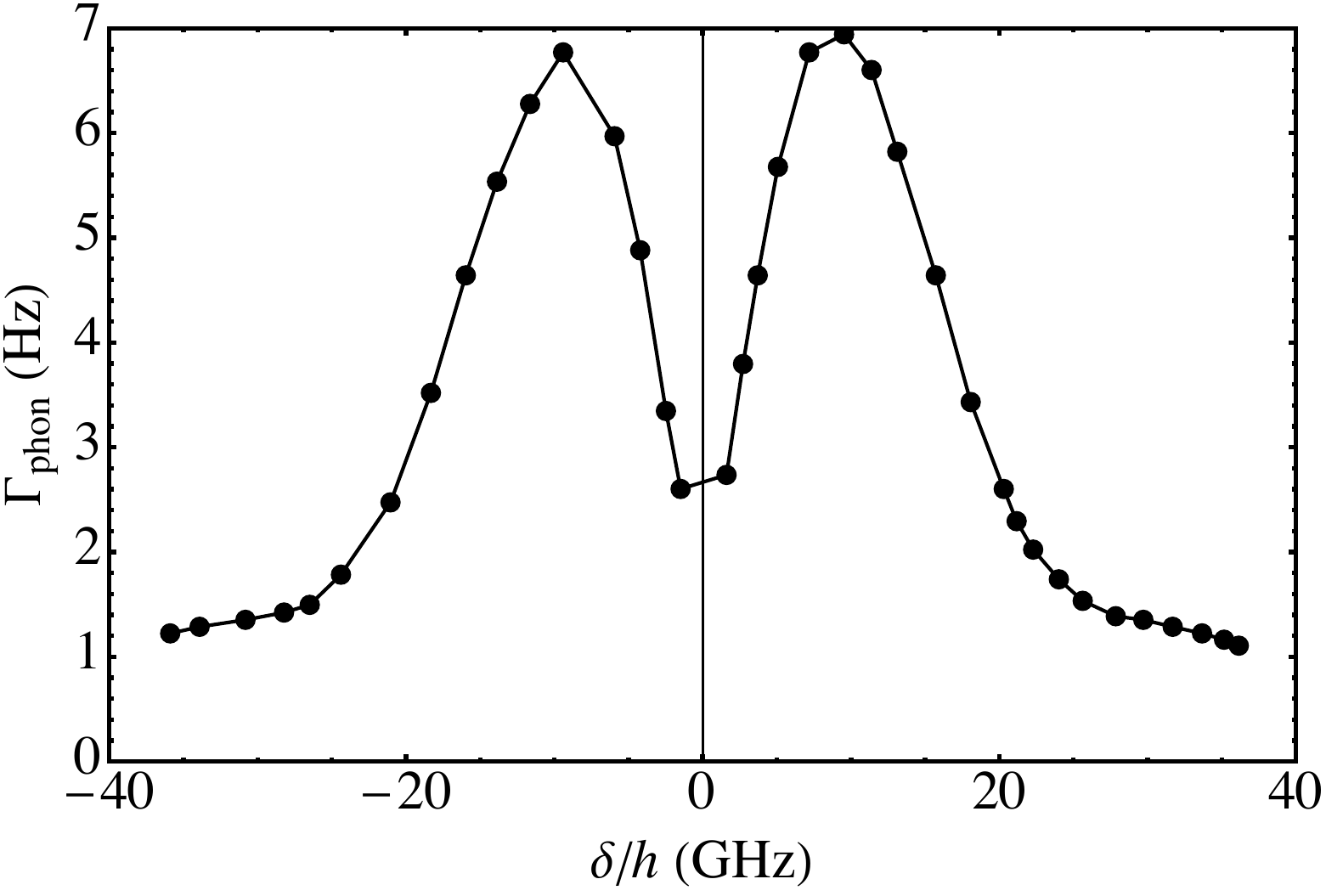}\\[12pt]
   (b)\phantom{\hspace{0.85\columnwidth}}\\[-12pt]
  \includegraphics[width=0.85\columnwidth]{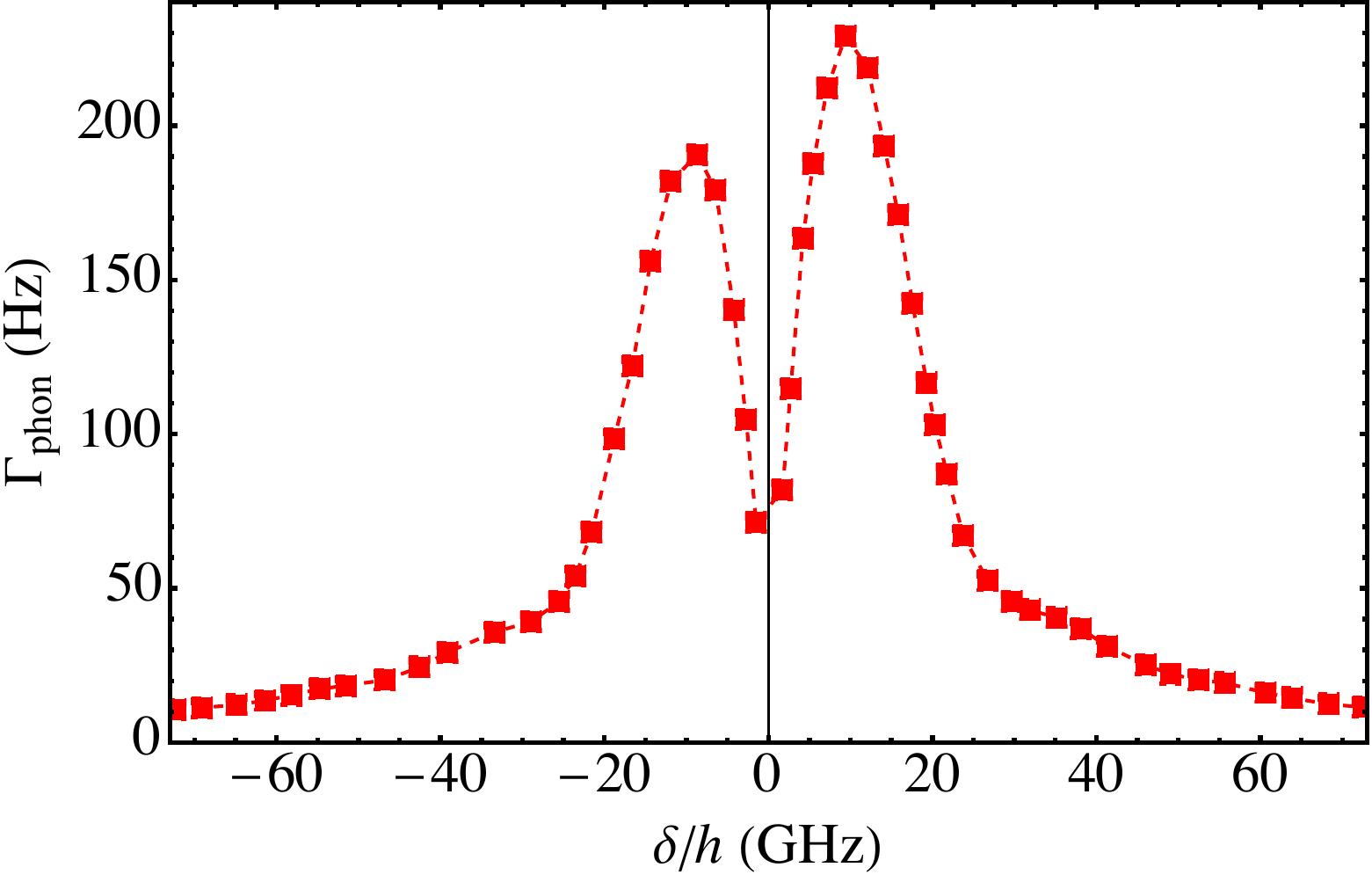}\\[12pt]
   (c)\phantom{\hspace{0.85\columnwidth}}\\
  \includegraphics[width=0.85
\columnwidth]{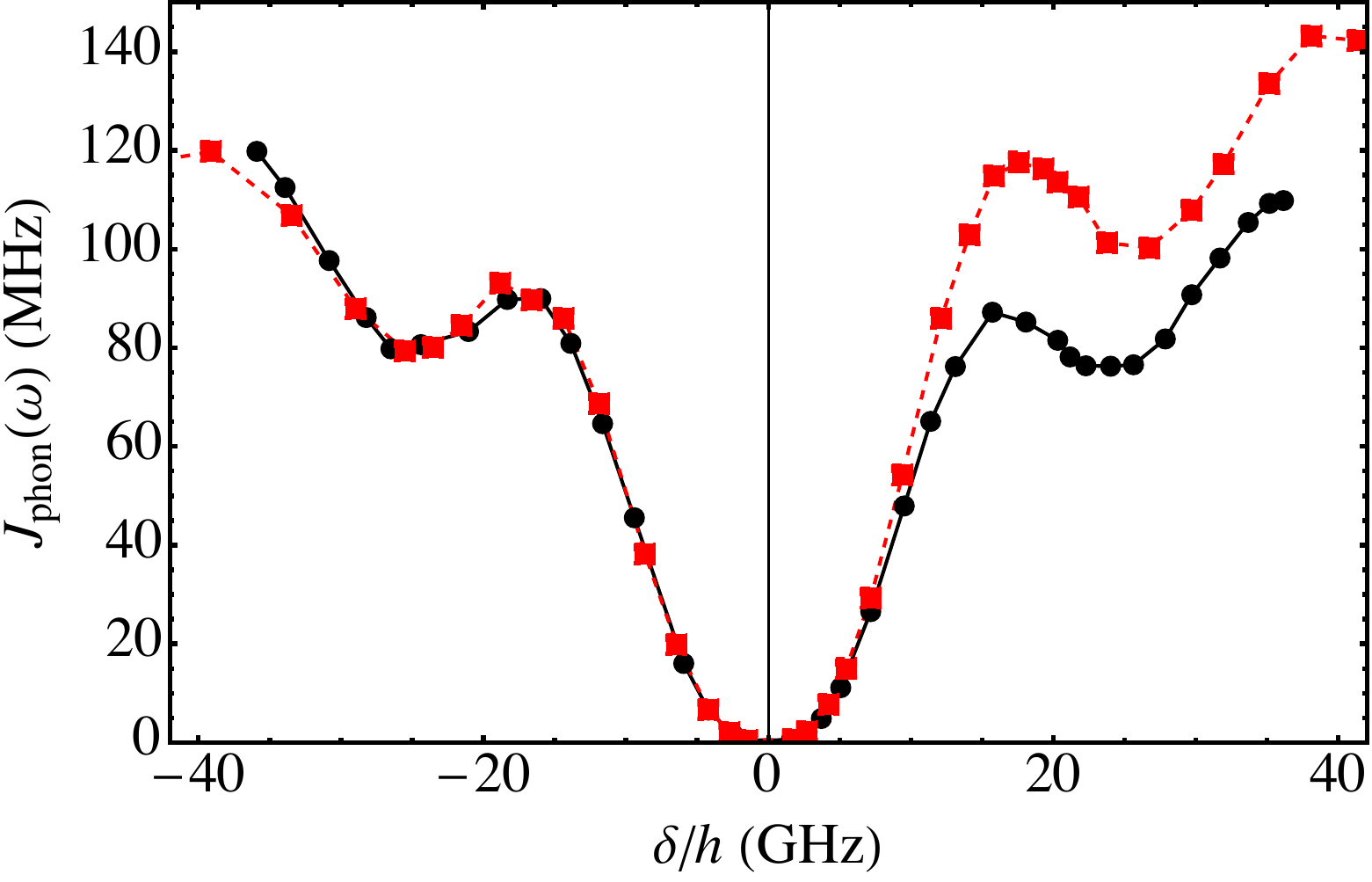}\\[3pt]
\caption{{\bf (a)} 
Phonon assisted decay rate $\Gamma_{rel}^{(phon)}$  of the DQD charge qubit 
measured in Ref.~\onlinecite{Hofmann:2020} 
with $t_{12} = 15$ neV.
{\bf (b)} 
Phonon assisted decay rate  $\Gamma_{rel}^{(phon)}$  of the DQD charge qubit 
measured in Ref.~\onlinecite{Hofmann:2020}  
with $t_{12} = 80$ neV. 
{\bf (c)} 
Extracted interaction spectral density $J_{phon}\left( \varepsilon\right)$ by inverting Eq.~(\ref{eq:Gamma_rel_1_app}).
}
\label{fig:11}
\end{figure}

In Fig.~\ref{fig:11}(a) and Fig.~\ref{fig:11}(b)
we report the experimental data for $\Gamma_{rel}^{(phon)} \left( \delta \right) $ 
measured in Ref.~\onlinecite{Hofmann:2020} for two different values of the tunneling 
amplitude $t_{12}$.
Using these curves, one can reconstruct the interaction spectral density 
associated to the phonons, $J_{phon}\left( \varepsilon\right)$, shown in Fig.~\ref{fig:11}c.
Although the latter function is expected to be independent of $t_{12}$, the discrepancies of the two curves in Fig.~\ref{fig:11}(c) for positive values of the detuning is due to measurement errors \cite{Hofmann:2020}.

%
%
%
%
%
%
%
%
%

In the diagram shown in Fig.~\ref{fig:8} of the main text, we report all the  possible photon  assisted decay 
processes  and the relevant phonon  assisted decays 
for the CPS formed by two DQDs.

We recall that the radiationless relaxation processes 
with no emitted photon 
$g_{R1} \rightarrow G_{1}$,  $g_{R2} \rightarrow G_{2}$,  $g_{L1} \rightarrow G_{1}$ and
$g_{L2} \rightarrow G_{2}$ are simply the relaxation processes occurring in one of the two QDs  as the photons state and the other QD state remain unchanged, namely
\begin{align*}
&{\left( g_{R1} \longrightarrow G_{1}  \right)}_{phon}=\\ 
& = \ket{L,+,\downarrow;R, -,\uparrow}\otimes{\ket{0,k_R}}\rightarrow \ket{L,-,\downarrow;R, -,\uparrow}\otimes{\ket{0,k_R}},\\
&=
\ket{L,+,\downarrow}\longrightarrow \ket{L,-,\downarrow}
\end{align*}
\begin{align*}
&{\left( g_{R2} \longrightarrow G_{2} \right)}_{phon} =\\
&=\ket{L,+,\uparrow;R, -,\downarrow}\otimes{\ket{0,k_R}}\rightarrow \ket{L,-,\uparrow;R, -,\downarrow}\otimes{\ket{0,k_R}},\\
&=
\ket{L,+,\uparrow}\longrightarrow \ket{L,-,\uparrow}
\end{align*}
\begin{align*}
&{\left( g_{L1} \longrightarrow G_{1L} \right)}_{phon} =\\
&=
\ket{L,-,\downarrow;R, +,\uparrow}\otimes{\ket{k_L,0}}\rightarrow \ket{L,-,\downarrow;R, -,\uparrow}\otimes{\ket{k_L,0}},\\
&=
\ket{R, +,\uparrow} \longrightarrow 
\ket{R, -,\uparrow} ,\\
\end{align*}
\begin{align*}
&{\left( g_{L2} \longrightarrow G_{2L} \right)}_{phon}= \\
&=\ket{L,-,\uparrow;R, +,\downarrow}\otimes{\ket{k_L,0}}\rightarrow \ket{L,-,\uparrow;R, -,\downarrow}\otimes{\ket{k_L,0}} \\
&=
\ket{R, +,\downarrow} \longrightarrow 
\ket{R, -,\downarrow}.
\end{align*}
Using Fermi's golden rule, we can write immediately the transition rates as
\begin{align}
W^{(phon)}_{ g_{R1} \rightarrow G_{1} }  &=  {\left(  \frac{ t_{12} }{\delta + \Delta \varepsilon_m}   \right)}^2 J_{phon} \left( \delta + \Delta\varepsilon_m  \right) \, , \label{eq:w_rate_1}\\ 
W^{(phon)}_{ g_{R2} \rightarrow G_{2} }  &=  {\left(  \frac{ t_{12} }{\delta - \Delta \varepsilon_m}   \right)}^2 J_{phon} \left( \delta - \Delta\varepsilon_m  \right) \, , \label{eq:w_rate_2}\\
W^{(phon)}_{ g_{L1} \rightarrow G_{1} }  &=  {\left(  \frac{ t_{12} }{\delta - \Delta \varepsilon_m}   \right)}^2 J_{phon} \left( \delta - \Delta\varepsilon_m  \right) \, , \label{eq:w_rate_3} \\
W^{(phon)}_{ g_{L2} \rightarrow G_{2} }  &=  {\left(  \frac{ t_{12} }{\delta + \Delta \varepsilon_m}   \right)}^2 J_{phon} \left( \delta + \Delta\varepsilon_m  \right)  \, .
\label{eq:w_rate_4}
\end{align}
In addition, there are other four assisted phonon decay processes, in which the electronic system relaxes from one of the two Andreev bound states $\ket{A_{+}}$ and $\ket{A_{-}}$  towards an uncorrelated state in which 
only one of the two DQDs has been relaxed 
\begin{align*}
&
{\left( A_{\pm} \longrightarrow g_{1R}  \right)}_{phon}
=
\\
&=\ket{A_{\pm}}\otimes\ket{0,0}\rightarrow \ket{L,+,\downarrow;R,-, \uparrow}\otimes\ket{0,0},\\
& 
{\left( A_{\pm} \rightarrow g_{2R} \right)}_{phon}
=\\
&=\ket{A_{\pm}} \otimes\ket{0,0}\rightarrow   \ket{L, +,\uparrow; R,-, \downarrow}\otimes\ket{0,0}  ,\\
& {\left( A_{\pm} \longrightarrow g_{1L} \right)}_{phon} =\\
&=\ket{A_{\pm}} \otimes\ket{0,0}\rightarrow  \ket{L, -,\downarrow;R,+, \uparrow} \otimes\ket{0,0}  , \\
%
& {\left( A_{\pm} \longrightarrow g_{2L}  \right)}_{phon} = \\
&=\ket{A_{\pm}}\otimes\ket{0,0} \rightarrow  \ket{L, -,\uparrow;R,+, \downarrow}\otimes\ket{0,0}   \, .
%
\end{align*}
In order to compute these rates we need the matrix elements of the operator 
$\hat{n}_{1,\nu} + e^{i q d} \,\, \hat{n}_{2,\nu}$.
For example,  for the relaxation in the right DQD we find
\begin{align}
& 
\bra{A_{\pm}} \big[ \hat{n}_{1,\nu} + e^{i q d} \hat{n}_{2,\nu}  \big] \ket{g_{1R}}
= \nonumber \\
&
-\frac{1}{2} \,\,   \left[  \frac{1 -  e^{i q d}}{2}   \right] \left( \frac{t_{12}}
{\delta+ \Delta\varepsilon_m} \right)  .
\end{align}
Since the factor $(1 -  e^{i q d})/2$ can be included in the definition of $J_{phon}(\varepsilon)$, 
in the limit $\Gamma_S \ll (\delta \pm \Delta\varepsilon_m)$, we find
\begin{equation}
W^{(phon)}_{\pm \rightarrow g_{1}} = \frac{1}{4} {\left| \frac{t_{12}}{\delta- \Delta\varepsilon_m} \right|}^2 
J_{phon}\left(  \delta- \Delta\varepsilon_m \right) \,  .\label{eq:w_rate_5}
\end{equation}
In a similar way, we obtain for the other rates
\begin{align}
W^{(phon)}_{\pm \rightarrow g_{2}} &= \frac{1}{4} 
{\left| \frac{t_{12}}{\delta+\Delta\varepsilon_m} \right|}^2 J_{phon}\left(  \delta+ \Delta\varepsilon_m \right) \, ,\label{eq:w_rate_6} \\
W^{(phon)}_{\pm \rightarrow g_{1}} &= \frac{1}{4} {\left| \frac{t_{12}}{\delta+ \Delta\varepsilon_m} \right|}^2 J_{phon}\left(  \delta+ \Delta\varepsilon_m \right) \, ,\label{eq:w_rate_7} \\
W^{(phon)}_{\pm \rightarrow g_{2}} &= \frac{1}{4} {\left| \frac{t_{12}}{\delta- \Delta\varepsilon_m} \right|}^2 J_{phon}\left(  \delta- \Delta\varepsilon_m \right) \, .  \label{eq:w_rate_8}
\end{align}
Comparing Eqs.~(\ref{eq:w_rate_1}), (\ref{eq:w_rate_2}), (\ref{eq:w_rate_3}), (\ref{eq:w_rate_4}) with Eqs.~(\ref{eq:w_rate_5}), (\ref{eq:w_rate_6}), (\ref{eq:w_rate_7}), (\ref{eq:w_rate_8}),
we notice an additional factor $1/4$ that stems from the superposition of the singlet state with the empty state (factor $1/\sqrt{2}$) and from the definition of the singlet state (factor $1/\sqrt{2}$ in $\ket{S_{++}}$).

%
%
%
%
\section{\addGR{Entanglement measure of the two photon state}}
\label{app:H}
\addGR{
In this Appendix,  
we quantify the entanglement present in the density matrix Eq.~(\ref{eq:rho_tot}) of Sec.~\ref{subsec:calculations}
which is a mixture of the four possible frequency-entangled photon pairs.
}

\addGR{
Specifically, we calculate the logarithmic negativity \cite{Vidal:2002}, which corresponds to 
an upper bound for the distillable entanglement.
This is based on the Peres-Horodecki criterion (or positive partial transpose) \cite{Peres:1996,Horodecki:1996}, 
which 
states that if the density matrix of a bipartite system is separable (i.e. no entanglement is present but only classical correlations), 
then the partial transpose density matrix with respect to one of the two subsystems, has positive eigenvalues. 
Hence, the amount of negativeness of the eigenvalues of the partial transpose 
can be considered as a measure of the nonseparability between two systems, 
viz. entangled states are present. 
The quantity that we consider is: 
\begin{equation}
E_{\mathcal{N}} 
= 
\mbox{log}_2
\left( 
1 + 2 \sum_{\lambda_k < 0 } \left| \lambda_k \left[ \hat{\rho}^{T_R} \right] \right| 
 \right)
\end{equation}
with $ \lambda_k \left[ \hat{\rho}^{T_R} \right]  = \lambda_k \left[ \hat{\rho}^{T_L} \right] $ the negative eigenvalues 
of the 
$\hat{\rho}^{T_R} $ and  $\hat{\rho}^{T_L} $, 
the partial transpose density matrices with respect to the right or left photon, respectively.
}
\mbox{}\\

\addGR{
We consider the density matrix 
$\hat{\rho} = p_{+} \hat{\rho}_{+}  +  p_{-} \hat{\rho}_{-} $
with $p_{+}+ p_{-}=1$ and 
$\hat{\rho}_{\pm}= \ket{\Psi_{\pm,1}}\bra{\Psi_{\pm,1}} +  \ket{\Psi_{\pm,2}}\bra{\Psi_{\pm,2}} $,
and the states normalized as 
$\left< \Psi_{\pm,i} | \Psi_{\pm,i} \right>=1/2$.
For the sake of convenience, we introduce a short-hand  notation for the frequencies and the states. We write the frequencies as
\begin{align}
\omega_{1} = \omega_{-} - \Gamma_S/\hbar  	\,, \quad  
\omega_{2} = \omega_{-}   				\,, \quad  
\omega_{3} = \omega_{-} + \Gamma_S/\hbar 
\\
\omega_{4} = \omega_{+} - \Gamma_S/\hbar  	\,, \quad 
\omega_{5} = \omega_{+}  				\,, \quad 
\omega_{6} = \omega_{+} + \Gamma_S/\hbar 
\end{align}
and the four frequency-entangled photon states as (see Fig~\ref{fig:NEW})
\begin{eqnarray}
\ket{\Psi_{+,2}} & =\frac{1}{2} \left(  \ket{2; 6} +  \ket{3; 5}  \right)	\\
\ket{\Psi_{+,1}} & =\frac{1}{2} \left(   \ket{6; 2} + \ket{5; 3}   \right) 	\\
\ket{\Psi_{-,2}}& =\frac{1}{2} \left(  \ket{1; 5} +  \ket{2; 4}  \right) 		\\
\ket{\Psi_{-,1}} & =\frac{1}{2} \left(  \ket{5; 1} +  \ket{4; 2}  \ \right) 
\end{eqnarray}
where $\ket{i;j}$ is the state with the left photon having frequency $\omega_i$ and the right one with $\omega_j$. Notice that $\langle i';j'\ket{i;j}=\delta_{i',i}\delta_{j',j}$.
}
%
%
%
%
%
\begin{figure}[btph]
 \includegraphics[width=1.\columnwidth]{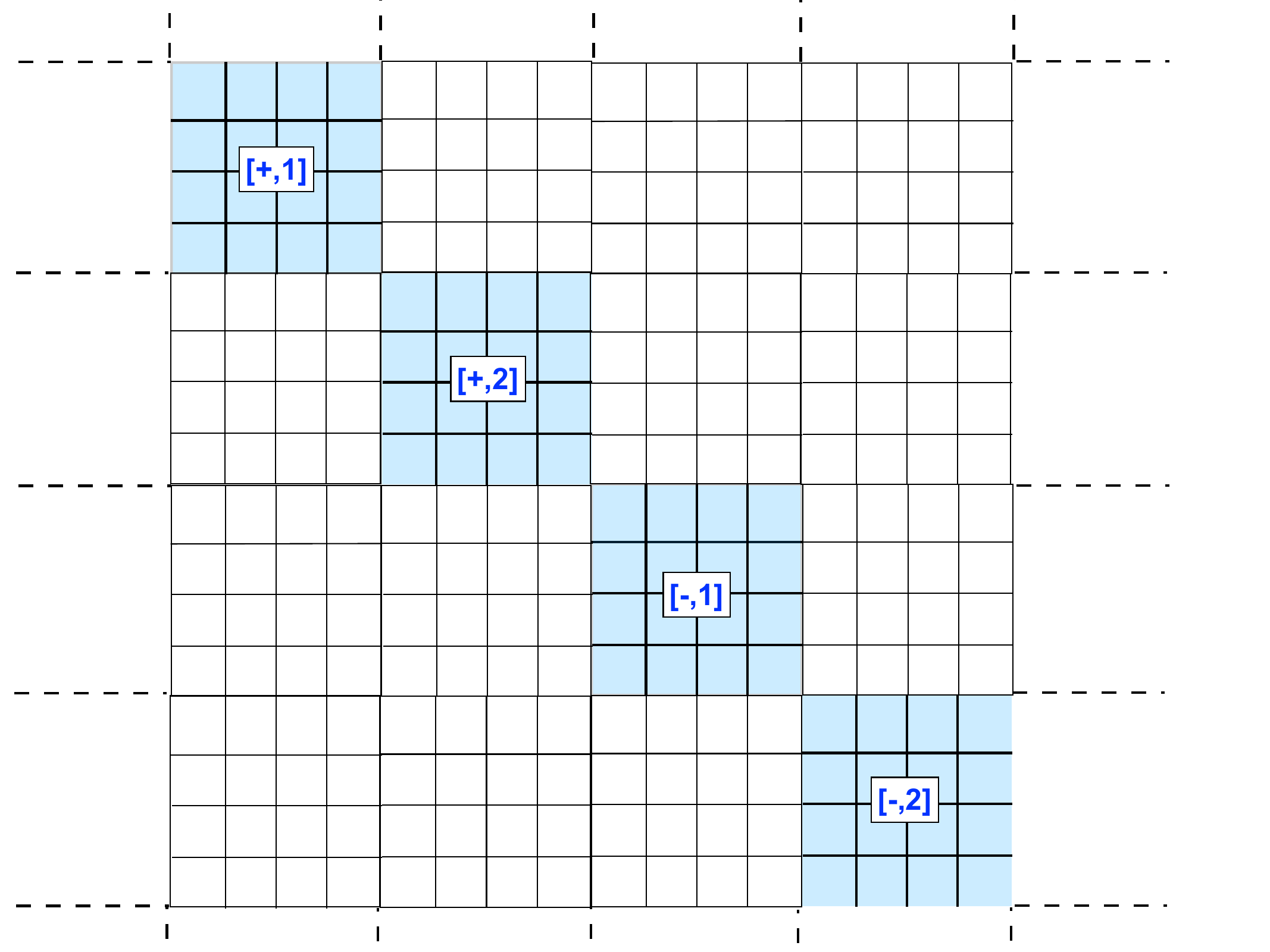}
\caption{
\addGR{
Ordering the two-photon states appropriately, the density matrix $\hat{\rho} = p_{+} \hat{\rho}_{+}  +  p_{-} \hat{\rho}_{-} $ 
can be brought in a block diagonal form (see text). 
Each block is associated with one $\ket{\Psi_{s,n}}$ of the four frequency-entangled photon states with indices $s=\pm$ and $n=1,2$.
}
}
\label{fig:block-diagonal}
\end{figure}
%
%
%
%
%

\addGR{
In the subspace given by all the possible combinations of the 6 frequencies, 
we have 36 possible states but the majority of them do not appear 
in $\hat{\rho}$.
Then, we can order the two photon states so that the density matrix is block diagonal, 
Fig.~\ref{fig:block-diagonal}.
This stems from the fact that
the four frequency-entangled photon states, $\ket{\Psi_{s,n}}$ with $s=\pm$ and $n=1,2$, do not have in common any two-frequency photon states, $\ket{i;j}$, see Fig.~\ref{fig:NEW}.
}

\begin{figure}[btph]
 \includegraphics[width=0.8\columnwidth]{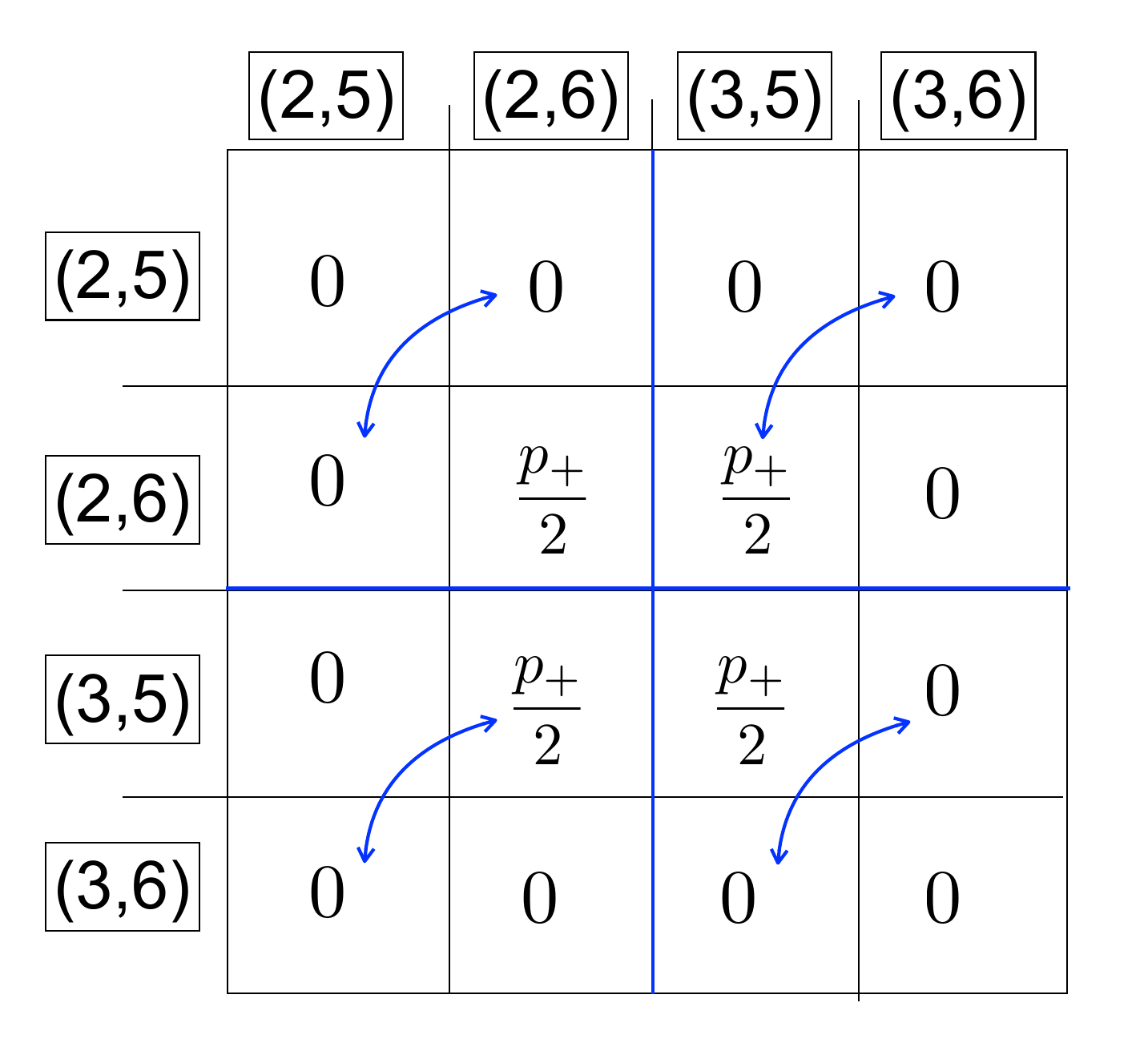}
\caption{
\addGR{
Example of the form of the block associated to the state $\ket{\Psi}_{+,2}$ in terms of the two photon states $\ket{i,j}$. 
In this case, there are four frequencies: the left photon can have frequencies $\omega_2$ and $\omega_3$, while the right photon can have frequencies $\omega_5$ and $\omega_6$. 
Curved arrows illustrate the process of exchanging matrix elements to generate the partial transpose.}
}
\label{fig:single-block}
\end{figure}

\addGR{
Then, the partial transposition with respect to the right photon can be carried out in each single block. The partial transposition with respect to the right photon is defined as $\rho_{ij,kl} \rightarrow \rho_{il,kj}$, with the notation $\bra{i;j} \hat{\rho}\ket{k;l}$.
It corresponds to the transposition of sub-blocks in which the indices for the left photon remain unchanged.
An example for the state $\ket{\Psi}_{+,2}$  is illustrated in Fig.~\ref{fig:single-block} where the sub-blocks are separated by blue lines.
}

\addGR{
The secular equations for the eigenvalues of the individual  blocks read 
\begin{equation}
\left( \lambda + \frac{p_{\pm}}{2}\right) {\left( \lambda - \frac{p_{\pm}}{2}\right)}^3 = 0
\end{equation}
which yields a single negative eigenvalue $p_{\pm}/2$.
The eigenvalues of the partial transpose are simply the eigenvalues of all the blocks.
Therefore,  the logarithmic negativity is given by
\begin{equation}
E_{\mathcal{N}} 
 =
 \mbox{log}_2
 \left( 
1  + 2  \left[ \frac{p_{+}}{2}   + \frac{p_{+}}{2}    +  \frac{p_{-}}{2}   +  \frac{p_{-}}{2}      \right]
 \right)
 =  \mbox{log}_2 3 \, .
\end{equation}
}

%
%
%
%
%
%
%
%
%

%
%
%
%
\section{\addGR{Characterization and readout of the DQD}}
\label{app:I}
\addGR{The readout of each double quantum dot (DQD) can be performed using several approaches, as outlined below.
We remind that knowing the absolute number of extra electrons or holes in the DQD is unnecessary; rather, achieving an effective few-particle configuration is sufficient. Our ``empty state'' is an effective vacuum (with filled orbital shells being irrelevant), as has been recently demonstrated in some spin qubit experiments \cite{DePalma2023,Leon:2020}.}

\addGR{
{\bf(1)}
One approach involves using an additional microwave resonator with tunable coupling to each DQD \cite{Stockklauser:2017,Scarlino2019}.  After calibration and determination of the stability phase diagram of the two DQDs, the readout resonator must be decoupled from the DQD to ensure that its presence does not interfere with the scheme for entangled photon emission.
}

\addGR{
{\bf(2)}
Another option involves placing a charge sensor, such as a quantum point contact, next to the DQD, as extensively demonstrated by the spin qubit community \cite{Hanson:2007,Burkard:2023}. 
This sensor can be used to fully characterize the absolute number of extra particles in each QD and to calibrate the tunneling rates of the DQD. During the photon pair generation stage, the charge sensor can be switched off to prevent any interference with the entangled photon emission.
}

\addGR{
{\bf(3)}
A last possible way is to use DC transport.
Circuit QED experiments implemented with QDs  have demonstrated that DC and RF measurements can take place simultaneously without too much of a compromise (see Refs.~\cite{Cottet:2017review,Burkard:2020,Frey:2011,Delbecq:2011,Petersson:2012,Basset:2013,Braakman:2014} for example). 
%
By performing DC transport through QDs, one can obtain the charge stability diagram of the DQD and the induced superconducting gap in the proximitized semiconducting lead.
Additionally, by applying a high magnetic field, one can also access the normal state of the same lead.
The crucial and challenging point in these experiments is to obtain a sufficiently wide range of  tunnel couplings simply by tuning  gate voltages. For example, recently Ref.~\cite{Dvir2023} 
successfully demonstrates that full pinch-off is achievable with InSb nanowires in which the wavefunction of the confined particles is primarily localized in the core of the nanowire rather than on its surface.
This means that, in our scheme, 
in addition to the central nanocontact S shown in Fig.~\ref{fig:1}, each DQD can be laterally coupled to an additional normal contact D through a highly tunable tunnel barrier.
By applying a strong magnetic field, the central nanocontact transitions to a normal state, enabling single-electron transport between it and the extra normal contacts under an applied bias voltage.
This setup enables the construction of the phase diagram of the DQD and provides access to both the ground and excited states at each point on the diagram, with corresponding gate voltage values.
Subsequently, the magnetic field can be reduced to restore the superconducting phase of the central nanocontact, allowing the characterization of the CPS and the Andreev bound state via DC spectroscopy measurements. Once this step is complete, the tunneling barrier between each DQD and the normal contacts can be significantly increased, rendering it effectively irrelevant.
}

%
%
%
%
%
%
%
%
\bibliography{references.bib}

\end{document}